\documentclass[epsfig,12pt]{article}
\usepackage{makeidx}
\usepackage{amsmath}
\usepackage{amsfonts}
\usepackage{amssymb}
\usepackage{graphicx}

\input epsf.sty
\newtheorem{theorem}{Theorem}
\newtheorem{acknowledgement}[theorem]{Acknowledgement}

\makeatletter \@addtoreset{equation}{section}
\renewcommand{\theequation}{\thesection.\arabic{equation}}
\input{tcilatex}
\begin{document}

\title{\vspace{-2.5cm}\rightline{\mbox{\small {LPHE-MS-1507}} \vspace
{1cm}} \textbf{MSSM-like from }$SU_{5}\times D_{4}$ \textbf{Models}}
\author{R.Ahl Laamara$^{1,2}$, M. Miskaoui$^{1,2}$, E.H Saidi$^{2,3\thanks{%
e-mail: h-saidi@fsr.ac.ma}}$ \\
{\small 1. LPHE-Modeling and Simulations, Faculty Of Sciences, Mohamed V
University, Rabat, Morocco}\\
{\small 2. Centre of Physics and Mathematics, CPM- Morocco}\\
{\small 3. International Centre for Theoretical Physics, Miramare, Trieste,
Italy }}
\maketitle

\begin{abstract}
Using finite discrete group characters and symmetry breaking by hyperflux as
well as constraints on top- quark family, we study minimal low energy
effective theory following from SU$_{5}\times D_{4}$ models embedded in
F-theory with non abelian flux. Matter curves spectrum of the models is
obtained from SU$_{5}\times S_{5}$ theory with monodromy $S_{5}$ by
performing two breakings; first from symmetric group $S_{5}$ to $S_{4}$
subsymmetry; and next to dihedral $D_{4}$ subgroup. As a consequence, and
depending on the ways of decomposing triplets of $S_{4}$, we end with three
types of $D_{4}$- models. Explicit constructions of these theories are given
and a MSSM- like spectrum is derived.\newline
\end{abstract}


\section{Introduction}

Recently, there has been an increasing interest in building $SU_{5}\times
\Gamma $ GUT models, with discrete symmetries $\Gamma $, embedded in
Calabi-Yau compactification of F-theory down to 4d space time \textrm{\cite%
{A1}-\cite{A10}}; and in looking for low energy minimal prototypes with
broken monodromies \textrm{\cite{B1}-\cite{B51}}. This class of
supersymmetric GUTs with discrete groups lead to quasi-realistic field
spectrum having quark and lepton mass matrices with properties fitting with
MSSM requirements. In the geometric engineering of these F-GUTs, splitting
spectral cover method together with Galois theory tools are used to generate
appropriate matter curves spectrum \textrm{\cite{C0}-\cite{C5}}; and a \emph{%
geometric} $Z_{2}$ parity has been also introduced to suppress unwanted
effects such as exotic couplings and undesired proton decay operators 
\textrm{\cite{D1,D2,D3,D4}}.\newline
In this paper, we develop another manner to deal with monodromy of F-GUT
that is different from the one proposed first in \textrm{\cite{B5},} and
further explored in \textrm{\cite{D2,E1,E2}}, where matter curves of the
same orbit of monodromy are identified. In our approach, we use the non
abelian flux conjecture of \textrm{\cite{B3,B4}} to think of monodromy group
of F- theory $SU_{5}$ models as a non abelian flavor symmetry $\Gamma $. Non
trivial irreducible representations of the non abelian discrete group $%
\Gamma $ are used to host the three generations of fundamental matter; a
feature that opens a window to build semi-realistic models with matter
curves distinguished from each other in accord with mass hierarchy and
mixing neutrino physics \textrm{\cite{U1,U2,U3}}. \newline
In this work, we study the family of supersymmetric $SU_{5}\times \Gamma
_{p}\times U\left( 1\right) ^{5-p}$ models in the framework of F-theory GUT;
with \emph{non abelian} monodromies $\Gamma _{p}$ contained in the
permutation group $\mathbb{S}_{5}$ \textrm{\cite{E1}-\cite{E11}}; and
analyse the realisation of low energy constraints under which one can
generate an effective field spectrum that resembles to MSSM. A list of main
constraints leading to a good low energy spectrum are described in section
5; it requires amongst others a tree- level Yukawa coupling for top-quark
family. To realise this condition with non abelian $\Gamma _{p}$, we
consider the case where $\Gamma _{p}$ is given by the order \emph{8}
dihedral group $\mathbb{D}_{4}$; this particular non abelian discrete
symmetry has representations which allow more flexibility in accommodating
matter generations. Recall that the non abelian alternating $\mathbb{A}_{4}$
group has no irreducible doublet as shown on the character relation $%
12=3^{2}+1^{2}+1^{2}+1^{2}$; and the irreducible representation of non
abelian $\mathbb{S}_{4}$ and $\mathbb{S}_{3}$, which can be respectively
read from $24=3^{2}+3^{2}+2^{2}+1^{2}+1^{2}$ and $6=2^{2}+1^{2}+1^{2}$, have
a doublet and two singlets. The non abelian dihedral group $\mathbb{D}_{4}$
however has representations $\boldsymbol{R}_{i}$ with dimensions, that can
be read from $8=2^{2}+1^{2}+1^{2}+1^{2}+1^{2}$, seemingly more attractable
phenomenologically; it has \emph{5} irreducible $\boldsymbol{R}_{i}$'s; four
singlets, indexed by their basis characters as $\mathbf{1}_{++},$ $\mathbf{1}%
_{+-},$ $\mathbf{1}_{-+},$ $\mathbf{1}_{--}$; and an irreducible doublet $%
\mathbf{2}_{00}$; offering therefore several pictures to accommodate the
three generations of matter of the electroweak theory; in particular more
freedom in accommodating top quark family. \newline
To deal with the engineering of $SU_{5}\times \mathbb{D}_{4}$- models, we
develop a \emph{new method} based on finite discrete group characters $\chi
_{_{\boldsymbol{R}_{i}}}$; avoiding as a consequence the complexity of
Galois theory approach. The latter is useful to study F- theory models with
the dihedral $\mathbb{D}_{4}$ and the alternating $\mathbb{A}_{4}$ subgroups
of $\mathbb{S}_{4}$ as they are not directly reached by the standard
splitting spectral cover method; they are obtained in Galois theory by
putting constraints on the discriminant of underlying spectral covers; and
introducing other monodromy invariant of the covers such a resolvent \textrm{%
\cite{B2,B3,D4}}. \newline
To derive the $\mathbb{D}_{4}$- matter curves spectrum in $SU_{5}\times 
\mathbb{D}_{4}$- models, we think of it in terms of a two steps descent from 
$\mathbb{S}_{5}$- theory; a first descent down to $\mathbb{S}_{4}$; and a
second one to $\mathbb{D}_{4}$ by turning on appropriate flux that will be
explicitly described in this work; see also appendix C. By studying all
scenarios of breaking the triplets $\mathbb{S}_{4}$- theory in terms of
irreducible $\mathbb{D}_{4}$- representations, we end with three kinds of $%
\mathbb{D}_{4}$- models; one having a field spectrum involving all $\mathbb{D%
}_{4}$- representations including doublet $\mathbf{2}_{00}$ (\emph{model I}%
); the second theory (\emph{model II}) has no doublet $\mathbf{2}_{00}$ nor
the singlet $\mathbf{1}_{--}$; and the third model has no $\mathbf{2}_{00}$;
but does have $\mathbf{1}_{--}$. We have studied the curves spectrum of the
three $\mathbb{D}_{4}$-models; and we have found that only model III allows
a tree level 3-couplings and exhibits phenomenologically interesting
features.\newline
The presentation is as follows: In section 2, we study the $SU_{5}\times 
\mathbb{S}_{5}$ model; and describes the picture of the two steps breaking $%
\mathbb{S}_{5}\rightarrow \mathbb{S}_{4}\rightarrow \mathbb{S}_{3}$ by using
standard methods. In section 3, we introduce our method; and we revisit the
construction of the $\mathbb{S}_{4}$- and $\mathbb{S}_{3}$- models from the
view of discrete group characters. In section 4, we use character group
method to build three $SU_{5}\times \mathbb{D}_{4}\times U_{1}^{\perp }$
models. In section 5, we solve basic conditions for deriving MSSM- like
spectrum from $SU_{5}\times \mathbb{D}_{4}\times U_{1}^{\perp }$ models. In
section 6, we conclude and make discussions. Last section is devoted to
three appendices: In appendix A, we give relations regarding group
characters. In appendix B, we report details on other results obtained in
this study; and in appendix C we exhibit the link between non abelian
monodromies and flavor symmetry.

\section{Spectral Covers in $SU_{5}\times \Gamma $ models}

In F-GUT models with $SU_{5}$ gauge symmetry, matter curves carry quantum
numbers in $SU_{5}\times SU_{5}^{\bot }$ bi-representations following from
the breaking of $E_{8}$ as given below%
\begin{equation}
\begin{tabular}{lll}
$\mathbf{248}$ & $\rightarrow $ & $\left( \mathbf{24},\mathbf{1}_{\perp
}\right) \oplus \left( \mathbf{1},\mathbf{24}_{\perp }\right) \oplus $ \\ 
&  & $\left( \mathbf{10},\mathbf{5}_{\perp }\right) \oplus \left( \overline{%
\mathbf{10}},\mathbf{\bar{5}}_{\perp }\right) \oplus $ \\ 
&  & $\left( \mathbf{\bar{5}},\mathbf{10}_{\perp }\right) \oplus \left( 
\mathbf{5},\overline{\mathbf{10}}_{\perp }\right) $%
\end{tabular}
\label{dec}
\end{equation}%
In this $SU_{5}$ theory, the perpendicular $SU_{5}^{\bot }$ is restricted to
its Cartan-Weyl subsymmetry $\left( U_{1}^{\bot }\right) ^{4}$, see appendix
C for some explicit details; and the matter content of the model is labeled
by five weights t$_{i}$ like%
\begin{equation}
\begin{tabular}{lllll}
$\mathbf{10}_{t_{i}},$ & $\overline{\mathbf{10}}_{-t_{i}},$ & $\mathbf{\bar{5%
}}_{t_{i}+t_{j}},$ & $\mathbf{5}_{-t_{i}-t_{j}},$ & $\mathbf{1}%
_{t_{i}-t_{j}} $%
\end{tabular}
\label{matter}
\end{equation}%
with traceless condition%
\begin{equation}
t_{1}+t_{2}+t_{3}+t_{4}+t_{5}=0  \label{cd}
\end{equation}%
The components of the five 10-plets $\mathbf{10}_{t_{i}}$ and those of the
ten 5-plets $\mathbf{\bar{5}}_{t_{i}+t_{j}}$ are related to each other by
monodromy symmetries $\Gamma $; offering a framework of approaching GUT -
models with discrete symmetries originating from geometric properties of the
elliptic Calabi-Yau fourfold $CY4$ which, naively, can be thought of as
given by the 4- $\dim $ complex space%
\begin{equation}
CY4\sim E\times \mathcal{B}_{3}
\end{equation}%
In this fibration, the complex 3- $\dim $ base $\mathcal{B}_{3}$ contains
the complex GUT surface $\mathcal{S}_{GUT}$ wrapped by 7-brane; and the
complex elliptic curve E fiber is as follows%
\begin{equation}
y^{2}=x^{3}+b_{5}xy+b_{4}x^{2}z+b_{3}yz^{2}+b_{2}xz^{3}+b_{0}z^{5}
\label{sl}
\end{equation}%
where the homology classes $\left[ x\right] ,$ $\left[ y\right] ,$ $\left[ z%
\right] $ and $\left[ b_{k}\right] ;$ associated with the holomorphic
sections $x,$ $y,$ $z$ and $b_{k},$ are expressed in terms of the Chern
class $c_{1}=c_{_{1}}\left( \mathcal{S}_{GUT}\right) $ of the tangent bundle
of the $\mathcal{S}_{GUT}$ surface; and the Chern class $-t$ of the normal
bundle $\mathcal{N}_{\mathcal{S}_{GUT}|\mathcal{B}_{3}}$ like%
\begin{equation}
\begin{tabular}{lllllllll}
$\left[ y\right] $ & $=$ & $3\left( c_{1}-t\right) $ &  & $,$ &  & $\left[ z%
\right] $ & $=$ & $-t$ \\ 
$\left[ x\right] $ & $=$ & $2\left( c_{1}-t\right) $ &  & $,$ &  & $\left[
b_{k}\right] $ & $=$ & $\left( 6c_{1}-t\right) -kc_{1}$%
\end{tabular}
\label{ct}
\end{equation}

\subsection{Matter curves in $SU_{5}\times \mathbb{S}_{5}$ model}

Matter curves of $SU_{5}\times U\left( 1\right) ^{5-k}\times \Gamma _{k}$
models live on GUT surface $\mathcal{S}_{GUT}$ with monodromy symmetries $%
\Gamma _{k}$ contained in $\mathbb{S}_{5}$, the Weyl group of $SU_{5}^{\bot
} $; see eq(\ref{gk}) of appendix C. In the case of $\Gamma _{5}=\mathbb{S}%
_{5} $; these curves organise into reducible multiplets\textrm{\footnote{%
An equivalent spectrum can be also given by using irreducible
representations of $\mathbb{S}_{5}$ and their characters; to fix ideas see
the analogous $\mathbb{S}_{4}$- and $\mathbb{S}_{3}$- models studied in
section 3.}} of $\mathbb{S}_{5}$ with the following characteristic
properties 
\begin{equation}
\begin{tabular}{|c|c|c|c|c|}
\hline
{\small matters curves} & {\small weights} & $\mathbb{S}_{5}$ {\small repres}
& {\small homology classes} & {\small holomorphic sections} \\ \hline
$10_{t_{{\small i}}}$ & $t_{{\small i}}$ & $5$ & $\eta -5c_{1}$ & $%
b_{5}=b_{0}\dprod \limits_{i=1}^{5}t_{i}\left. 
\begin{array}{c}
\text{ \ } \\ 
\text{ \  \ }%
\end{array}%
\right. $ \\ \hline
$\bar{5}_{t_{{\small i}}+t_{j}}$ & $t_{{\small i}}+t_{j}$ & $10$ & $\eta
^{\prime }-10c_{1}$ & $d_{10}=d_{0}\dprod \limits_{j>i=1}^{5}T_{ij}\left. 
\begin{array}{c}
\text{ \ } \\ 
\text{ \  \ }%
\end{array}%
\right. $ \\ \hline
$1_{t_{{\small i}}-t_{j}}$ & $t_{{\small i}}-t_{j}$ & $20$ & $\eta ^{\prime
\prime }-20c_{1}$ & $g_{20}=g_{0}\dprod \limits_{i\neq j=1}^{5}S_{ij}\left. 
\begin{array}{c}
\text{ \ } \\ 
\text{ \  \ }%
\end{array}%
\right. $ \\ \hline
\end{tabular}
\label{1}
\end{equation}%
\begin{equation*}
\end{equation*}%
where the $t_{i}$' s as above; $T_{ij}=t_{i}+t_{j}$ with $i<j$; and $%
S_{ij}=t_{i}-t_{j}$ with $i\neq j$. These $t_{i}$'s, $T_{ij}$'s; and $S_{ij}$%
's are respectively interpreted as the simple zeros of the spectral covers $%
\mathcal{C}_{5}=0$ describing ten-plets$,$ $\mathcal{C}_{10}=0$ describing
five-pelts and $\mathcal{C}_{20}=0$ for flavon singlets \textrm{\cite{X0}-%
\cite{X5}} 
\begin{equation}
\begin{tabular}{lllll}
$\mathcal{C}_{5}$ & $=$ & $b_{0}\dprod \limits_{i=1}^{5}\left(
s-t_{i}\right) $ & $\equiv $ & $b_{0}\dprod \limits_{i=1}^{5}s_{i}$ \\ 
&  &  &  &  \\ 
$\mathcal{C}_{10}$ & $=$ & $d_{0}\dprod \limits_{j>i=1}^{5}\left(
s-T_{ij}\right) $ & $\equiv $ & $d_{0}\dprod \limits_{j>i=1}^{5}s_{ij}$ \\ 
&  &  &  &  \\ 
$\mathcal{C}_{20}$ & $=$ & $g_{0}\dprod \limits_{i\neq j}^{5}\left(
s-S_{ij}\right) $ & $\equiv $ & $g_{0}\dprod \limits_{i\neq
j}^{5}s_{ij}^{\prime }$%
\end{tabular}
\label{22}
\end{equation}%
\begin{equation*}
\end{equation*}%
The homology classes of the complex curves in (\ref{1}) are nicely obtained
by defining the spectral covers in terms of the usual holomorphic sections;
for the 5-sheeted covering\ of $\mathcal{S}_{GUT}$, we have 
\begin{equation}
\mathcal{C}_{5}=b_{0}s^{5}+b_{1}s^{4}+b_{2}s^{3}+b_{3}s^{2}+b_{4}s+b_{5}=0
\label{C5}
\end{equation}%
with $b_{1}=0$ due to traceless condition; and homology classes of the
complex holomorphic sections $b_{k}$ as follows 
\begin{equation}
\begin{tabular}{|l|l|}
\hline
holomorphic sections \  \  \  & \  \  \ homology classes \  \  \  \\ \hline
$\  \  \  \  \  \  \  \  \ s$ & $\  \  \  \  \  \  \  \  \ -c_{1}$ \\ \hline
$\  \  \  \  \  \  \  \  \ b_{k}$ & $\  \  \  \  \  \  \  \  \  \eta -kc_{1}$ \\ \hline
\end{tabular}%
\end{equation}%
with canonical homology class $\eta $ given by 
\begin{equation}
\eta =6c_{1}-t
\end{equation}%
with $c_{1}$ and $-t\ $as in eqs(\ref{ct}). From these relations, the
homology class $\left[ 10_{t_{{\small i}}}\right] =\left[ \left. \mathcal{C}%
_{5}\right \vert _{s=0}\right] $ is given by $\left[ b_{5}\right] $; by
using $b_{5}=b_{0}\dprod \nolimits_{i=1}^{5}t_{i}$, we have $\left[ b_{5}%
\right] =\eta -5c_{1}$ in agreement with (\ref{ct}). For the 10-sheeted
covering, we have 
\begin{equation}
\mathcal{C}_{10}=\tsum \limits_{k=0}^{10}d_{k}s^{10-k}  \label{c10}
\end{equation}%
and leads to the homology class $\left[ d_{10}\right] =\eta ^{\prime
}-10c_{1}$ where, due to $d_{0}=b_{0}^{3}$, the class $\eta ^{\prime }$ can
be related to the canonical $\eta $ of the 5- sheeted cover like $3\eta $.
Similar relation can be written down for singlets%
\begin{equation}
\mathcal{C}_{20}=\dsum \limits_{k=0}^{20}g_{k}s^{20-k}  \label{c11}
\end{equation}%
leading to $\left[ g_{20}\right] =\eta ^{\prime \prime }-20c_{1}$ with the
property $\eta ^{\prime \prime }=9\eta $. \newline
For later use, we consider together with (\ref{1}) the so called geometric $%
Z_{2}$ parity of \textrm{\cite{B51}}; but as approached in \textrm{\cite%
{B2,B3} }in dealing with local models. For simplicity, we use a short way to
introduce this parity by requiring, up to an overall phase, invariance of $%
\mathcal{C}_{5}=0,$ $\mathcal{C}_{10}=0,$ $\mathcal{C}_{20}=0$ under the
following transformations along the spectral fiber; see \textrm{\cite%
{B2,B3,B4} }for explicit details, 
\begin{equation}
\begin{tabular}{lll}
$s_{i}^{\prime }$ & $=$ & $e^{-i\phi }s_{i}$ \\ 
$b_{k}^{\prime }$ & $=$ & $e^{i\left[ \beta +(5-k)\phi \right] }b_{k}$ \\ 
$d_{k}^{\prime }$ & $=$ & $e^{i\left[ \gamma +(10-k)\phi \right] }d_{k}$ \\ 
$g_{k}^{\prime }$ & $=$ & $e^{i\left[ \delta +(20-k)\phi \right] }g_{k}$%
\end{tabular}
\label{Y1}
\end{equation}%
Under this phase change, the spectral covers eqns transform like%
\begin{equation}
\begin{tabular}{lll}
$\mathcal{C}_{5}^{\prime }$ & $=$ & $e^{i\beta }\mathcal{C}_{5}$ \\ 
$\mathcal{C}_{10}^{\prime }$ & $=$ & $e^{i\gamma }\mathcal{C}_{10}$ \\ 
$\mathcal{C}_{20}^{\prime }$ & $=$ & $e^{i\delta }\mathcal{C}_{20}$%
\end{tabular}
\label{Y2}
\end{equation}%
Focussing on 10-plets, and equating above $\mathcal{C}_{5}^{\prime }$ with
the one deduced from construction of \textrm{\cite{B4}} namely $\mathcal{C}%
_{5}^{\prime }=e^{i\left( \zeta -\phi \right) }\mathcal{C}_{5}$; we learn
that we should have $\beta =\zeta -\phi ;$ and therefore $b_{k}^{\prime
}=e^{i\left[ \zeta +(k-6)\phi \right] }b_{k}$. For the particular choice $%
\phi =\pi $, we have $s_{i}^{\prime }=-s_{i}$ and 
\begin{equation}
b_{k}^{\prime }=\left( -\right) ^{k}e^{i\zeta }b_{k}
\end{equation}%
If we put $\zeta =0$, we get $\left( b_{0}^{\prime },b_{5}^{\prime }\right)
=\left( +b_{0},-b_{5}\right) $; while by taking $\zeta =\pi $, we have $%
\left( b_{0}^{\prime },b_{5}^{\prime }\right) =\left( -b_{0},+b_{5}\right) $%
; below we set $\zeta =\pi $. To get the parity of the holomorphic sections $%
d_{k}$ and g$_{k}$ of eqs (\ref{22}), we use their relationships with the $%
b_{k}$\ coefficients. By help of the relations $%
d_{10}=b_{3}^{2}b_{4}-b_{2}b_{3}b_{5}+b_{0}b_{5}^{2}$ and $%
g_{20}=256b_{4}^{5}b_{0}^{4}+...$, it follows that $Z_{2}(d_{10})\sim
Z_{2}(b_{3}^{2}b_{4})$ and $Z_{2}(g_{20})=Z_{2}(b_{4}^{5}b_{0}^{4});$ so we
have \textrm{\cite{D2,E1,E2}} 
\begin{equation}
\begin{tabular}{lllll}
$Z_{2}(d_{10})=-1$ & , & $Z_{2}(g_{20})=-1$ & , & $Z_{2}(b_{5})=+1$ \\ 
$Z_{2}(d_{0})=-1$ & , & $Z_{2}(g_{0})=-1$ & , & $Z_{2}(b_{0})=-1$%
\end{tabular}
\label{d10}
\end{equation}%
in agreement with the homology class properties $\eta ^{\prime }=3\eta $ and 
$\eta ^{\prime \prime }=9\eta $.

\subsection{Models with broken $\mathbb{S}_{5}$}

To engineer matter curves with monodromy $\Gamma _{k}\subset \mathbb{S}_{5}$%
; we generally use spectral cover splitting method combined with constraints
inspired from Galois theory \textrm{\cite{B2,B3,B4,D1,D2}}. In this study,
we develop a new method without need of the involved tools of Galois group
theory; our approach uses characters $\mathrm{\chi }_{\mathbf{R}}\left(
g\right) $ of discrete group representations; and relies directly the roots
of the spectral covers. To illustrate the method; but also for later use, we
first study the two interesting cases by using the standard method:

$\bullet $ $\Gamma _{4}=\mathbb{S}_{4}\subset \mathbb{S}_{5},$

$\bullet $ $\Gamma _{3}=\mathbb{S}_{3}\subset \mathbb{S}_{5}$. \newline
The case $\Gamma _{4}=\mathbb{D}_{4}$ requires more tools; it will be
studied later after revisiting $\mathbb{S}_{4}$- and $\mathbb{S}_{3}$-
models from the view of characters of their representations.

\subsubsection{$\mathbb{S}_{4}$- model in standard approach}

To engineer the breaking of $\mathbb{S}_{5}$ down to $\mathbb{S}_{4}$, we
proceed as follows: First, we use $\mathbb{S}_{5}$- invariance to rewrite
the holomorphic polynomial $\mathcal{C}_{5}$ like%
\begin{equation}
\mathcal{C}_{5}=\frac{b_{0}}{5!}\dsum \limits_{\sigma \in \Gamma }\dprod
\limits_{i=1}^{5}\left( s-t_{\sigma \left( i\right) }\right)
\end{equation}%
and similarly for $\mathcal{C}_{10}$ and $\mathcal{C}_{20}$. To break $%
\mathbb{S}_{5}$ down to $\mathbb{S}_{4}$, we impose a condition fixing one
of the weight \textrm{\cite{S1}}; for example 
\begin{equation}
\sigma \left( t_{5}\right) =t_{5}\qquad \Leftrightarrow \qquad \sigma \left(
5\right) =5
\end{equation}%
This requirement breaks $\mathbb{S}_{5}$ down to one of the five possible $%
\mathbb{S}_{4}$ subgroups living inside $\mathbb{S}_{5}$; and leads to the
following features:\newline
$\left( \mathbf{a}\right) $ the traceless condition (\ref{cd}) of the
orthogonal $SU_{5}^{\bot }$ is solved as $t_{5}=-\left(
t_{1}+t_{2}+t_{3}+t_{4}\right) $; it is manifestly $\mathbb{S}_{4}$-
invariant. To deal with this $t_{5}$ weight, we shall think about the
breaking of $\mathbb{S}_{5}$ down to $\mathbb{S}_{4}$ in terms of the
descent of the symmetry $SU_{5}\times U\left( 1\right) ^{5-k}\times \Gamma
_{k}$ from k=5 to k=4 as follows \textrm{\cite{F1,F2}} 
\begin{equation}
\begin{tabular}{lll}
$SU_{5}\times U\left( 1\right) ^{5-5}\times \mathbb{S}_{5}$ & $\qquad
\rightarrow \qquad $ & $SU_{5}\times U\left( 1\right) ^{5-4}\times \mathbb{S}%
_{4}$ \\ 
& $\qquad \sim \qquad $ & $SU_{5}\times \mathbb{S}_{4}\times U\left(
1\right) $%
\end{tabular}%
\end{equation}%
$\left( \mathbf{b}\right) $ the spectral covers $\mathcal{C}_{5}$ and $%
\mathcal{C}_{10}$ split as the product of two factors: $\left( \mathbf{%
\alpha }\right) $ the spectral cover $\mathcal{C}_{5}$ factorises like $%
\mathcal{C}_{4}\times \mathcal{C}_{1}$ with%
\begin{equation}
\mathcal{C}_{4}=A_{0}\dprod \limits_{i=1}^{4}\left( s-t_{i}\right) \qquad
,\qquad \mathcal{C}_{1}=a_{0}\left( s-t_{5}\right)  \label{ff}
\end{equation}%
and\textrm{\footnote{%
\ The holomorphic sections $A_{l}$ and $a_{m}$ eqs(\ref{ff}) are directly
derived by expanding the factorised forms of the spectral covers $\mathcal{C}%
_{4}$ and $\mathcal{C}_{1}$; we will not give these details here; for
example the relevant $A_{4}$ and $a_{1}$ are given by $A_{4}=A_{0}\dprod%
\nolimits_{i=1}^{4}t_{i}$ and \ $a_{1}=-a_{0}t_{5}.$}}%
\begin{equation}
\begin{tabular}{lll}
$b_{0}$ & $=$ & $A_{0}\times a_{0}$ \\ 
$b_{5}$ & $=$ & $A_{4}\times a_{1}$%
\end{tabular}
\label{Y11}
\end{equation}%
together with the transformations following from (\ref{Y1}-\ref{Y2}). Notice
that the above factorisations put conditions on the field\textrm{\ }$%
\mathcal{K}$\textrm{\ }where live the holomorphic sections; a feature that
is also predicted by Galois theory\textrm{\  \cite{D3,D4}. }As a naive
illustration, we use the comparison with arithmetics in the set\ of integers 
$\mathcal{Z}$; an integer number like $6$\ can be factorised in\textrm{\ }$%
\mathcal{Z}$\textrm{\ }as\textrm{\ }$6=2\times 3$; while a prime integer
like $5$ has no factorisation. \newline
By using $\mathcal{C}_{5}^{\prime }=\mathcal{C}_{4}^{\prime }\times \mathcal{%
C}_{1}^{\prime }$ and equating $e^{i\left( \zeta -\phi \right) }\left( 
\mathcal{C}_{4}\times \mathcal{C}_{1}\right) $ with $\left( e^{i\xi }%
\mathcal{C}_{4}\right) \times \left( e^{i\psi }\mathcal{C}_{1}\right) $; it
follows that $\zeta -\phi =\xi +\psi $; and 
\begin{equation}
\begin{tabular}{lll}
$A_{4}^{\prime }$ & $=$ & $e^{i\xi }A_{4}$ \\ 
$a_{1}^{\prime }$ & $=$ & $e^{i\left( \zeta -\xi -\phi \right) }a_{1}$%
\end{tabular}%
\end{equation}%
from which we learn that $A_{4}$ and $a_{1}$ sections transform differently;
and then $Z_{2}\left( b_{4}\right) =Z_{2}\left( A_{4}\right) \times
Z_{2}\left( a_{1}\right) .$ $\left( \mathbf{\beta }\right) $ the $\mathcal{C}%
_{10}$ splits in turns like $\mathcal{\tilde{C}}_{6}\times \mathcal{\tilde{C}%
}_{4}$ with%
\begin{equation}
\mathcal{C}_{6}=\tilde{A}_{0}\dprod \limits_{j>i=1}^{4}\left(
s-T_{ij}\right) \qquad ,\qquad \mathcal{C}_{4}=\tilde{a}_{0}\dprod
\limits_{i=1}^{4}\left( s-T_{i5}\right)
\end{equation}%
and%
\begin{equation}
\begin{tabular}{lll}
$d_{0}$ & $=$ & $\tilde{A}_{0}\times \tilde{a}_{0}$ \\ 
$d_{10}$ & $=$ & $\tilde{A}_{6}\times \tilde{a}_{4}$%
\end{tabular}%
\end{equation}%
\begin{equation*}
\end{equation*}%
as well as $\mathcal{\tilde{C}}_{6}=e^{2i\tilde{\xi}}\mathcal{\tilde{C}}_{6}$
and $\mathcal{\tilde{C}}_{4}=e^{2i\tilde{\psi}}\mathcal{\tilde{C}}_{4}$ with 
$\tilde{\xi}+\tilde{\psi}=\tilde{\zeta}-\phi .$ \  \newline
Under the above splitting, the spectrum (\ref{1}) decomposes in terms of
reducible $\mathbb{S}_{4}$ multiplets as follows%
\begin{equation}
\begin{tabular}{|c|c|c|c|c|c|c|c|}
\hline
{\small curves} & {\small weights} & ${\small S}_{4}$ & ${\small U}%
_{1}^{\perp }$ & {\small homology} & {\small sections} & {\small Z}$_{2}$ & 
{\small U}${\small (1)}_{Y}${\small \ flux} \\ \hline
${\small 10}_{t_{{\small i}}}$ & ${\small t}_{{\small i}}$ & ${\small 4}$ & $%
{\small 0}$ & ${\small \eta -4c}_{1}{\small +\chi }$ & $A_{4}$ & ${\small %
\varkappa }_{4}$ & ${\small N}$ \\ \hline
${\small 10}_{t_{{\small 5}}}$ & ${\small t}_{{\small 5}}$ & ${\small 1}$ & $%
{\small 1}$ & ${\small -\chi -c}_{1}$ & $a_{1}$ & ${\small \varkappa }_{1}$
& ${\small -N}$ \\ \hline \hline
${\small 5}_{t_{{\small i}}+t_{j}}$ & ${\small t}_{{\small i}}{\small +t}%
_{j} $ & ${\small 6}$ & ${\small 0}$ & ${\small \eta }^{\prime }{\small -6c}%
_{1}{\small +\tilde{\chi}}$ & $\tilde{A}_{6}$ & $\tilde{\varkappa}_{6}$ & $%
{\small N}$ \\ \hline
${\small 5}_{t_{{\small i}}+t_{5}}$ & ${\small t}_{{\small i}}{\small +t}%
_{5} $ & ${\small 4}$ & ${\small 1}$ & ${\small -\tilde{\chi}-4c}_{1}$ & $%
{\small \tilde{a}}_{4}$ & $\tilde{\varkappa}_{4}$ & ${\small -N}$ \\ \hline
\end{tabular}
\label{S4}
\end{equation}%
with%
\begin{equation}
\begin{tabular}{lllllll}
$A_{4}$ & $=A_{0}\dprod \limits_{i=1}^{4}t_{i}$ &  & , &  & $\tilde{A}_{6}$
& $=\tilde{A}_{0}\dprod \limits_{j>i=1}^{4}T_{ij}$ \\ 
$a_{1}$ & $=a_{0}t_{5}$ &  & , &  & $\tilde{a}_{4}$ & $=\tilde{a}_{0}\dprod
\limits_{i=1}^{4}T_{i5}$%
\end{tabular}
\label{aa}
\end{equation}%
and where $\varkappa _{i}$ and $\tilde{\varkappa}_{k}$ refer to Z$_{2}$
parities; for instance 
\begin{equation}
\begin{tabular}{lllllllll}
$\varkappa _{4}$ & $=$ & $Z_{2}\left( A_{4}\right) $ &  & , &  & $\tilde{%
\varkappa}_{6}$ & $=$ & $Z_{2}\left( \tilde{A}_{6}\right) $ \\ 
$\varkappa _{1}$ & $=$ & $Z_{2}\left( a_{1}\right) $ &  & , &  & $\tilde{%
\varkappa}_{4}$ & $=$ & $Z_{2}\left( \tilde{a}_{4}\right) $ \\ 
$\varkappa _{4}\varkappa _{1}$ & $=$ & $Z_{2}\left( b_{5}\right) $ &  & , & 
& $\tilde{\varkappa}_{4}\tilde{\varkappa}_{6}$ & $=$ & $Z_{2}\left(
d_{10}\right) $%
\end{tabular}
\label{pa}
\end{equation}%
\begin{equation*}
\end{equation*}%
The last column of eq(\ref{S4}) refers to the hyperflux of the U$\left(
1\right) _{Y}$ gauge field strength; it breaks $SU_{5}$ gauge symmetry down
to standard model gauge invariance; and also pierces the matter curves of
the model as shown on table.

\subsubsection{$\mathbb{S}_{3}$- model in standard approach}

The breaking of $\mathbb{S}_{5}$ down to $\mathbb{S}_{3}$ may be obtained
from above $\mathbb{S}_{4}$ model by further breaking $\mathbb{S}_{4}$ down
to $\mathbb{S}_{3}$; this corresponds to $SU_{5}\times U\left( 1\right)
^{5-5}\times \mathbb{S}_{5}$ $\rightarrow $ $SU_{5}\times U\left( 1\right)
^{5-3}\times \mathbb{S}_{3}$. This can be realised by fixing one of the four
t$_{i}$ roots; say t$_{4}$; so that the breaking pattern is given by%
\begin{equation}
\begin{tabular}{lll}
$SU_{5}\times U\left( 1\right) ^{5-5}\times \mathbb{S}_{5}$ & $\qquad
\rightarrow \qquad $ & $SU_{5}\times U\left( 1\right) ^{5-3}\times \mathbb{S}%
_{3}$ \\ 
& $\qquad \sim \qquad $ & $SU_{5}\times \mathbb{S}_{3}\times U\left(
1\right) ^{2}$%
\end{tabular}%
\end{equation}%
Setting $U\left( 1\right) ^{2}=U_{1}^{\perp }{\small \times }U_{1}^{\perp }$%
, the previous $\mathbb{S}_{4}$ spectrum decomposes into reducible $\mathbb{S%
}_{3}$ multiplets as follows, 
\begin{equation}
\begin{tabular}{|c|c|c|c|c|c|}
\hline
{\small curves} & ${\small S}_{3}$ & ${\small U}_{1}^{\perp }{\small \times U%
}_{1}^{\perp }$ & {\small homology} & {\small section} & {\small U}${\small %
(1)}_{Y}${\small \ flux} \\ \hline
${\small 10}_{t_{{\small i}}}$ & ${\small 3}$ & $({\small 0,0)}$ & ${\small %
\eta -3c}_{1}{\small -\chi -\chi }^{\prime }$ & ${\small A}_{3}^{\prime }$ & 
${\small -N-P}$ \\ 
${\small 10}_{t_{{\small 4}}}$ & ${\small 1}$ & $({\small 1,0)}$ & ${\small %
\chi }^{\prime }{\small -c}_{1}$ & ${\small A}_{1}^{\prime }$ & ${\small P}$
\\ 
${\small 10}_{t_{{\small 5}}}$ & ${\small 1}$ & $({\small 0,1)}$ & ${\small %
\chi -c}_{1}$ & ${\small a}_{1}$ & ${\small N}$ \\ \hline \hline
${\small 5}_{t_{{\small i}}+t_{j}}$ & ${\small 3}$ & $({\small 0,0)}$ & $%
{\small \eta }^{\prime }{\small -3c}_{1}{\small -\tilde{\chi}-\tilde{\chi}}%
^{\prime }$ & ${\small \tilde{A}}_{3}^{\prime }$ & ${\small -N-P}$ \\ 
${\small 5}_{t_{{\small i}}+t_{4}}$ & ${\small 3}$ & $({\small 1,0)}$ & $%
{\small \tilde{\chi}}^{\prime }{\small -3c}_{1}$ & ${\small \tilde{A}}%
_{3}^{\prime \prime }$ & ${\small P}$ \\ 
${\small 5}_{t_{{\small i}}+t_{5}}$ & ${\small 3}$ & $({\small 0,1)}$ & $%
{\small \tilde{\chi}-3c}_{1}{\small -\tilde{\chi}}^{\prime }$ & ${\small 
\tilde{a}}_{3}^{\prime }$ & ${\small N-P}$ \\ 
${\small 5}_{t_{{\small 4}}+t_{5}}$ & ${\small 1}$ & $({\small 1,1)}$ & $%
{\small \tilde{\chi}}^{\prime }{\small -c}_{1}$ & ${\small \tilde{a}}%
_{1}^{\prime \prime }$ & ${\small P}$ \\ \hline
\end{tabular}
\label{T3}
\end{equation}%
with%
\begin{equation}
b_{5}=\left( A_{3}^{\prime }A_{1}^{\prime }\right) \times a_{1}\qquad
,\qquad d_{10}=\left( \tilde{A}_{3}^{\prime }\tilde{A}_{3}^{\prime \prime
}\right) \times \left( \tilde{a}_{3}^{\prime }\tilde{a}_{1}^{\prime \prime
}\right)
\end{equation}%
where $A_{3}^{\prime },$ $A_{1}^{\prime },$ $a_{1}$ and $\tilde{A}%
_{3}^{\prime },$ $\tilde{A}_{3}^{\prime \prime },$ $\tilde{a}_{3}^{\prime },$
$\tilde{a}_{1}^{\prime \prime }$ are given by relations of form as in (\ref%
{aa}). An extra column for $Z_{2}$- parity can be also added as in (\ref{S4}%
) with the property%
\begin{equation}
\begin{tabular}{lll}
$Z_{2}\left( b_{5}\right) $ & $=$ & $Z_{2}(A_{3}^{\prime })\times
Z_{2}(A_{1}^{\prime })\times Z_{2}(a_{1})$ \\ 
$Z_{2}\left( d_{10}\right) $ & $=$ & $Z_{2}(\tilde{A}_{3}^{\prime })\times
Z_{2}(\tilde{A}_{3}^{\prime \prime })\times Z_{2}(\tilde{a}_{3}^{\prime
})\times Z_{2}(\tilde{a}_{1}^{\prime \prime })$%
\end{tabular}%
\end{equation}%
Observe also that here we have two new homology class cycles $\chi $ and $%
\chi ^{\prime }$ with%
\begin{equation}
\int_{\chi }\mathcal{F}_{X}=N\qquad ,\qquad \int_{\chi ^{\prime }}\mathcal{F}%
_{X}=P  \label{np}
\end{equation}%
The non zero $P$ is responsible for the second splitting; this is because
the breaking of $\mathbb{S}_{5}$ down to $\mathbb{S}_{3}$ has been
undertaken into two stages: first $\mathbb{S}_{5}\rightarrow \mathbb{S}_{4}$%
; and second $\mathbb{S}_{4}\rightarrow \mathbb{S}_{3}$. In what follows we
extend this idea to the breaking pattern of $\mathbb{S}_{5}$ down to $%
\mathbb{D}_{4}$.

\section{Revisiting $\mathbb{S}_{4}$ and $\mathbb{S}_{3}$- models}

In this section, we develop tools towards the study of the breaking of $%
\mathbb{S}_{5}$ monodromy down to its $\mathbb{D}_{4}$ sub-symmetry. To our
knowledge these tools, have not been used before; even for $\mathbb{S}_{n}$
permutation groups; so we begin by revisiting the $\mathbb{S}_{4}$- and $%
\mathbb{S}_{3}$- models from the view of characters of their irreducible
representations; and turn in next section to develop the $\mathbb{D}_{4}$
theory.

\subsection{$SU_{5}\times \mathbb{S}_{4}\times U_{1}^{\perp }$ model}

In the canonical $t_{i}$-weight basis, the matter spectrum of $\mathbb{S}%
_{4} $- model is given by (\ref{S4}); there matter curves are organised into 
\emph{reducible} multiplets of $\mathbb{S}_{4}\times U_{1}^{\perp }$. Below,
we give another manner to approach the spectrum of $\mathbb{S}_{4}$- model. 
\newline
By help of the standard relation $24=1^{2}+1^{2}+2^{2}+3^{2}+3^{2}$ showing
that $\mathbb{S}_{4}$\ has $\emph{5}$ \emph{irreducible} representations $%
\boldsymbol{R}_{i}$ and \emph{5 conjugacy} classes $\mathfrak{C}_{i}$ 
\textrm{\cite{E8,E9,E10,E11}}; and by using properties of the irreducible $%
\boldsymbol{R}_{i}$ representations of $\mathbb{S}_{4}$ given in appendix;
eq(\ref{S4}) may be expressed in terms of the $\boldsymbol{R}_{i}$'s and
their $\mathrm{\chi }_{R}^{\left( a,b,c\right) }$ characters as follows%
\begin{equation}
\begin{tabular}{|c|c|c|c|c|c|c|}
\hline
{\small curves} & {\small weights} & {\small Irrep }$S_{4}$ & $\mathrm{\chi }%
_{R}^{\left( a,b,c\right) }$ & ${\small U}_{1}^{\perp }$ & {\small homology}
& {\small U}${\small (1)}_{Y}${\small \ flux} \\ \hline
$\left. 
\begin{array}{c}
{\small 10}_{x_{{\small i}}} \\ 
{\small 10}_{x_{{\small 4}}} \\ 
{\small 10}_{t_{5}}%
\end{array}%
\right. $ & $\left. 
\begin{array}{c}
{\small x}_{i} \\ 
{\small x}_{4} \\ 
{\small t}_{5}%
\end{array}%
\right. $ & $\left. 
\begin{array}{c}
\mathbf{3} \\ 
\mathbf{1} \\ 
\mathbf{1}%
\end{array}%
\right. $ & $\left. 
\begin{array}{c}
\left( 1,0,-1\right) \\ 
\left( 1,1,1\right) \\ 
\left( 1,1,1\right)%
\end{array}%
\right. $ & $\left. 
\begin{array}{c}
{\small 0} \\ 
{\small 0} \\ 
{\small 1}%
\end{array}%
\right. $ & $\left. 
\begin{array}{c}
{\small \eta -3c}_{1} \\ 
{\small \chi -c}_{1} \\ 
{\small -\chi -c}_{1}%
\end{array}%
\right. $ & $\left. 
\begin{array}{c}
{\small 0} \\ 
{\small N} \\ 
{\small -N}%
\end{array}%
\right. $ \\ \hline
$\left. 
\begin{array}{c}
{\small 5}_{X_{{\small ij}}} \\ 
{\small 5}_{X_{{\small i4}}} \\ 
{\small 5}_{X_{i5}} \\ 
{\small 5}_{X_{45}}%
\end{array}%
\right. $ & $\left. 
\begin{array}{c}
{\small X}_{{\small ij}} \\ 
{\small X}_{{\small i4}} \\ 
{\small X}_{i5} \\ 
{\small X}_{45}%
\end{array}%
\right. $ & $\left. 
\begin{array}{c}
\mathbf{3}^{\prime } \\ 
\mathbf{3} \\ 
\mathbf{3} \\ 
\mathbf{1}%
\end{array}%
\right. $ & $\left. 
\begin{array}{c}
\left( -1,0,1\right) \\ 
\left( 1,0,-1\right) \\ 
\left( 1,0,-1\right) \\ 
\left( 1,1,1\right)%
\end{array}%
\right. $ & $\left. 
\begin{array}{c}
{\small 0} \\ 
{\small 0} \\ 
{\small 0} \\ 
{\small 1}%
\end{array}%
\right. $ & $\left. 
\begin{array}{c}
{\small \eta }^{\prime }{\small -3c}_{1} \\ 
{\small -3c}_{1}+{\small \chi }^{\prime } \\ 
{\small -3c}_{1}{\small -\chi }^{\prime } \\ 
{\small -c}_{1}%
\end{array}%
\right. $ & $\left. 
\begin{array}{c}
{\small 0} \\ 
{\small N} \\ 
{\small -N} \\ 
{\small 0}%
\end{array}%
\right. $ \\ \hline
\end{tabular}
\label{T2}
\end{equation}%
\begin{equation*}
\end{equation*}

Notice that $\mathbb{S}_{4}$ has three generators denoted here by $\left(
a,b,c\right) $ and chosen as given by 2-, 3- and 4-cycles; they obey amongst
others the cyclic properties $a^{2}=b^{3}=c^{4}=I_{id}$; these three
generators are non commuting permutation operators making extraction of full
information from them a difficult task; but part of these information is
given their $\mathrm{\chi }_{R}^{\left( a,b,c\right) }$'s; these characters
are real numbers as collected in following table \textrm{\cite{E8,E9,E10,E11}%
}, 
\begin{equation}
\begin{tabular}{|l|l|l|l|l|l|}
\hline
$\mathrm{\chi }_{ij}$ & $\mathrm{\chi }_{_{\boldsymbol{I}}}$ & $\mathrm{\chi 
}_{_{\boldsymbol{3}^{\prime }}}$ & $\mathrm{\chi }_{_{\boldsymbol{2}}}$ & $%
\mathrm{\chi }_{_{\boldsymbol{3}}}$ & $\mathrm{\chi }_{_{\epsilon }}$ \\ 
\hline
$a$ & $1$ & $-1$ & $\  \ 0$ & $\  \ 1$ & $-1$ \\ \hline
$b$ & $1$ & $\  \ 0$ & $-1$ & $\  \ 0$ & $\  \ 1$ \\ \hline
$c$ & $1$ & $\  \ 1$ & $\  \ 0$ & $-1$ & $-1$ \\ \hline
\end{tabular}%
\end{equation}%
\begin{equation*}
\end{equation*}%
Notice also that the $\mathbf{4}$- and $\mathbf{6}$- representations of $%
\mathbb{S}_{4},$ which have been used in the canonical formulation of
section 2, are decomposed in (\ref{T2}) as direct sums of irreducible
components as follows:%
\begin{equation}
\begin{tabular}{lll}
$\mathbf{4}_{\left( 2,1,0\right) }$ & $\mathbf{=}$ & $\mathbf{1}_{\left(
1,1,1\right) }\oplus \mathbf{3}_{\left( 1,0,-1\right) }$ \\ 
$\mathbf{6}_{\left( 0,0,0\right) }$ & $\mathbf{=}$ & $\mathbf{3}_{\left(
1,0,-1\right) }\oplus \mathbf{3}_{\left( -1,0,1\right) }^{\prime }$%
\end{tabular}%
\end{equation}%
Notice\ moreover that the previous $t_{i}$- weights are now replaced by new
quantities $x_{i}$ given by some linear combinations of the $t_{i}$'s fixed
by representation theory of $\mathbb{S}_{4}$. One of these weights; say $%
x_{4}$, is given by the usual completely $\mathbb{S}_{4}$- symmetric term 
\begin{equation}
x_{4}\sim \left( t_{1}+t_{2}+t_{3}+t_{4}\right)  \label{x4}
\end{equation}%
transforming in the trivial representation of $\mathbb{S}_{4}$; the three
other $x_{i}$ are given by some orthogonal linear combinations of the four $%
t_{i}$'s that we express as follows%
\begin{equation}
x_{i}=\alpha _{i}t_{1}+\beta _{i}t_{2}+\gamma _{i}t_{3}+\delta _{i}t_{4}
\label{m}
\end{equation}%
These three weights transform as an irreducible triplet of $\mathbb{S}_{4}$;
but seen that we have two kinds of 3-$\dim $ representations in $\mathbb{S}%
_{4}$ namely $\mathbf{3}$ and $\mathbf{3}^{\prime }$, the explicit
expressions of (\ref{m}) depend in which of the two representations the $%
x_{i}$'s are sitting; details are reported in appendix where one also finds
the relationships $t_{\mu }=U_{\mu \rho }x_{\rho }$ and $t_{\mu }\pm t_{\nu
}=\left( U_{\mu \rho }\pm U_{\nu \rho }\right) x_{\rho }$. Notice finally
that the explicit expressions of ${\small X}_{{\small \mu \nu }}$ weights in
(\ref{T2}) are not needed in our approach; their role will be played by the
characters of the representations.

\subsection{$SU_{5}\times \mathbb{S}_{3}\times \left( U_{1}^{\perp }\right)
^{2}$ model}

The spectrum of GUT- curves of the $SU_{5}\times \mathbb{S}_{3}\times \left(
U_{1}^{\perp }\right) ^{2}$ model follows from the spectrum of the $%
SU_{5}\times \mathbb{S}_{5}$ theory by using splitting spectral method. By
working in the canonical basis for t$_{i}$- weights, this spectrum,
expressed in terms of reducible multiplets, is given by (\ref{T3}).\ Here,
we revisit the $SU_{5}\times \mathbb{S}_{3}\times \left( U_{1}^{\perp
}\right) ^{2}$ curves spectrum by using irreducible representations of $%
\mathbb{S}_{3}$ and their characters.\textrm{\newline
}We start by recalling that $\mathbb{S}_{3}$ has three irreducible
representations as shown of the usual character relation $6=1^{2}+1^{2\prime
}+2^{2}$ linking the order of $\mathbb{S}_{3}$ to the squared dimensions of
its irreducible representations; these irreducible representations are
nicely described in terms of Young diagrams \textrm{\cite{E11}}%
\begin{equation}
\mathbf{1}:\text{ \  \ }%
\begin{tabular}{|l|l|l|}
\hline
&  &  \\ \hline
\end{tabular}%
\qquad ,\qquad \mathbf{2}:\text{ \  \ }%
\begin{tabular}{|l|l}
\hline
& \multicolumn{1}{|l|}{} \\ \hline
& \  \  \\ \cline{1-1}
\end{tabular}%
\qquad ,\qquad \mathbf{1}^{\prime }:\text{ \  \ }%
\begin{tabular}{|l|}
\hline
\\ \hline
\\ \hline
\\ \hline
\end{tabular}%
\end{equation}%
The group $\mathbb{S}_{3}$ is a non abelian discrete group; it has two non
commuting generators $\left( a,b\right) $ satisfying $a^{2}=b^{3}=1$ with
characters as follows%
\begin{equation}
\begin{tabular}{|l|l|l|l|}
\hline
$\mathrm{\chi }_{R}$ & $\mathrm{\chi }_{_{\boldsymbol{I}}}$ & $\mathrm{\chi }%
_{_{\boldsymbol{2}}}$ & $\mathrm{\chi }_{_{\epsilon }}$ \\ \hline
$\mathrm{a}$ & $\  \ 1$ & $\ 0$ & $-1$ \\ \hline
$\mathrm{b}$ & $\  \ 1$ & $-1$ & $\  \ 1$ \\ \hline
\end{tabular}%
\end{equation}%
The spectrum of matter curves in the $\mathbb{S}_{3}$-model is obtained here
by starting from the $\mathbb{S}_{4}$\ spectrum $(t_{1},t_{2},t_{3})$\ (\ref%
{T3}); and then breaking $\mathbb{S}_{4}$\ monodromy to $\mathbb{S}%
_{3}\times \mathbb{S}_{1}$\ . We find

\begin{equation}
\begin{tabular}{|c|c|c|c|c|c|c|}
\hline
{\small curves} & {\small weights} & {\small Irrep }$S_{3}$ & $\mathrm{\chi }%
_{R}^{\left( a,b\right) }$ & ${\small U}_{1}^{\perp }$ & {\small homology} & 
{\small U}${\small (1)}_{Y}${\small \ flux} \\ \hline
$\left. 
\begin{array}{c}
{\small 10}_{x_{{\small i}}} \\ 
{\small 10}_{x_{3}} \\ 
{\small 10}_{x_{{\small 4}}} \\ 
{\small 10}_{t_{5}}%
\end{array}%
\right. $ & $\left. 
\begin{array}{c}
{\small x}_{i} \\ 
x_{3} \\ 
{\small x}_{4} \\ 
{\small t}_{5}%
\end{array}%
\right. $ & $\left. 
\begin{array}{c}
\mathbf{2} \\ 
\mathbf{1} \\ 
\mathbf{1} \\ 
\mathbf{1}%
\end{array}%
\right. $ & $\left. 
\begin{array}{c}
{\small (0,-1)} \\ 
{\small (1,1)} \\ 
{\small (1,1)} \\ 
{\small (1,1)}%
\end{array}%
\right. $ & $\left. 
\begin{array}{c}
{\small 0} \\ 
{\small 0} \\ 
{\small 0} \\ 
{\small 1}%
\end{array}%
\right. $ & $\left. 
\begin{array}{c}
{\small \eta -2c}_{1}{\small -\chi ^{\prime }} \\ 
{\small -\chi -c}_{1} \\ 
{\small \chi }^{\prime }{\small -c}_{1} \\ 
{\small \chi -c}_{1}%
\end{array}%
\right. $ & $\left. 
\begin{array}{c}
{\small -P} \\ 
{\small -N} \\ 
{\small P} \\ 
{\small N}%
\end{array}%
\right. $ \\ \hline
$\left. 
\begin{array}{c}
{\small 5}_{X_{{\small ij}}} \\ 
{\small 5}_{X_{{\small i3}}} \\ 
{\small 5}_{X_{{\small i4}}} \\ 
{\small 5}_{X_{{\small 34}}} \\ 
{\small 5}_{X_{i5}} \\ 
{\small 5}_{X_{35}} \\ 
{\small 5}_{X_{45}}%
\end{array}%
\right. $ & $\left. 
\begin{array}{c}
{\small X}_{{\small ij}} \\ 
{\small X}_{{\small i3}} \\ 
{\small X}_{{\small i4}} \\ 
X_{{\small 34}} \\ 
{\small X}_{i5} \\ 
X_{35} \\ 
{\small X}_{45}%
\end{array}%
\right. $ & $\left. 
\begin{array}{c}
\mathbf{2} \\ 
\mathbf{1} \\ 
\mathbf{2} \\ 
\mathbf{1} \\ 
\mathbf{2} \\ 
\mathbf{1} \\ 
\mathbf{1}%
\end{array}%
\right. $ & $\left. 
\begin{array}{c}
{\small (0,-1)} \\ 
{\small (-1,1)} \\ 
{\small (0,-1)} \\ 
{\small (1,1)} \\ 
{\small (0,-1)} \\ 
{\small (1,1)} \\ 
{\small (1,1)}%
\end{array}%
\right. $ & $\left. 
\begin{array}{c}
{\small 0} \\ 
{\small 0} \\ 
{\small 0} \\ 
{\small 0} \\ 
{\small 0} \\ 
{\small 0} \\ 
{\small 1}%
\end{array}%
\right. $ & $\left. 
\begin{array}{c}
{\small \eta }^{\prime }{\small -2c}_{1} \\ 
{\small -c}_{1}{\small -\chi }^{{\small \prime }}-{\small \chi } \\ 
{\small -2c}_{1} \\ 
{\small \chi }^{{\small \prime }}{\small -c}_{1} \\ 
{\small -2c}_{1} \\ 
{\small -c}_{1}{\small -\chi }^{{\small \prime }}{\small +\chi } \\ 
{\small \chi ^{\prime }-c}_{1}%
\end{array}%
\right. $ & $\left. 
\begin{array}{c}
{\small 0} \\ 
{\small -P-N} \\ 
{\small 0} \\ 
{\small P} \\ 
0 \\ 
{\small N-P} \\ 
{\small P}%
\end{array}%
\right. $ \\ \hline
\end{tabular}%
\end{equation}%
\begin{equation*}
\end{equation*}%
where the integers P and N are as in eq(\ref{np}).

\section{$SU_{5}\times \mathbb{D}_{4}$ models}

First notice that the engineering of the $SU_{5}\times \mathbb{D}_{4}\times
U_{1}^{\perp }$ theory has been recently studied in \textrm{\cite{B4} }by
using Galois theory; but here we use a method based on characters of the
irreducible representations of $\mathbb{D}_{4}$; and finds at the end that
there are in fact three kinds of $SU_{5}\times \mathbb{D}_{4}\times
U_{1}^{\perp }$ models; they are explicitly constructed in this section. To
that purpose, we first review useful aspects on characters of the dihedral
group; then we turn to construct the three $\mathbb{D}_{4}\times
U_{1}^{\perp }$ models.

\subsection{Characters in $\mathbb{D}_{4}$ models}

The dihedral $\mathbb{D}_{4}$ is an order \emph{8} subgroup of $\mathbb{S}%
_{4}$ with no 3-cycles; there are three kinds of such subgroups inside $%
\mathbb{S}_{4};$ an example of $\mathbb{D}_{4}$ subgroup is the one having
the following elements%
\begin{equation}
\begin{tabular}{lllllll}
$I_{id}$ & $,$ & $\left. 
\begin{array}{c}
\left( 24\right) \\ 
\left( 13\right)%
\end{array}%
\right. $ & $,$ & $\left. 
\begin{array}{c}
\left( 13\right) \left( 24\right) \\ 
\left( 12\right) \left( 34\right) \\ 
\left( 14\right) \left( 23\right)%
\end{array}%
\right. $ & , & $\left. 
\begin{array}{c}
\left( 1234\right) \\ 
\left( 1432\right)%
\end{array}%
\right. $%
\end{tabular}
\label{D4}
\end{equation}%
\begin{equation*}
\end{equation*}%
with non commuting generators $a=\left \langle \left( 24\right)
\right
\rangle $ and $b=\left \langle \left( 1234\right) \right \rangle $
satisfying $a^{2}=b^{4}=I$ and $aba=b^{3}$. The two other $\mathbb{D}%
_{4}^{\prime }$ and $\mathbb{D}_{4}^{\prime \prime }$ have similar contents;
but with other transpositions and 4-cycles. In terms of $\left( a,b\right) $
generators, the eight elements (\ref{D4}) of the dihedral $\mathbb{D}_{4}$
reads as%
\begin{equation}
\begin{tabular}{lllllll}
$I_{id}$ & $,$ & $\left. 
\begin{array}{c}
a \\ 
b^{2}a%
\end{array}%
\right. $ & , & $\left. 
\begin{array}{c}
b^{2} \\ 
ab \\ 
ba%
\end{array}%
\right. $ & , & $\left. 
\begin{array}{c}
b \\ 
b^{3}%
\end{array}%
\right. $%
\end{tabular}%
\end{equation}%
they form \emph{5} conjugacy classes as follows%
\begin{equation}
\begin{tabular}{lllll}
$\mathfrak{C}_{1}=\left \{ I_{id}\right \} $ & $,$ & $\mathfrak{C}_{2}=\left
\{ b^{2}\right \} $ & $,$ & $\mathfrak{C}_{3}=\left \{ b,b^{3}\right \} $ \\ 
$\mathfrak{C}_{4}=\left \{ a\right \} $ & $,$ & $\mathfrak{C}_{5}=\left \{
ab\right \} $ &  & 
\end{tabular}
\label{c}
\end{equation}%
The dihedral group $\mathbb{D}_{4}$ has also \emph{5} irreducible
representations $\boldsymbol{R}_{i}$; this can be directly learnt on the
character formula $8=1_{1}^{2}+1_{2}^{2}+1_{3}^{2}+1_{4}^{2}+2^{2}$, linking
the order of $\mathbb{D}_{4}$ with the sum of $d_{i}^{2}$, the squares of
the dimensions $d_{i}$ of the irreducible $\boldsymbol{R}_{i}$
representations of $\mathbb{D}_{4}$. So, the order \emph{8} dihedral group
has four irreducible representations with 1-$\dim $; and a fifth irreducible 
$\mathbb{D}_{4}$- representation with 2-$\dim $ \textrm{\cite{E11}}. The
character table of $\mathbb{D}_{4}$ representations is given by 
\begin{equation}
\begin{tabular}{|l|l|l|l|l|l|l|}
\hline
$\mathfrak{C}_{i}$\TEXTsymbol{\backslash}$\mathrm{\chi }_{\boldsymbol{R}%
_{j}} $ & $\  \  \mathrm{\chi }_{_{\mathbf{1}_{1}}}$ & $\  \  \mathrm{\chi }_{_{%
\mathbf{1}_{2}}}$ & $\  \  \mathrm{\chi }_{_{\mathbf{1}_{3}}}$ & $\  \  \mathrm{%
\chi }_{\mathbf{1}_{4}}$ & $\  \  \mathrm{\chi }_{_{2}}$ & number \\ \hline
$\mathfrak{C}_{1}$ & $\  \ 1\  \ $ & $\  \ 1\  \ $ & $\  \ 1\  \ $ & $\  \ 1\  \ $ & 
$\  \ 2\  \ $ & $\  \ 1\  \ $ \\ \hline
$\mathfrak{C}_{2}$ & $\  \ 1$ & $\  \ 1$ & $\  \ 1$ & $\  \ 1$ & $-2$ & $\  \ 1$
\\ \hline
$\mathfrak{C}_{3}$ & $\  \ 1$ & $\  \ 1$ & $-1$ & $-1$ & $\  \ 0$ & $\  \ 2$ \\ 
\hline
$\mathfrak{C}_{4}$ & $\  \ 1$ & $-1$ & $\  \ 1$ & $-1$ & $\  \ 0$ & $\  \ 2$ \\ 
\hline
$\mathfrak{C}_{5}$ & $\  \ 1$ & $-1$ & $-1$ & $\  \ 1$ & $\  \ 0$ & $\  \ 2$ \\ 
\hline
\end{tabular}%
\end{equation}%
\begin{equation*}
\end{equation*}%
from which we learn the following characters of the $\left( a,b\right) $
generators 
\begin{equation}
\begin{tabular}{|l|l|l|l|l|l|}
\hline
$\mathfrak{\chi }_{ij}^{\left( g\right) }$ & $\  \  \mathrm{\chi }_{_{\mathbf{1%
}_{1}}}$ & $\  \  \mathrm{\chi }_{_{\mathbf{1}_{2}}}$ & $\  \  \mathrm{\chi }_{_{%
\mathbf{1}_{3}}}$ & $\  \  \mathrm{\chi }_{\mathbf{1}_{4}}$ & $\  \  \mathrm{%
\chi }_{_{2}}$ \\ \hline
$a$ & $\  \ 1$ & $-1$ & $\  \ 1$ & $-1$ & $\  \ 0$ \\ \hline
$b$ & $\  \ 1$ & $\  \ 1$ & $-1$ & $-1$ & $\  \ 0$ \\ \hline
\end{tabular}
\label{cr}
\end{equation}%
\begin{equation*}
\end{equation*}%
For other features see \textrm{\cite{E10}}. With these tools at hand, we
turn to engineer the $SU_{5}\times \mathbb{D}_{4}\times U_{1}^{\perp }$
models with dihedral monodromy symmetry.

\subsection{Three $\mathbb{D}_{4}$- models}

As in the case of $\mathbb{S}_{3}$ monodromy, the breaking of $\mathbb{S}%
_{4} $ down to $\mathbb{D}_{4}$ is induced by non zero flux\textrm{\ }%
piercing the curves of the $SU_{5}\times \mathbb{S}_{4}\times U_{1}^{\perp }$
model. Using properties from the character table of $\mathbb{D}_{4}$, we
distinguish three kinds of models depending on the way the $\mathbb{S}_{4}$-
irreducible triplets have been pierced; there are three possibilities and
are as described in what follows:\  \ 

\subsubsection{First case: $\mathbf{3}=\mathbf{1}_{+,-}\oplus \mathbf{2}%
_{0,0}$}

In this model, the various irreducible triplets of $\mathbb{S}_{4}$; in
particular those involved in:

\begin{description}
\item[$\left( i\right) $] the five 10-plets namely $\mathbf{5}=\mathbf{1}%
\oplus \mathbf{3}\oplus \mathbf{1}_{t_{5}},$ and

\item[$\left( ii\right) $] the ten 5-plets which includes the four 10-plets
charged under U$_{1}^{\perp }$ namely $\mathbf{4}_{t_{5}}=\mathbf{1}%
_{t_{5}}\oplus \mathbf{3}_{t_{5}},$ and the six uncharged 10-plets given by $%
\mathbf{6}=\mathbf{3}\oplus \mathbf{3}^{\prime }$,
\end{description}

\  \  \  \  \newline
are decomposed as sums of two singlets $\mathbf{1}_{p,q}+\mathbf{1}%
_{p^{\prime },q^{\prime }}$ and a doublet $\mathbf{2}_{0,0}$. The character
properties of the $\mathbb{D}_{4}$- representations indicate that the
decompositions of the triplets should be as 
\begin{equation}
\begin{tabular}{lll}
$\left. \mathbf{3}\right \vert _{\mathbb{D}_{4}}$ & $=$ & $\mathbf{1}%
_{+,-}\oplus \mathbf{2}_{0,0}$ \\ 
$\left. \mathbf{3}^{\prime }\right \vert _{\mathbb{D}_{4}}$ & $=$ & $\mathbf{%
1}_{-,+}\oplus \mathbf{2}_{0,0}$%
\end{tabular}
\label{w1}
\end{equation}

By substituting these relations back into the restricted spectrum resulting
from (\ref{T2}), we end with the following $SU_{5}\times \mathbb{D}%
_{4}\times U_{1}^{\perp }$ spectrum

\begin{itemize}
\item \emph{five 10-plets}%
\begin{equation}
\begin{tabular}{|c|c|c|c|c|c|c|}
\hline
curves & weights & $\mathbb{D}_{4}$ & $\mathrm{\chi }_{_{\mathbf{R}%
}}^{\left( a,b\right) }$ & $U_{1}^{\perp }$ & homology & U$\left( 1\right)
_{Y}$ flux \\ \hline
$\left. 
\begin{array}{c}
10_{y_{{\small i}}} \\ 
10_{y_{{\small 3}}} \\ 
10_{y_{{\small 4}}} \\ 
10_{t_{5}}%
\end{array}%
\right. \ $ & $\left. 
\begin{array}{c}
y_{i} \\ 
y_{{\small 3}} \\ 
y_{4} \\ 
t_{5}%
\end{array}%
\right. $ & $\left. 
\begin{array}{c}
\mathbf{2} \\ 
\mathbf{1} \\ 
\mathbf{1} \\ 
\mathbf{1}%
\end{array}%
\right. $ & $\left. 
\begin{array}{c}
\left( 0,0\right) \\ 
\left( 1,-1\right) \\ 
\left( 1,1\right) \\ 
\left( 1,1\right)%
\end{array}%
\right. $ & $\left. 
\begin{array}{c}
0 \\ 
0 \\ 
0 \\ 
1%
\end{array}%
\right. $ & $\left. 
\begin{array}{c}
\eta -2c_{1}-\varphi \\ 
-c_{1} \\ 
{\small \chi }^{{\small \prime }}-c_{1} \\ 
\chi -c_{1}%
\end{array}%
\right. $ & $\left. 
\begin{array}{c}
-N-P \\ 
0 \\ 
P \\ 
N%
\end{array}%
\right. $ \\ \hline
\end{tabular}
\label{m1}
\end{equation}%
\begin{equation*}
\end{equation*}%
where $\mathrm{\chi }_{_{\mathbf{R}}}^{\left( a,b\right) }$ stands for the
character of the generators in the $\mathbf{R}$ representation; $\varphi
=\chi +\chi ^{{\small \prime }}$, and the integers N and P as in eqs(\ref{np}%
). Notice that the multiplets $10_{y_{{\small 4}}}$ and $10_{t_{5}}$
transform in the same trivial $\mathbb{D}_{4}$- representation; but having
different t$_{5}$- charges; the $10_{y_{{\small 3}}}$ transforms also as a
singlet; but with character $\left( 1,-1\right) $; it is a good candidate
for accommodating the top-quark family.

\item \emph{ten 5-plets}%
\begin{equation}
\begin{tabular}{|c|c|c|c|c|c|c|}
\hline
curves & weight & $\mathbb{D}_{4}$ & $\mathrm{\chi }_{_{\mathbf{R}}}^{\left(
a,b\right) }$ & $U_{1}^{\perp }$ & {\small homology} & U$\left( 1\right)
_{Y} $ flux \\ \hline
$\left. 
\begin{array}{c}
5_{Y_{{\small i3}}} \\ 
5_{Y_{{\small 12}}} \\ 
5_{Y_{{\small i4}}} \\ 
5_{Y_{{\small 34}}} \\ 
5_{Y_{i5}} \\ 
5_{Y_{35}} \\ 
5_{Y_{45}}%
\end{array}%
\right. \ $ & $\left. 
\begin{array}{c}
{\small Y}_{{\small i3}} \\ 
{\small Y}_{{\small 12}} \\ 
{\small Y}_{{\small i4}} \\ 
{\small Y}_{{\small 34}} \\ 
{\small Y}_{i5} \\ 
{\small Y}_{35} \\ 
{\small Y}_{45}%
\end{array}%
\right. $ & $\left. 
\begin{array}{c}
\mathbf{2} \\ 
\mathbf{1} \\ 
\mathbf{2} \\ 
\mathbf{1} \\ 
\mathbf{2} \\ 
\mathbf{1} \\ 
\mathbf{1}%
\end{array}%
\right. $ & $\left. 
\begin{array}{c}
{\small (0,0)} \\ 
{\small (-1,1)} \\ 
{\small (0,0)} \\ 
{\small (1,-1)} \\ 
{\small (0,0)} \\ 
{\small (1,-1)} \\ 
{\small (1,1)}%
\end{array}%
\right. $ & $\left. 
\begin{array}{c}
0 \\ 
0 \\ 
0 \\ 
0 \\ 
1 \\ 
1 \\ 
1%
\end{array}%
\right. $ & $\left. 
\begin{array}{c}
{\small \eta }^{\prime }{\small -2c}_{1}{\small +\varphi } \\ 
-{\small \chi -c}_{1} \\ 
-{\small \chi }^{{\small \prime }}{\small -c}_{1} \\ 
{\small -2c}_{1} \\ 
-{\small \chi }^{{\small \prime }}{\small -c}_{1} \\ 
-{\small \chi -c}_{1} \\ 
{\small \varphi -2c}_{1}%
\end{array}%
\right. $ & $\left. 
\begin{array}{c}
{\small N+P} \\ 
-{\small N} \\ 
-{\small P} \\ 
{\small 0} \\ 
-{\small P} \\ 
-{\small N} \\ 
{\small N+P}%
\end{array}%
\right. $ \\ \hline
\end{tabular}
\label{4t}
\end{equation}%
\begin{equation*}
\end{equation*}%
where we have set $\varphi =\chi +\chi ^{\prime }$. From this table, we
learn that among the ten 5-plets, two sit in the $1_{+,-}$ representation
with character $\left( 1,-1\right) $; but with differen t$_{5}$ charges; one
in $1_{-,+}$ with character $\left( -1,1\right) $ with no t$_{5}$ charge;
and a fourth in the trivial representation of $\mathbb{D}_{4}$ with a unit t$%
_{5}$ charge.\ 

\item \emph{flavons}\newline
Among the \emph{24} flavons of the $SU_{5}\times \mathbb{S}_{4}\times
U_{1}^{\perp }$ model, there are \emph{20} ones charged under $\mathbb{D}%
_{4} $ monodromy symmetry; but because of hermitic feature, they can be
organised into $\mathbf{10}\oplus \mathbf{10}^{\prime }$ subsets with
opposite $\mathbb{D}_{4}$ characters and opposite t$_{5}$ charges. Moreover
due to \ reducibility of the 10-$\dim $ multiplet as $\mathbf{10}=\mathbf{4}%
_{t_{5}}\oplus \mathbf{6}$, which is also equal to $\left( \mathbf{1}%
_{t_{5}}\oplus \mathbf{3}_{t_{5}}\right) \oplus \left( \mathbf{3}\oplus 
\mathbf{3}^{\prime }\right) $; and therefore to the direct sum $\mathbf{1}%
_{+,+}^{t_{5}}\oplus \left( \mathbf{1}_{+,-}^{t_{5}}\oplus \mathbf{2}%
_{0,0}^{t_{5}}\right) $ plus $\left( \mathbf{1}_{+,-}\oplus \mathbf{2}%
_{0,0}\right) +\left( \mathbf{1}_{-,+}\oplus \mathbf{2}_{0,0}\right) $; one
ends with: $\left( \alpha \right) $ flavons doublets $\vartheta ^{i},$ $%
\vartheta _{t_{5}}^{i}$ having character $\left( 0,0\right) $ with and
without $t_{5}$ charges; and $\left( \beta \right) $ flavon singlets having
characters $\left( \pm 1,\pm 1\right) $ with and without $t_{5}$ charges;
they are as collected below.%
\begin{equation}
\begin{tabular}{|c|c|c|c|c|}
\hline
curves & weights & $\mathbb{D}_{4}$ irrep & $\mathrm{\chi }_{_{\mathbf{R}%
}}^{\left( a,b\right) }$ character & $t_{5}$ charge \\ \hline
$\left. 
\begin{array}{c}
1_{\pm Z_{{\small i3}}} \\ 
1_{\pm Z_{{\small 12}}} \\ 
1_{\pm Z_{{\small i4}}} \\ 
1_{\pm Z_{{\small 34}}} \\ 
1_{\pm Z_{i5}} \\ 
1_{\pm Z_{35}} \\ 
1_{\pm Z_{45}}%
\end{array}%
\right. $ & $\left. 
\begin{array}{c}
\pm Z_{{\small i3}} \\ 
\pm Z_{{\small 12}} \\ 
\pm Z_{{\small i4}} \\ 
\pm Z_{{\small 34}} \\ 
\pm Z_{i5} \\ 
\pm Z_{35} \\ 
\pm Z_{45}%
\end{array}%
\right. $ & $\left. 
\begin{array}{c}
\mathbf{2} \\ 
\mathbf{1} \\ 
\mathbf{2} \\ 
\mathbf{1} \\ 
\mathbf{2} \\ 
\mathbf{1} \\ 
\mathbf{1}%
\end{array}%
\right. $ & $\left. 
\begin{array}{c}
{\small (0,0)} \\ 
\pm {\small (-1,1)} \\ 
{\small (0,0)} \\ 
\pm {\small (1,-1)} \\ 
{\small (0,0)} \\ 
\pm {\small (1,-1)} \\ 
\pm {\small (1,1)}%
\end{array}%
\right. $ & $\left. 
\begin{array}{c}
0 \\ 
0 \\ 
0 \\ 
0 \\ 
\mp 1 \\ 
\mp 1 \\ 
\mp 1%
\end{array}%
\right. $ \\ \hline
\end{tabular}%
\end{equation}
\end{itemize}

\subsubsection{Second case: $\mathbf{3}=\mathbf{1}_{+,-}\oplus \mathbf{1}%
_{+,-}\oplus \mathbf{1}_{-,+}$}

This is a completely reducible model; under restriction to dihedral
subsymmetry, the $\mathbf{3}$ and $\mathbf{3}^{\prime }$ triplets of $%
\mathbb{S}_{4}$ are decomposed as follows 
\begin{equation}
\begin{tabular}{lll}
$\left. \mathbf{3}\right \vert _{\mathbb{D}_{4}}$ & $=$ & $\mathbf{1}%
_{+,-}\oplus \mathbf{1}_{+,-}\oplus \mathbf{1}_{-,+}$ \\ 
$\left. \mathbf{3}^{\prime }\right \vert _{\mathbb{D}_{4}}$ & $=$ & $\mathbf{%
1}_{-,+}\oplus \mathbf{1}_{-,+}\oplus \mathbf{1}_{+,-}$%
\end{tabular}
\label{w2}
\end{equation}%
by substituting these decompositions back into the spectrum of $SU_{5}\times 
\mathbb{S}_{4}\times U_{1}^{\perp }$- theory given by (\ref{T2}), we obtain
the curves spectrum of the second $SU_{5}\times \mathbb{D}_{4}\times
U_{1}^{\perp }$ - model:

\begin{itemize}
\item \emph{five 10-plets}\newline
The spectrum of the 10-plets in the $\mathbb{D}_{4}$- model II can be also
deduced from (\ref{m1}) by splitting the $\mathbf{2}_{0,0}$ doublet as $%
\mathbf{1}_{+,-}\oplus \mathbf{1}_{-,+}$; we have 
\begin{equation}
\begin{tabular}{|c|c|c|c|c|c|}
\hline
curves & $\mathbb{D}_{4}$ irrep & {\small character} & $U_{1}^{\perp }$ & 
homology & U$\left( 1\right) _{Y}$ flux \\ \hline
$\left. 
\begin{array}{c}
10_{+,-} \\ 
10_{-,+} \\ 
10_{+,-} \\ 
10_{+,+} \\ 
10_{+,+}^{t_{5}}%
\end{array}%
\right. \ $ & $\left. 
\begin{array}{c}
\mathbf{1}_{+,-} \\ 
\mathbf{1}_{-,+} \\ 
\mathbf{1}_{+,-} \\ 
\mathbf{1}_{+,+} \\ 
\mathbf{1}_{+,+}%
\end{array}%
\right. $ & $\left. 
\begin{array}{c}
\left( 1,-1\right) \\ 
\left( -1,1\right) \\ 
\left( 1,-1\right) \\ 
\left( 1,1\right) \\ 
\left( 1,1\right)%
\end{array}%
\right. $ & $\left. 
\begin{array}{c}
0 \\ 
0 \\ 
0 \\ 
0 \\ 
1%
\end{array}%
\right. $ & $\left. 
\begin{array}{c}
\eta -c_{1}-\chi -{\small \chi }^{{\small \prime }} \\ 
-c_{1} \\ 
-c_{1} \\ 
{\small \chi }^{{\small \prime }}-c_{1} \\ 
\chi -c_{1}%
\end{array}%
\right. $ & $\left. 
\begin{array}{c}
-N-P \\ 
0 \\ 
0 \\ 
P \\ 
N%
\end{array}%
\right. $ \\ \hline
\end{tabular}
\label{m2}
\end{equation}%
\begin{equation*}
\end{equation*}%
Here we have two matter multiplets namely $10_{+,+}$ and $10_{+,+}^{t_{5}}$;
they transform in the same trivial $\mathbb{D}_{4}$- representation with
character $\left( 1,1\right) $; but having different t$_{5}$- charges. We
also have two $10_{+,-}$ multiplets transforming in $1_{+,-}$ with character 
$\left( 1,-1\right) $; but with different fluxes; and one multiplet $%
10_{-,+} $ with character $\left( -1,1\right) $; it will be interpreted in
appendix B as the one accommodating the top-quark family.

\item \emph{ten 5-plets}%
\begin{equation}
\begin{tabular}{|c|c|c|c|c|c|}
\hline
curves & $\mathbb{D}_{4}$ irrep & $\mathrm{\chi }_{_{\mathbf{R}}}^{\left(
a,b\right) }$ & $U_{1}^{\perp }$ & {\small homology} & U$\left( 1\right)
_{Y} $ flux \\ \hline
$\left. 
\begin{array}{c}
5_{+,-} \\ 
5_{-,+} \\ 
5_{-,+} \\ 
5_{+,-} \\ 
5_{-,+} \\ 
5_{+,-} \\ 
5_{+,-}^{t_{5}} \\ 
5_{-,+}^{t_{5}} \\ 
5_{+,-}^{t_{5}} \\ 
5_{+,+}^{t_{5}}%
\end{array}%
\right. \ $ & $\left. 
\begin{array}{c}
\mathbf{1}_{+,-} \\ 
\mathbf{1}_{-,+} \\ 
\mathbf{1}_{-,+} \\ 
\mathbf{1}_{+,-} \\ 
\mathbf{1}_{-,+} \\ 
\mathbf{1}_{+,-} \\ 
\mathbf{1}_{+,-} \\ 
\mathbf{1}_{-,+} \\ 
\mathbf{1}_{+,-} \\ 
\mathbf{1}_{+,+}%
\end{array}%
\right. $ & $\left. 
\begin{array}{c}
{\small (1,-1)} \\ 
{\small (-1,1)} \\ 
{\small (-1,1)} \\ 
{\small (1,-1)} \\ 
{\small (-1,1)} \\ 
{\small (1,-1)} \\ 
{\small (1,-1)} \\ 
{\small (-1,1)} \\ 
{\small (1,-1)} \\ 
{\small (1,1)}%
\end{array}%
\right. $ & $\left. 
\begin{array}{c}
0 \\ 
0 \\ 
0 \\ 
0 \\ 
0 \\ 
0 \\ 
1 \\ 
1 \\ 
1 \\ 
1%
\end{array}%
\right. $ & $\left. 
\begin{array}{c}
{\small \eta }^{\prime }{\small -c}_{1}{\small +\chi +\chi }^{{\small \prime 
}} \\ 
{\small -c}_{1} \\ 
-{\small \chi -c}_{1} \\ 
-{\small \chi }^{{\small \prime }}{\small -c}_{1} \\ 
{\small -c}_{1} \\ 
{\small -c}_{1} \\ 
-{\small \chi }^{{\small \prime }}{\small -c}_{1} \\ 
{\small -c}_{1} \\ 
-{\small \chi -c}_{1} \\ 
{\small -2c}_{1}{\small +\chi +\chi }^{{\small \prime }}%
\end{array}%
\right. $ & $\left. 
\begin{array}{c}
{\small N+P} \\ 
0 \\ 
-{\small N} \\ 
-{\small P} \\ 
0 \\ 
{\small 0} \\ 
-{\small P} \\ 
0 \\ 
-{\small N} \\ 
{\small N+P}%
\end{array}%
\right. $ \\ \hline
\end{tabular}%
\left. {}\right.  \label{5t}
\end{equation}%
\begin{equation*}
\end{equation*}%
where\ N and P as in eqs(\ref{np}). \newline
In this model, there is no flavon doublets; there are only singlet flavons
transforming in the representations $\mathbf{1}_{+,+},$ $\mathbf{1}_{-,-},$ $%
\mathbf{1}_{+,-}$, $\mathbf{1}_{-,+}$ with and without t$_{5}$ charges; they
are denoted in what follows as $\vartheta _{p,q}$ and $\vartheta _{p,q}^{\pm
t_{5}}$ with $p,q=\pm 1$.
\end{itemize}

\subsubsection{Third case: $\mathbf{3}=\mathbf{1}_{+,+}\oplus \mathbf{1}%
_{-,-}\oplus \mathbf{1}_{+,-}$}

This $\mathbb{D}_{4}$- model differs from the previous one by the characters
of the singlets; since in this case the $\mathbb{S}_{4}$- triplets $\left. 
\mathbf{3}\right \vert _{\mathbb{S}_{4}}$ and $\left. \mathbf{3}^{\prime
}\right \vert _{\mathbb{S}_{4}}$ are decomposed in terms of irreducible
representations of $\mathbb{D}_{4}$ like%
\begin{equation}
\begin{tabular}{lll}
$\left. \mathbf{3}\right \vert _{\mathbb{D}_{4}}$ & $=$ & $\mathbf{1}%
_{+,+}\oplus \mathbf{1}_{-,-}\oplus \mathbf{1}_{+,-}$ \\ 
$\left. \mathbf{3}^{\prime }\right \vert _{\mathbb{D}_{4}}$ & $=$ & $\mathbf{%
1}_{+,+}\oplus \mathbf{1}_{-,-}\oplus \mathbf{1}_{-,+}$%
\end{tabular}
\label{w3}
\end{equation}%
Substituting these relationships back into (\ref{T2}), we get the curve
spectrum of the third model namely:

\begin{itemize}
\item \emph{five 10-plets}%
\begin{equation}
\begin{tabular}{|c|c|c|c|c|c|}
\hline
curves & $\mathbb{D}_{4}$ irrep & character & $U_{1}^{\perp }$ & homology & U%
$\left( 1\right) _{Y}$ flux \\ \hline
$\left. 
\begin{array}{c}
10_{+,+} \\ 
10_{-,-} \\ 
10_{+,-} \\ 
10_{+,+} \\ 
10_{+,+}^{t_{5}}%
\end{array}%
\right. \ $ & $\left. 
\begin{array}{c}
\mathbf{1}_{+,+} \\ 
\mathbf{1}_{-,-} \\ 
\mathbf{1}_{+,-} \\ 
\mathbf{1}_{+,+} \\ 
\mathbf{1}_{+,+}%
\end{array}%
\right. $ & $\left. 
\begin{array}{c}
\left( 1,1\right) \\ 
\left( -1,-1\right) \\ 
\left( 1,-1\right) \\ 
\left( 1,1\right) \\ 
\left( 1,1\right)%
\end{array}%
\right. $ & $\left. 
\begin{array}{c}
0 \\ 
0 \\ 
0 \\ 
0 \\ 
1%
\end{array}%
\right. $ & $\left. 
\begin{array}{c}
\eta -c_{1}-\varphi \\ 
-c_{1} \\ 
-c_{1} \\ 
\chi ^{{\small \prime }}-c_{1} \\ 
\chi -c_{1}%
\end{array}%
\right. $ & $\left. 
\begin{array}{c}
-N-P \\ 
0 \\ 
0 \\ 
P \\ 
N%
\end{array}%
\right. $ \\ \hline
\end{tabular}
\label{m3}
\end{equation}%
\begin{equation*}
\end{equation*}%
Here we have three $10_{p,q}$ matter multiplets in the trivial $\mathbb{D}%
_{4}$- representation with character $\left( p,q\right) =\left( 1,1\right) $%
; one of them namely $10_{+,+}^{t_{5}}$ having a t$_{5}$ charge and the two
others not. A fourth curve $10_{+,-}$ in $1_{+,-}$ without t$_{5}$ charge
nor a flux; and a fifth $10_{-,-}$ in $1_{-,-}$ with no t$_{5}$ but carrying
a flux.\newline

\item \emph{ten 5-plets}%
\begin{equation}
\begin{tabular}{|c|c|c|c|c|c|}
\hline
curves & $\mathbb{D}_{4}$ irrep & $\mathrm{\chi }_{_{\mathbf{R}}}^{\left(
a,b\right) }$ & $U_{1}^{\perp }$ & {\small homology} & {\small U}${\small (1)%
}_{{\small Y}}${\small \ flux} \\ \hline
$\left. 
\begin{array}{c}
5_{+,+} \\ 
5_{-,-} \\ 
5_{-,+} \\ 
5_{+,+} \\ 
5_{-,-} \\ 
5_{+,-} \\ 
5_{+,+}^{t_{5}} \\ 
5_{-,-}^{t_{5}} \\ 
5_{+,-}^{t_{5}} \\ 
5_{+,+}^{t_{5}}%
\end{array}%
\right. \ $ & $\left. 
\begin{array}{c}
\mathbf{1}_{+,+} \\ 
\mathbf{1}_{-,-} \\ 
\mathbf{1}_{-,+} \\ 
\mathbf{1}_{+,+} \\ 
\mathbf{1}_{-,-} \\ 
\mathbf{1}_{+,-} \\ 
\mathbf{1}_{+,+} \\ 
\mathbf{1}_{-,-} \\ 
\mathbf{1}_{+,-} \\ 
\mathbf{1}_{+,+}%
\end{array}%
\right. $ & $\left. 
\begin{array}{c}
{\small (1,1)} \\ 
{\small (-1,-1)} \\ 
{\small (-1,1)} \\ 
{\small (1,1)} \\ 
{\small (-1,-1)} \\ 
{\small (1,-1)} \\ 
{\small (1,1)} \\ 
{\small (-1,-1)} \\ 
{\small (1,-1)} \\ 
{\small (1,1)}%
\end{array}%
\right. $ & $\left. 
\begin{array}{c}
0 \\ 
0 \\ 
0 \\ 
0 \\ 
0 \\ 
0 \\ 
1 \\ 
1 \\ 
1 \\ 
1%
\end{array}%
\right. $ & $\left. 
\begin{array}{c}
{\small \eta }^{\prime }{\small -c}_{1}{\small +\chi +\chi }^{{\small \prime 
}} \\ 
-{\small \kappa }_{1}{\small \chi }^{{\small \prime }}{\small -c}_{1} \\ 
-{\small \chi -c}_{1} \\ 
-{\small \kappa }_{2}{\small \chi }^{{\small \prime }}{\small -c}_{1} \\ 
{\small -c}_{1} \\ 
{\small -c}_{1} \\ 
-{\small \kappa }_{1}{\small \chi }^{{\small \prime }}{\small -c}_{1} \\ 
-{\small \kappa }_{2}{\small \chi }^{{\small \prime }}{\small -c}_{1} \\ 
-{\small \chi -c}_{1} \\ 
{\small -2c}_{1}{\small +\chi +\chi }^{{\small \prime }}%
\end{array}%
\right. $ & $\left. 
\begin{array}{c}
{\small N+P} \\ 
-{\small \kappa }_{1}{\small P} \\ 
-{\small N} \\ 
-{\small \kappa }_{2}{\small P} \\ 
0 \\ 
{\small 0} \\ 
-{\small \kappa }_{1}{\small P} \\ 
-{\small \kappa }_{2}{\small P} \\ 
-{\small N} \\ 
{\small N+P}%
\end{array}%
\right. $ \\ \hline
\end{tabular}%
\left. {}\right.  \label{6t}
\end{equation}%
\begin{equation*}
\end{equation*}%
with $\kappa _{1}+\kappa _{2}=1$ whose values will be fixed by the
derivation of MSSM. The ten 5-plets $5_{p,q}$ splits as follows: \emph{4}
with $p=q=1$; the two $5_{+,+}^{t_{5}}$ having a t$_{5}$ charge and the two
others $5_{+,+}$ chargeless; the U$_{1}^{\perp }$ charges and the $\left(
N,P\right) $ fluxes allow to distinguish the four. There are also \emph{3}
types of $5_{-,-}$- plets; two $5_{+,-}$ and one $5_{-,+}$. This model has
no flavon doublets; there are only singlet flavons $\vartheta _{p,q}$ and $%
\vartheta _{p,q}^{\pm t_{5}}$ with $p,q=\pm 1$.
\end{itemize}

\section{MSSM like spectrum}

First, we describe the breaking of the SU$_{5}\times \mathbb{D}_{4}\times
U_{1}^{\perp }$ theory down to supersymmetric standard model; then we study
the derivation of the spectrum of MSSM like model with $\mathbb{D}_{4}$
monodromy; and where the heaviest top-quark family is singled out.

\subsection{Breaking gauge symmetry}

Gauge symmetry is broken by U$\left( 1\right) _{Y}$ hyperflux; by assuming
doublet- triplet splitting produced by $N$ units of U$\left( 1\right) _{Y}$,
but still preserving $\mathbb{D}_{4}\times U_{1}^{\perp }$, the 10-plets and
5-plets get decomposed into irreducible representations of standard model
symmetry. The 5-plets of the SU$_{5}\times \mathbb{D}_{4}\times U_{1}^{\perp
}$ models with multiplicity $M_{5}$ split as \textrm{\cite{F3,F4}}%
\begin{equation}
\begin{tabular}{lll}
$n_{\left( 3,1\right) _{-1/3}}-n_{\left( \bar{3},1\right) _{+1/3}}$ & $=$ & $%
M_{5}$ \\ 
$n_{\left( 1,2\right) _{+1/2}}-n_{\left( 1,2\right) _{-1/2}}$ & $=$ & $%
M_{5}+N$%
\end{tabular}
\label{5}
\end{equation}%
\begin{equation*}
\end{equation*}%
leading to a difference between number of triplets and doublets in the low
energy MSSM effective theory. These two relations are important since for $%
N\neq 0$ the correlation is some how relaxed; by choosing 
\begin{equation}
M_{5}^{\left( Higgs\right) }=0
\end{equation}%
the coloured triplet-antitriplet fields $\left( 3,1\right) _{-1/3}$ and $%
\left( \bar{3},1\right) _{+1/3}$ in the Higgs matter curve come in pair that
form heavy massive states; which decouple at low energy. Moreover, by making
particular choices of the $M_{5}^{\left( matter\right) }$ multiplicities, we
can also have the desired matter curve properties for accommodating fermion
families; in particular the chirality property $n_{\left( 1,2\right)
_{+1/2}}\neq n_{\left( 1,2\right) _{-1/2}}$ which is induced by hyperflux.
Furthermore, due to the flux, we also have different numbers of down quarks $%
d_{L}^{c}$ and lepton doublets $L$.\newline
For the 10- plets of the GUT- model with multiplicity $M_{10}$, we have the
following decompositions \textrm{\cite{D2,F5,F6}} 
\begin{equation}
\begin{tabular}{lll}
$n_{\left( 3,2\right) _{+1/6}}-n_{\left( \bar{3},2\right) _{-1/6}}$ & $=$ & $%
M_{10}$ \\ 
$n_{\left( \bar{3},1\right) _{-2/3}}-n_{\left( 3,1\right) _{+2/3}}$ & $=$ & $%
M_{10}-N$ \\ 
$n_{\left( 1,1\right) _{+1}}-n_{\left( 1,1\right) _{-1}}$ & $=$ & $M_{10}+N$%
\end{tabular}
\label{10}
\end{equation}%
\begin{equation*}
\end{equation*}%
The first relation with $M_{10}\neq 0$ generates up-quark chirality since
the number $n_{\left( 3,2\right) _{+1/6}}$ of $Q_{L}=\left( 3,2\right)
_{+1/6}$ representations differs from the number $n_{\left( \bar{3},2\right)
_{+1/6}}$ of $\bar{Q}_{L}=\left( \bar{3},2\right) _{-1/6}$. With non zero
units of hyperflux, the two extra relations leads to the other desired
splitting; the second relation leads for $N\neq 0$ to lifting the
multiplicities between $Q=\left( 3,2\right) _{+1/6}$ and $u^{c}=\left( \bar{3%
},1\right) _{-2/3}$ while the third relation ensures the chirality property
of $e_{L}^{c}$.\newline
In what follows, we study the derivation of an effective matter curve
spectrum that resembles to the field content of MSSM. In addition to three
families and 
\begin{equation}
\sum M_{10}+\sum M_{5}=0  \label{15}
\end{equation}%
as well as total hyperflux conservation%
\begin{equation}
\sum_{fluxes}N_{i}=0
\end{equation}%
we demand the following:

$\bullet $ only a tree- level Yukawa coupling is allowed; and is given by
the top-quark family,

$\bullet $ the heaviest third generation is the least family affected by
hyperflux,

$\bullet $ MSSM matter generations are in $\mathbb{D}_{4}\times U_{1}^{\perp
}$ representations,

$\bullet $ no dimension 4 and 5 proton decay operators are allowed,

$\bullet $\ no $\mu $- term at a tree level,

$\bullet $ two Higgs doublets H$_{u}$ and H$_{d}$ as required by MSSM.

\subsection{Building the spectrum}

Seen that there are three\ possible $SU_{5}\times \mathbb{D}_{4}\times
U_{1}^{\perp }$ models, we focus on the first model with curve spectrum
given by eqs(\ref{m1}-\ref{4t}); and consider first the 10-plets; then turn
after to 5-plets. Results regarding the two other models II and III are
reported in appendix B.

\subsubsection{Ten-plets sector in $\mathbb{D}_{4}$- model I}

The five 10-plets of the $\mathbb{D}_{4}$ model carry different quantum
numbers with respect to $\mathbb{D}_{4}\times U_{1}^{\perp }$
representations, different hyperflux units $\left( N,P\right) ;$ and
different $M_{10}^{\left( n\right) }$ multiplicities satisfying the
properties (\ref{10}). By thinking about $\sum M_{10}$ as given by the
number of MSSM generations%
\begin{equation}
\sum M_{10}=3  \label{51}
\end{equation}%
and taking into account that the two components of the $10_{i}$- doublet are
monodromy equivalent; it follows that one of the five 10-plets should be
disregarded; at least at a tree level analysis. Moreover, using the property
that top- quark 10-plet should be a $\mathbb{D}_{4}$- singlet; one may
choose the $M_{10}^{\left( n\right) }$'s as in following table,

\begin{equation}
\begin{tabular}{|c|c|c|c|c|c|}
\hline
curves & $\mathbb{D}_{4}$ irrep & $U_{1}^{\perp }$ & {\small U}${\small (1)}%
_{{\small Y}}${\small \ }flux & \multicolumn{2}{|c|}{multiplicity} \\ \hline
$\left. 
\begin{array}{c}
10_{i} \\ 
10_{3} \\ 
10_{4} \\ 
10_{5}%
\end{array}%
\right. $ & $\left. 
\begin{array}{c}
\mathbf{2}_{0,0} \\ 
\mathbf{1}_{+,-} \\ 
\mathbf{1}_{++} \\ 
\mathbf{1}_{++}%
\end{array}%
\right. $ & $\left. 
\begin{array}{c}
0 \\ 
0 \\ 
0 \\ 
1%
\end{array}%
\right. $ & $\left. 
\begin{array}{c}
-N-P \\ 
0 \\ 
P \\ 
N%
\end{array}%
\right. $ & $\left. 
\begin{array}{c}
M_{10}^{\left( a\right) } \\ 
M_{10}^{\left( 3\right) } \\ 
M_{10}^{\left( 4\right) } \\ 
M_{10}^{\left( 5\right) }%
\end{array}%
\right. $ & $\left. 
\begin{array}{c}
2 \\ 
1 \\ 
0 \\ 
0%
\end{array}%
\right. $ \\ \hline
\end{tabular}%
\end{equation}%
\begin{equation*}
\end{equation*}%
where chiral modes of $10_{4}$ have been ejected ($M_{10}^{\left( 4\right)
}=0$). Notice that the top- quark generation can a priori be taken in any
one of the three $\mathbb{D}_{4}$- singlets; that is either $10_{3}$ or $%
10_{4}$; or $10_{5}$; the basic difference between these $\mathbb{D}_{4}$-
singlets is given by t$_{5}$ charge and hyperflux. But the choice of the $%
10_{3}$-multiplet looks be the natural one as it is unaffected by hyperflux,
a desired property for MSSM and beyond; and has no t$_{5}$ charge%
\begin{equation}
\mathbf{10}_{3}=\left( Q_{L},U_{L}^{c},e_{L}^{c}\right) \equiv \mathbf{10}%
_{+,-}
\end{equation}%
This multiplet captures also an interesting signature of $\mathbb{D}_{4}$
monodromy in the sense it behaves as a $\mathbb{D}_{4}$-singlet $1_{+,-}$
with non trivial character $\left( +1,-1\right) $. The importance of this
feature at modeling level is twice: $\left( \mathbf{i}\right) $ first it
fixes the quantum numbers of the $5_{H_{u}}$ Higgs representation as a $%
\mathbb{D}_{4}$- singlet $5_{p,q}$ as shown on the tree level top- quark
Yukawa coupling 
\begin{equation}
10_{+,-}\otimes 10_{+,-}\otimes 5_{H_{u}}  \label{3y}
\end{equation}%
Monodromy invariance of (\ref{3y}) under $\mathbb{D}_{4}\times U_{1}^{\perp
} $ requires $5_{H_{u}}$ in the trivial representation with no t$_{5}$
charge; i.e: $5_{H_{u}}\sim 1_{+,+}$. However, an inspection of the
characters of the U$_{1}^{\perp }$ chargeless 5-plets revels that there is
no $\left( 5_{+,+}\right) _{t_{5}=0}$ in the spectrum of the the $\mathbb{D}%
_{4}\times U_{1}^{\perp }$- models I and II constructed above. To bypass
this constraint, we realise the role of the Higgs $5_{H_{u}}$ by allowing
VEVs to come from flavons as well; in other words by thinking of $5_{H_{u}}$
as follows%
\begin{equation}
5_{H_{u}}\text{ \  \  \ }\rightarrow \text{ \  \  \ }5_{p,q}\otimes \vartheta
_{p^{\prime },q^{\prime }}\qquad with\qquad pp^{\prime }=1,\text{ }%
qq^{\prime }=1  \label{pq}
\end{equation}%
where $\vartheta _{p^{\prime },q^{\prime }}$ stands for a flavon in the
representation $1_{p^{\prime },q^{\prime }}$.\newline
$\left( \mathbf{ii}\right) $ second it gives an important tool to
distinguish between matter and Higgs in the 5-plets sector as manifestly
exhibited by the tri-coupling $10_{+,-}\otimes \bar{5}^{M}\otimes \bar{5}%
_{H_{d}}$. This interaction requires matter $\bar{5}_{3}^{M}$ and Higgs $%
\bar{5}_{H_{d}}$ to be in different $\mathbb{D}_{4}$- singlets $1_{p,q}$ and 
$1_{p^{\prime },q^{\prime }}$ with $pp^{\prime }=1$ and $qq^{\prime }=-1$;
see discussion given later on. \newline
By choosing the hyperflux units as $N=P=1$; and using (\ref{10}) we obtain
the matter content 
\begin{equation}
\begin{tabular}{|c|c|c|c|c|c|}
\hline
curves & $\mathbb{D}_{4}$ & $U_{1}^{\perp }$ & flux & matter content & Z$%
_{2} $ parity \\ \hline
$\left. 
\begin{array}{c}
10_{i} \\ 
10_{3} \\ 
10_{4} \\ 
10_{5}%
\end{array}%
\right. $ & $\left. 
\begin{array}{c}
\mathbf{2}_{0,0} \\ 
\mathbf{1}_{+,-} \\ 
\mathbf{1}_{+,+} \\ 
\mathbf{1}_{+,+}%
\end{array}%
\right. $ & $\left. 
\begin{array}{c}
0 \\ 
0 \\ 
0 \\ 
1%
\end{array}%
\right. $ & $\left. 
\begin{array}{c}
-2 \\ 
0 \\ 
1 \\ 
1%
\end{array}%
\right. $ & $\left. 
\begin{array}{c}
2Q_{L}\oplus 4e_{L}^{c} \\ 
Q_{L}\oplus U_{L}^{c}\oplus e_{L}^{c} \\ 
U_{L}^{c}\ominus e_{L}^{c} \\ 
U_{L}^{c}\ominus e_{L}^{c}%
\end{array}%
\right. $ & $%
\begin{array}{ccc}
\varkappa _{43} & = & - \\ 
\varkappa _{42} & = & - \\ 
\varkappa _{41} & = & + \\ 
\varkappa _{1} & = & +%
\end{array}%
$ \\ \hline
\end{tabular}
\label{3u}
\end{equation}%
\begin{equation*}
\end{equation*}%
Notice that by following \textrm{\cite{B4}} using Galois theory, the
10-plets have been attributed $\mathbb{Z}_{2}$ parity charges as reported by
the last column of above table. In our formulation these parities correspond
to $s_{i}$ $\rightarrow -s_{i}$ and $\varkappa _{1}$ and $\varkappa
_{4}=\varkappa _{41}\varkappa _{42}\varkappa _{43}$ as in eq(\ref{pa}); by
help of (\ref{Y1}) and (\ref{Y11}) we obtain 
\begin{equation}
\begin{tabular}{lllll}
$\mathbb{Z}_{2}\left( b_{5}\right) $ & $=+1$ & , & $\mathbb{Z}_{2}\left(
b_{0}\right) $ & $=-1$ \\ 
$\mathbb{Z}_{2}\left( d_{10}\right) $ & $=-1$ & , & $\mathbb{Z}_{2}\left(
d_{0}\right) $ & $=-1$%
\end{tabular}
\label{d}
\end{equation}%
in agreement with (\ref{d10}).

\subsubsection{Five-plets sector}

Like for 10-plets, the ten 5-plets carry different quantum numbers of $%
\mathbb{D}_{4}\times U_{1}^{\perp }$ representations, hyperflux units $%
\left( N,P\right) $ and $M_{5}^{\left( n\right) }$ multiplicities as in (\ref%
{5}). To have a matter curve spectrum that resembles to MSSM, we choose the $%
M_{5}^{\left( n\right) }$'s and the hyperflux as

\begin{equation}
\begin{tabular}{|c|c|c|c|c|c|}
\hline
curves & $\mathbb{D}_{4}$ irrep & $U_{1}^{\perp }$ & homology & flux & 
multiplicity \\ \hline
$\left. 
\begin{array}{c}
5_{Y_{{\small i3}}} \\ 
5_{Y_{{\small 12}}} \\ 
5_{Y_{{\small i4}}} \\ 
5_{Y_{{\small 34}}} \\ 
5_{Y_{i5}} \\ 
5_{Y_{35}} \\ 
5_{Y_{45}}%
\end{array}%
\right. \ $ & $\left. 
\begin{array}{c}
\mathbf{2}_{{\small 0,0}} \\ 
\mathbf{1}_{{\small -,+}} \\ 
\mathbf{2}_{{\small 0,0}} \\ 
\mathbf{1}_{{\small +,-}} \\ 
\mathbf{2}_{{\small 0,0}} \\ 
\mathbf{1}_{{\small +,-}} \\ 
\mathbf{1}_{{\small +,+}}%
\end{array}%
\right. $ & $\left. 
\begin{array}{c}
{\small 0} \\ 
{\small 0} \\ 
{\small 0} \\ 
{\small 0} \\ 
{\small -1} \\ 
{\small -1} \\ 
{\small -1}%
\end{array}%
\right. $ & $\left. 
\begin{array}{c}
{\small \eta }^{\prime }{\small -2c}_{1}{\small -\chi }^{\prime }{\small %
+\xi }^{\prime } \\ 
{\small \chi }^{\prime }{\small -c}_{1} \\ 
{\small \xi }^{\prime }{\small -c}_{1} \\ 
{\small -2c}_{1} \\ 
{\small \xi }^{\prime }{\small -c}_{1} \\ 
{\small \chi }^{\prime }{\small -c}_{1} \\ 
{\small -2c}_{1}{\small -\chi }^{\prime }{\small -\xi }^{\prime }%
\end{array}%
\right. $ & $\left. 
\begin{array}{c}
{\small -N-P} \\ 
{\small N} \\ 
{\small P} \\ 
{\small 0} \\ 
{\small P} \\ 
{\small N} \\ 
{\small -N-P}%
\end{array}%
\right. $ & $\left. 
\begin{array}{c}
{\small M}_{{\small 5}}^{{\small (1)}} \\ 
{\small M}_{{\small 5}}^{{\small (2)}} \\ 
{\small M}_{{\small 5}}^{{\small (3)}} \\ 
{\small M}_{{\small 5}}^{{\small (4)}} \\ 
{\small M}_{{\small 5}}^{{\small (5)}} \\ 
{\small M}_{{\small 5}}^{{\small (6)}} \\ 
{\small M}_{{\small 5}}^{{\small (7)}}%
\end{array}%
\right. $ \\ \hline
\end{tabular}%
\end{equation}%
\begin{equation*}
\end{equation*}%
where $\chi ^{\prime }$ and $\xi ^{\prime }$ are two classes playing similar
role as in the case of breaking $\mathbb{S}_{5}$ monodromy down to $\mathbb{S%
}_{3}$. By using (\ref{15}-\ref{51}), we have%
\begin{equation}
\sum M_{5}=-\sum M_{10}=-3
\end{equation}%
and thinking of this number as $\sum M_{5}=3-6$, a possible configuration
for a MSSM like spectrum is given by 
\begin{equation}
\left. 
\begin{array}{ccc}
{\small M}_{{\small 5}}^{{\small (1)}} & = & {\small 2} \\ 
{\small M}_{{\small 5}}^{{\small (2)}} & = & {\small 0} \\ 
{\small M}_{{\small 5}}^{{\small (3)}} & = & -4 \\ 
{\small M}_{{\small 5}}^{{\small (4)}} & = & {\small 0} \\ 
{\small M}_{{\small 5}}^{{\small (6)}} & = & {\small 1} \\ 
{\small M}_{{\small 5}}^{{\small (7)}} & = & -{\small 2}%
\end{array}%
\right.
\end{equation}%
\begin{equation*}
\end{equation*}%
By choosing the hyperflux as $N=P=1$, and putting back into above table, we
obtain, after relabeling, the 5-plets 
\begin{equation}
\begin{tabular}{|c|c|c|c|c|c|c|}
\hline
curves & $\mathbb{D}_{4}$ & $U_{1}^{\perp }$ & flux & $M_{5}^{\left(
n\right) }$ & matter & parity \\ \hline
$\left. 
\begin{array}{c}
\left( 5_{{\small i}}^{M}\right) _{0} \\ 
\left( 5_{{\small -,+}}^{{\small H}_{{\small u}}}\right) _{0} \\ 
\left( 5_{{\small +,-}}^{M}\right) _{0} \\ 
\left( 5_{{\small +,+}}^{{\small H}_{d}}\right) _{-t_{5}} \\ 
\left( 5_{{\small +,-}}^{M}\right) _{-t_{5}} \\ 
\left( 5_{{\small i}}^{M}\right) _{-t_{5}} \\ 
\left( 5_{{\small i}}^{M}\right) _{0}%
\end{array}%
\right. \ $ & $\left. 
\begin{array}{c}
\mathbf{2}_{{\small 0,0}} \\ 
\mathbf{1}_{{\small -,+}} \\ 
\mathbf{1}_{{\small +,-}} \\ 
\mathbf{1}_{{\small +,+}} \\ 
\mathbf{1}_{{\small +,-}} \\ 
\mathbf{2}_{{\small 0,0}} \\ 
\mathbf{2}_{{\small 0,0}}%
\end{array}%
\right. $ & $\left. 
\begin{array}{c}
0 \\ 
0 \\ 
0 \\ 
-1 \\ 
-1 \\ 
-1 \\ 
0%
\end{array}%
\right. $ & $\left. 
\begin{array}{c}
-2 \\ 
1 \\ 
1 \\ 
1 \\ 
1 \\ 
-2 \\ 
0%
\end{array}%
\right. $ & $%
\begin{array}{c}
{\small 2} \\ 
{\small 0} \\ 
-4 \\ 
{\small 0} \\ 
{\small 1} \\ 
-{\small 2} \\ 
{\small 0}%
\end{array}%
$ & $\left. 
\begin{array}{c}
{\small 2\bar{d}}_{L}^{c} \\ 
{\small H}_{u} \\ 
{\small -4\bar{d}}_{L}^{c}{\small -3\bar{L}} \\ 
{\small -H}_{{\small d}} \\ 
{\small \bar{d}}_{L}^{c} \\ 
{\small -2\bar{d}}_{L}^{c} \\ 
{\small 0}%
\end{array}%
\right. $ & $\left. 
\begin{array}{ccc}
{\small \tilde{\varkappa}}_{61} & = & + \\ 
{\small \tilde{\varkappa}}_{62} & = & + \\ 
{\small \tilde{\varkappa}}_{63} & = & - \\ 
{\small \tilde{\varkappa}}_{41} & = & + \\ 
{\small \tilde{\varkappa}}_{42} & = & + \\ 
{\small \tilde{\varkappa}}_{43} & = & + \\ 
{\small \tilde{\varkappa}}_{64} & = & +%
\end{array}%
\right. $ \\ \hline
\end{tabular}
\label{hu}
\end{equation}%
\begin{equation*}
\end{equation*}%
From this table we learn that the up-Higgs 5-plet $\left( 5_{{\small -,+}}^{%
{\small H}_{{\small u}}}\right) _{0}$ has a character equal to $\left(
-1,+1\right) $ and no t$_{5}$ charge; by substituting in (\ref{pq}), we
obtain $5_{H_{u}}\sim \left( 5_{{\small -,+}}^{{\small H}_{{\small u}%
}}\right) _{0}\otimes \left( \vartheta _{-,+}\right) _{0}$. \newline
We also learn that the 5-plet $(5_{{\small +,-}}^{M})_{0}$ is the least
multiplet affected by hyperflux; and because of our assumptions, it is the
candidate for matter $\bar{5}_{3}^{M}$; the partner of $10_{3}$ in the
underlying SO$_{10}$ GUT-model. With this choice, the down-type quarks
tri-coupling for the third family namely $10_{3}\otimes \bar{5}%
_{3}^{M}\otimes \bar{5}_{{}}^{H_{d}}$; and which we rewrite like 
\begin{equation}
10_{+,-}\otimes \bar{5}_{p,q}^{M}\otimes \bar{5}_{p^{\prime },q^{\prime
}}^{H_{d}}\qquad with\qquad pp^{\prime }=1,qq^{\prime }=-1  \label{5h}
\end{equation}%
This coupling requires the matter $\bar{5}_{3}^{M}$ and the down-Higgs $\bar{%
5}^{H_{d}}$ multiplets to belong to different $\mathbb{D}_{4}$ singlets seen
that $10_{3}$ is in $\mathbf{1}_{+,-}$ representation. However, the
candidates $(\bar{5}_{-{\small ,-}}^{{\small H}_{d}})_{t_{5}}$ and $\bar{5}%
_{3}^{M}\equiv $ $(\bar{5}_{-,{\small +}}^{M})_{0}$ are ruled out because of
the non conservation of t$_{5}$ charge. Nevertheless, a typical diagonal
mass term of third family may be generated by using a flavon $\vartheta
_{-t_{5}}^{+}$ carrying $-1$ unit charge under $U_{1}^{\perp }$ and
transforming as a trivial $\mathbb{D}_{4}$ singlet. This leads to the
realisation $\bar{5}^{H_{d}}\sim (\bar{5}_{-{\small ,-}}^{{\small H}%
_{d}})_{t_{5}}(\vartheta _{++})_{-t_{5}}$; and then to 
\begin{equation}
\left( 10_{+,-}\right) _{0}\otimes \left( \bar{5}_{-,{\small +}}^{M}\right)
_{0}\otimes \left( \bar{5}_{-{\small ,-}}^{{\small H}_{d}}\right)
_{t_{5}}\otimes \left( \vartheta _{++}\right) _{-t_{5}}  \label{40}
\end{equation}%
Non diagonal 4-order coupling superpotentials with one $\left(
10_{+,-}\right) _{0}$ are as follows\textrm{\footnote{%
a complete classification requires also use Z$_{2}$ partity; see \cite{B4}.}}%
\begin{equation}
\begin{tabular}{l}
$\left( 10_{+,-}\right) _{0}\otimes \left( 10_{+,+}\right) _{0}\otimes
\left( 5_{{\small -,+}}^{{\small H}_{{\small u}}}\right) _{0}\otimes \left(
\vartheta _{-,-}\right) _{0}$ \\ 
$\left( 10_{+,-}\right) _{0}\otimes \left( 10_{+,+}\right) _{t_{5}}\otimes
\left( 5_{{\small -,+}}^{{\small H}_{{\small u}}}\right) _{0}\otimes \left(
\vartheta _{-,-}\right) _{-t_{5}}$ \\ 
$\left( 10_{+,-}\right) _{0}\otimes \left( 5_{{\small -,+}}^{{\small H}_{%
{\small u}}}\right) _{0}\otimes \left( 10_{0,0}^{i}\right) _{0}\otimes
\left( \vartheta _{0,0}^{i}\right) _{0}$ \\ 
$\left( 10_{+,-}\right) _{0}\otimes \left( 10_{+,-}\right) _{0}\otimes
\left( 5_{{\small -,+}}^{{\small H}_{{\small u}}}\right) _{0}\otimes \left(
\vartheta _{-,+}\right) _{0}$ \\ 
$\left( 10_{+,-}\right) _{0}\otimes \left( 5_{{\small -,+}}^{{\small H}_{%
{\small u}}}\right) _{0}\otimes \left( 10_{0,0}^{i}\right) _{0}\left(
\vartheta _{0,0}^{i}\right) _{0}$ \\ 
$\left( 10_{+,+}\right) _{0}\otimes \left( 5_{{\small -,+}}^{{\small H}_{%
{\small u}}}\right) _{0}\otimes \left( 10_{0,0}^{i}\right) _{0}\otimes
\left( \vartheta _{0,0}^{i}\right) _{0}$ \\ 
$\left( 10_{+,+}\right) _{t_{5}}\otimes \left( 5_{{\small -,+}}^{{\small H}_{%
{\small u}}}\right) _{0}\otimes \left( 10_{0,0}^{i}\right) _{0}\otimes
\left( \vartheta _{0,0}^{i}\right) _{-t_{5}}$%
\end{tabular}%
\end{equation}%
Below, we discuss some properties of these couplings.

\subsection{More on couplings in $\mathbb{D}_{4}$ model I}

First, we study the quark sector; and turn after to the case of leptons.

\subsubsection{Quark sector}

From the view of supersymmetric standard model with $SU\left( 3\right)
\times SU_{L}\left( 2\right) \times U_{Y}\left( 1\right) $ gauge symmetry;
and denoting the triplet and doublet components of the Higss 5-plets $%
5^{H_{x}}=3^{H_{x}}\oplus 2^{H_{x}}$ respectively like $D_{x}\oplus H_{x}$,
the usual tree level up/down-type Yukawa couplings in $SU_{5}$ model split
like 
\begin{equation}
\begin{tabular}{lllll}
$10^{M}.10^{M}.5^{H_{u}}$ &  & $\rightarrow $ &  & $%
Qu^{c}H_{u}+u^{c}e^{c}D_{u}^{c}+QQD_{u}^{c}$ \\ 
$10^{M}.\overline{5}^{M}.\overline{5}^{H_{d}}$ &  & $\rightarrow $ &  & $%
Qd^{c}H_{d}+e^{c}LH_{d}+QD_{d}^{c}L$%
\end{tabular}
\label{a}
\end{equation}%
They involve up/down Higgs triplets $D_{u}^{c}$\ and $D_{d}^{c},$ which are
exotic to MSSM; but with the hyperflux $U_{Y}\left( 1\right) $ choice we
have made in $SU_{5}\times \mathbb{D}_{4}\times U_{1}^{\perp }$ model (\ref%
{3u},\ref{hu}), they are removed; therefore we have%
\begin{equation}
\begin{tabular}{lllll}
$10^{M}.10^{M}.5^{H_{u}}$ &  & $\rightarrow $ &  & $Qu^{c}H_{u}\ $ \\ 
$10^{M}.\overline{5}^{M}.\overline{5}^{H_{d}}$ &  & $\rightarrow $ &  & $%
Qd^{c}H_{d}+e^{c}LH_{d}\ $%
\end{tabular}%
\end{equation}%
with right hand sides capturing same monodromy representations as left hand
sides; that is $Q$, $u^{c}$ same $\mathbb{D}_{4}\times U_{1}^{\perp }$
representations as $10^{M};$ and so on. In what follows, we study each of
these terms separately by taking into account $\vartheta _{p,q}$ flavon
contributions up to order four couplings; some of these flavons are
interpreted as right neutrinos; they will be discussed at proper time.

\  \  \  \  \ 

$\bullet $ \emph{Up-type Yukawa couplings}\newline
Because of the $\mathbb{D}_{4}\times U_{1}^{\perp }$ monodromy charge of the
up-Higgs 5-plet like $(5_{-,+}^{H_{u}})_{0}$, there is no monodromy
invariant 3-coupling type $10^{M}.10^{M}.5^{H_{u}}$. As shown by eq(\ref{pq}%
), one needs to go to higher orders by implementing flavons with quantum
numbers depending on the monodromy representation of the 10-plets. Indeed,
by focussing on the third generation $10_{3}^{M}\equiv (10_{+,-})_{0}$; we
can distinguish diagonal and non diagonal interactions; an inspection of $%
\mathbb{D}_{4}$ quantum numbers of matter and Higgs multiplets reveals that
we need $\mathbb{D}_{4}$- charged flavons to have monodromy invariant
superpotentials as shown below%
\begin{equation}
W_{top}^{(4)}=\alpha _{3}Tr[(10_{+,-})_{0}\otimes (10_{+,-})_{0}\otimes
(5_{-,+}^{H_{u}})_{0}\otimes (\vartheta _{-,+})_{0}]
\end{equation}%
By restricting to VEVs $\left \langle \vartheta _{-,+}\right \rangle =\rho
_{0} $ and $\left \langle H_{u}\right \rangle =v_{u}$; this non
renormalisable coupling leads to the top quark mass term $%
m_{t}Q_{3}u_{3}^{c} $ with $m_{t}$ equal to $\alpha _{3}v_{u}\rho _{0}$.
Such a term should be thought of as a particular contribution to a general
up-quark mass terms $u_{i}^{c}M^{ij}u_{j}$ with 3$\times $3 mass matrix as
follows%
\begin{equation}
M_{u,c,t}=v_{u}\left( 
\begin{array}{ccc}
\ast & \ast & \ast \\ 
\ast & \ast & \ast \\ 
\ast & \ast & \alpha _{3}\rho _{0}%
\end{array}%
\right)
\end{equation}%
where the ($\ast $)'s refer to contributions coming from other terms
including non diagonal couplings; one of them is%
\begin{equation}
Tr[(10_{+,-})_{0}\otimes (5_{-,+}^{H_{u}})_{0}\otimes (10_{0,0})_{0}\otimes
(\vartheta _{0,0})_{0}^{\prime }]  \label{yc}
\end{equation}%
it involves a 10-plet doublet $(10_{0,0})_{0}\equiv \left( 10_{i}\right)
_{0} $ and a flavon doublet $(\vartheta _{0,0})_{0}^{\prime }\equiv \left(
\vartheta _{i}\right) _{0}^{\prime }$ with VEVs $(\rho _{1},\rho _{2})$; the
latter $(\vartheta _{0,0})_{0}^{\prime }$ will be combined the 10$_{i}$-plet
doublet like $(10_{0,0})_{0}\otimes (\vartheta _{0,0})_{0}^{\prime }$ to
make a scalar. Indeed, the tensor product can be reduced as direct sum over
irreducible representations of $\mathbb{D}_{4}$ having amongst others the $%
\mathbb{D}_{4}$-component 
\begin{equation}
S_{-,-}=\left. (10_{0,0})_{0}\otimes (\vartheta _{0,0})_{0}^{\prime }\right
\vert _{-,-}
\end{equation}%
with $(-,-)$ charge character. This negative charge is needed to compensate
the $(-,-)$ charge coming from $(10_{+,-})_{0}\otimes (5_{-,+}^{H_{u}})_{0}$%
. Restricting to quarks, this reduction corresponds to $(10_{0,0})_{0}%
\otimes (\vartheta _{0,0})_{0}^{\prime }\rightarrow Q_{i}\otimes \rho _{i}$
with%
\begin{equation}
\left. Q_{i}\otimes \rho _{i}\right \vert _{(-,-)}=Q_{1}\rho _{2}-Q_{2}\rho
_{1}
\end{equation}%
Putting back into (\ref{yc}), and thinking of $S_{-,-}$ in terms of the
linear combination $\alpha _{2}(Q_{1}\rho _{2}-Q_{2}\rho _{1})$ of quarks,
we obtain $\alpha _{2}v_{u}(Q_{1}\rho _{2}-Q_{2}\rho _{1})u_{3}^{c}$; which
can be put into the form $u_{i}^{c}M^{ij}u_{j}$ with mass matrix as%
\begin{equation}
M_{u,c,t}=v_{u}\left( 
\begin{array}{ccc}
\ast & \ast & \alpha _{2}\rho _{2} \\ 
\ast & \ast & -\alpha _{2}\rho _{1} \\ 
\ast & \ast & \alpha _{3}\rho _{0}%
\end{array}%
\right)  \label{Mu}
\end{equation}%
One can continue to fill this mass matrix by using the VEV's of other
flavons; however to do that, one needs to rule out couplings with those
flavons describing right neutrinos $\nu _{i}^{c}$. Extending ideas from 
\textrm{\cite{B4}, }the 3 generations of the right handed neutrinos $\nu
_{i}^{c}$ in $SU_{5}\times \mathbb{D}_{4}\times U_{1}^{\perp }$ model should
be as 
\begin{equation}
\begin{tabular}{lllll}
$\nu _{3}^{c}$ &  & $\rightarrow $ &  & $\left( \vartheta _{+,-}\right) _{0}$
\\ 
$(\nu _{1}^{c},\nu _{2}^{c})^{\top }$ &  & $\rightarrow $ &  & $\left(
\vartheta _{0,0}\right) _{0}$%
\end{tabular}
\label{nn}
\end{equation}%
with the following features among the set of 15 flavons of the model%
\begin{equation}
\begin{tabular}{|c|c|c|c|c|c|}
\hline
flavons & $SU_{5}$ & $\mathbb{D}_{4}$ irrep & $U_{1}^{\perp }$ & Z$_{2}$
Parity & VEV \\ \hline
$\left. 
\begin{array}{c}
\left( \vartheta _{0,0}\right) _{0}^{\prime } \\ 
\left( \vartheta _{-,+}\right) _{0} \\ 
\left( \vartheta _{0,0}\right) _{\pm t_{5}} \\ 
\left( \vartheta _{+,-}\right) _{\pm t_{5}} \\ 
\left( \vartheta _{+,+}\right) _{\pm t_{5}}%
\end{array}%
\right. $ & $\left. 
\begin{array}{c}
1 \\ 
1 \\ 
1 \\ 
1 \\ 
1%
\end{array}%
\right. $ & $\left. 
\begin{array}{c}
2 \\ 
1 \\ 
2 \\ 
1 \\ 
1%
\end{array}%
\right. $ & $\left. 
\begin{array}{c}
0 \\ 
0 \\ 
\pm 1 \\ 
\pm 1 \\ 
\pm 1%
\end{array}%
\right. $ & $\left. 
\begin{array}{c}
+ \\ 
+ \\ 
+ \\ 
+ \\ 
\mp%
\end{array}%
\right. $ & $\left. 
\begin{array}{c}
(\rho _{1},\rho _{2})^{\top } \\ 
\rho _{0} \\ 
(\sigma _{1},\sigma _{2})^{\top } \\ 
- \\ 
\omega%
\end{array}%
\right. $ \\ \hline
$\left. 
\begin{array}{ccc}
\left( \vartheta _{0,0}\right) _{0} & = & (\nu _{1}^{c},\nu _{2}^{c})^{\top }
\\ 
\left( \vartheta _{+,-}\right) _{0} & = & \nu _{3}^{c}%
\end{array}%
\right. $ & $\left. 
\begin{array}{c}
1 \\ 
1%
\end{array}%
\right. $ & $\left. 
\begin{array}{c}
2 \\ 
1%
\end{array}%
\right. $ & $\left. 
\begin{array}{c}
0 \\ 
0%
\end{array}%
\right. $ & $\left. 
\begin{array}{c}
- \\ 
-%
\end{array}%
\right. $ & $\left. 
\begin{array}{c}
- \\ 
-%
\end{array}%
\right. $ \\ \hline
\end{tabular}
\label{vev}
\end{equation}%
\begin{equation*}
\end{equation*}%
Therefore, the contribution to (\ref{Mu}) coming from the diagonal couplings
of the doublets $(10_{0,0})_{0}$ follows from 
\begin{equation}
W^{(4)}=Tr[(5_{-,+}^{H_{u}})_{0}\otimes \left. \lbrack (10_{0,0})_{0}\otimes
(10_{0,0})_{0}]\right \vert _{p,q}\otimes (\vartheta _{-,+})_{0}]
\end{equation}%
However, though monodromy invariant, this couplings cannot generate the mass
term $mQ_{1,2}u_{1,2}^{c}$ since the matter curve $(10_{0,0})_{0}$ don't
contain the quark $u_{1,2}^{c}$; so the mass matrix (\ref{Mu}) for the
up-type quarks is%
\begin{equation}
M_{u,c,t}=v_{u}\left( 
\begin{array}{ccc}
0 & 0 & \alpha _{2}\rho _{2} \\ 
0 & 0 & -\alpha _{2}\rho _{1} \\ 
0 & 0 & \alpha _{3}\rho _{0}%
\end{array}%
\right)  \label{mu}
\end{equation}%
it is a rank one matrix; it gives mass to the third generation (top-quark);
while the two first generations are massless.

\  \  \  \  \ 

\emph{masses for lighter families}\newline
The rank one property of above mass matrix (\ref{mu}) is a known feature in
GUT models building including F-Theory constructions; see for instance 
\textrm{\cite{E5,EE0,EE1,Z0}}. To generate masses for the up- quarks in the
first two generations, different approaches have been used in literature: $%
\left( i\right) $ approach based on flux corrections using non perturbative
effects \textrm{\cite{C0}} or non commutative geometry \textrm{\cite{C1}};
and $\left( ii\right) $ method using $\delta W$ deformations of the GUT
superpotential $W$ by higher order chiral operators\textrm{\cite%
{B2,EE0,EE1,Z0,Z1,Z2}}. Following the second way of doing, masses to the two
lighter families are generated by higher dimensional operators corrections
that are invariant under $\mathbb{D}_{4}$\ symmetry and Z$_{2}$\ parity.
This invariance requirement leads to involve 6- and 7-dimensional chiral
operators which contribute to the up- quark mass matrix as follows%
\begin{equation}
\delta W=\sum_{i=1}^{5}x_{i}\delta W_{i}
\end{equation}%
\textrm{\ }with%
\begin{equation}
\begin{tabular}{lll}
$\delta W_{1}$ & $=$ & $\left( {\small 10}_{{\small 0,0}}^{{\small i}%
}\right) _{{\small 0}}{\small \otimes }\left( {\small 10}_{{\small +,+}%
}\right) _{{\small 0}}{\small \otimes }\left( {\small 5}_{{\small -,+}}^{%
{\small H}_{{\small u}}}\right) _{{\small 0}}{\small \otimes }\left( {\small %
\vartheta }_{{\small 0,0}}^{i}\right) _{{\small -t}_{5}}{\small \otimes }%
\left( {\small \vartheta }_{{\small +,+}}\right) _{{\small t}_{5}}$ \\ 
$\delta W_{2}$ & $=$ & $\left( {\small 10}_{{\small 0,0}}^{i}\right) _{0}%
{\small \otimes }\left( {\small 10}_{{\small +,+}}\right) _{t_{5}}{\small %
\otimes }\left( {\small 5}_{{\small -,+}}^{{\small H}_{{\small u}}}\right)
_{0}{\small \otimes }\left( {\small \vartheta }_{{\small 0,0}}^{i}\right) _{%
{\small -t}_{5}}{\small \otimes }\left( {\small \vartheta }_{{\small 0,0}%
}^{i}\right) _{{\small -t}_{5}}{\small \otimes }\left( {\small \vartheta }_{%
{\small +,+}}\right) _{{\small t}_{5}}$ \\ 
$\delta W_{3}$ & $=$ & $\left( {\small 10}_{{\small 0,0}}^{i}\right) _{%
{\small 0}}{\small \otimes }\left( {\small 10}_{{\small +,+}}\right) _{t_{5}}%
{\small \otimes }\left( {\small 5}_{{\small -,+}}^{{\small H}_{{\small u}%
}}\right) _{0}{\small \otimes }\left( {\small \vartheta }_{{\small 0,0}}^{%
{\small i}}\right) _{{\small -t}_{5}}{\small \otimes }\left( {\small %
\vartheta }_{+,-}\right) _{{\small -t}_{5}}{\small \otimes }\left( {\small %
\vartheta }_{{\small +,+}}\right) _{{\small t}_{5}}$%
\end{tabular}
\label{o1}
\end{equation}%
and%
\begin{equation}
\begin{tabular}{lll}
$\delta W_{4}$ & $=$ & $\left( {\small 10}_{{\small +,-}}\right) _{{\small 0}%
}{\small \otimes }\left( {\small 10}_{{\small +,+}}\right) _{{\small 0}}%
{\small \otimes }\left( {\small 5}_{{\small -,+}}^{{\small H}_{{\small u}%
}}\right) _{{\small 0}}{\small \otimes }\left( {\small \vartheta }_{{\small %
-,+}}\right) _{{\small 0}}{\small \otimes }\left( {\small \vartheta }_{%
{\small +,-}}\right) _{{\small -t}_{5}}{\small \otimes }\left( {\small %
\vartheta }_{{\small +,+}}\right) _{{\small t}_{5}}$ \\ 
$\delta W_{5}$ & $=$ & $\left( {\small 10}_{{\small +,-}}\right) _{{\small 0}%
}{\small \otimes }\left( {\small 10}_{{\small +,+}}\right) _{{\small t}_{5}}%
{\small \otimes }\left( {\small 5}_{{\small -,+}}^{{\small H}_{{\small u}%
}}\right) _{{\small 0}}{\small \otimes }\left( {\small \vartheta }_{{\small %
0,0}}^{i}\right) _{{\small -t}_{5}}^{{\small 2}}{\small \otimes }\left( 
{\small \vartheta }_{{\small +,+}}\right) _{{\small t}_{5}}$%
\end{tabular}
\label{o2}
\end{equation}%
Notice that the adjunction of $\left( \vartheta _{+,+}\right) _{t_{5}}$
chiral superfield is required by invariance under Z$_{2}$ parity. Using this
deformation, a higher rank up- quark mass matrix is obtained as usual by
giving VEVs to flavons as in in (\ref{vev}) and $\left \langle \left(
\vartheta _{+,-}\right) _{-t_{5}}\right \rangle =\varphi $. By calculating
the product of the operators in eqs(\ref{o1}-\ref{o2}) using $\mathbb{D}_{4}$%
\ fusion rules, we obtain%
\begin{equation*}
\begin{tabular}{lll}
$x_{1}\delta W_{1}$ & $=$ & $\left( {\small 10}_{{\small 0,0}}^{i}\right) _{%
{\small 0}}\otimes \left( {\small 10}_{{\small +,+}}\right) _{{\small 0}%
}\otimes \left( {\small 5}_{{\small -,+}}^{{\small H}_{{\small u}}}\right) _{%
{\small 0}}\otimes \left( {\small \vartheta }_{{\small 0,0}}^{i}\right) _{%
{\small -t}_{5}}\otimes \left( {\small \vartheta }_{{\small +,+}}\right) _{%
{\small t}_{5}}$ \\ 
& $=$ & $x_{1}v_{u}(Q_{1}\sigma _{1}-Q_{2}\sigma _{2})u_{2}^{c}\omega $%
\end{tabular}%
\end{equation*}%
and%
\begin{equation}
\begin{tabular}{lll}
$x_{2}\delta W_{2}$ & $=$ & $\left( {\small 10}_{{\small 0,0}}^{i}\right) _{%
{\small 0}}{\small \otimes }\left( {\small 10}_{{\small +,+}}\right) _{%
{\small t}_{5}}{\small \otimes }\left( {\small 5}_{{\small -,+}}^{{\small H}%
_{{\small u}}}\right) _{{\small 0}}{\small \otimes }\left( {\small \vartheta 
}_{{\small 0,0}}^{i}\right) _{{\small -t}_{5}}{\small \otimes }\left( 
{\small \vartheta }_{{\small +,-}}\right) _{{\small -t}_{5}}{\small \otimes }%
\left( {\small \vartheta }_{+,+}\right) _{{\small t}_{5}}$ \\ 
& $=$ & $x_{2}v_{u}(Q_{1}\sigma _{2}-Q_{2}\sigma _{1})u_{1}^{c}\omega
\varphi $%
\end{tabular}
\label{711}
\end{equation}%
The operator%
\begin{equation*}
\left( {\small 10}_{0,0}^{i}\right) _{{\small 0}}{\small \otimes }\left( 
{\small 10}_{+,+}\right) _{{\small t}_{5}}{\small \otimes }\left( {\small 5}%
_{{\small -,+}}^{{\small H}_{{\small u}}}\right) _{{\small 0}}{\small %
\otimes }\left( {\small \vartheta }_{0,0}^{i}\right) _{{\small -t}_{5}}%
{\small \otimes }\left( {\small \vartheta }_{0,0}^{i}\right) _{{\small -t}%
_{5}}{\small \otimes }\left( {\small \vartheta }_{+,+}\right) _{{\small t}%
_{5}}
\end{equation*}%
contributes in the up-quark mass matrix (\ref{mu}) as a correction to the
matrix elements $m_{1,1}$\ and $m_{1,2}$; it has the same role as the higher
operator (\ref{711}); so we will not take it into account in the quark mass
matrix. Expanding the remaining operators by help of the $\mathbb{D}_{4}$\
rules, we have%
\begin{equation*}
\begin{tabular}{lll}
$x_{4}\delta W_{4}$ & $=$ & $\left( {\small 10}_{{\small +,-}}\right) _{%
{\small 0}}\otimes \left( {\small 10}_{{\small +,+}}\right) _{{\small 0}%
}\otimes \left( {\small 5}_{{\small -,+}}^{{\small H}_{{\small u}}}\right) _{%
{\small 0}}\otimes \left( {\small \vartheta }_{{\small -,+}}\right) _{%
{\small 0}}\otimes \left( {\small \vartheta }_{{\small +,-}}\right) _{%
{\small -t}_{5}}\otimes \left( {\small \vartheta }_{{\small +,+}}\right) _{%
{\small t}_{5}}$ \\ 
& $=$ & $x_{4}v_{u}\rho _{0}\varphi \omega Q_{3}u_{2}^{c}$%
\end{tabular}%
\end{equation*}%
and%
\begin{equation*}
\begin{tabular}{lll}
$x_{5}\delta W_{5}$ & $=$ & $\left( {\small 10}_{{\small +,-}}\right) _{%
{\small 0}}\otimes \left( {\small 10}_{{\small +,+}}\right) _{{\small t}%
_{5}}\otimes \left( {\small 5}_{{\small -,+}}^{{\small H}_{{\small u}%
}}\right) _{{\small 0}}\otimes \left( {\small \vartheta }_{{\small 0,0}%
}^{i}\right) _{{\small -t}_{5}}^{{\small 2}}\otimes \left( {\small \vartheta 
}_{{\small +,+}}\right) _{{\small t}_{5}}$ \\ 
& $=$ & $x_{5}v_{u}\omega Q_{3}u_{1}^{c}\left( \sigma _{1}\sigma _{2}-\sigma
_{2}\sigma _{1}\right) =0$%
\end{tabular}%
\end{equation*}%
Summing up all contributions, we end with the following up-quark matrix%
\begin{equation}
M_{u,c,t}=v_{u}\left( 
\begin{array}{ccc}
x_{2}\sigma _{1}\omega \varphi & x_{1}\omega & \alpha _{2}\rho _{2} \\ 
-x_{2}\sigma _{1}\omega \varphi & -x_{1}\omega & -\alpha _{2}\rho _{1} \\ 
0 & x_{4}\rho _{0}\varphi \omega & \alpha _{4}\rho _{0}%
\end{array}%
\right)
\end{equation}%
$\bullet $ \emph{Down-type Yukawa}\newline
Following the same procedure as in up-Higgs type coupling, we can build
invariant operators for the down-type Yukawa%
\begin{equation}
\begin{tabular}{lll}
& $(10_{+,-})_{0}\otimes (\overline{5}_{+,-}^{M})_{0}\otimes (\overline{5}%
_{+,+}^{H_{d}})_{t_{5}}\otimes \left( \vartheta _{+,+}\right) _{-t_{5}}$ & 
\\ 
& $(\overline{5}_{+,-}^{M})_{0}\otimes (\overline{5}_{+,+}^{H_{d}})_{t_{5}}%
\otimes (10_{0,0})_{0}\otimes \left( \vartheta _{0,0}\right) _{-t_{5}}$ & 
\end{tabular}
\label{yu}
\end{equation}%
Restricting VEV of down Higgs $\left \langle H_{d}\right \rangle =v_{d}$,
and using the flavons VEVs as in(\ref{vev}) as well as taking into account
multiplicities, the first coupling gives a mass term of the form $%
m_{i}d_{i}^{c}Q_{3}$ with $m_{i}=\omega v_{d}y_{_{3,i}}$ where $y_{_{3,i}}$
are coupling constants. For the second term, we need to reduce $%
(10_{0,0})_{0}\otimes \left( \vartheta _{0,0}\right) _{-t_{5}}$ into
irreducible $\mathbb{D}_{4}$ representations; and restricts to the component 
$S_{(+,-)}=\left. Q_{i}\otimes \sigma _{i}\right \vert _{(+,-)}$ with%
\begin{equation}
S_{(+,-)}=Q_{1}\sigma _{1}+Q_{2}\sigma _{2}
\end{equation}%
So the couplings in eqs(\ref{yu}) may expressed like 
\begin{equation}
y_{3,i}Q_{3}d_{i}^{c}\omega v_{d}+y_{1,i}(Q_{1}\sigma _{1}+Q_{2}\sigma
_{2})d_{i}^{c}v_{d}
\end{equation}%
leading to the mass matrix 
\begin{equation}
m_{d,s,b}=v_{d}\left( 
\begin{array}{ccc}
y_{1,1}\sigma _{1} & y_{1,2}\sigma _{1} & y_{1,3}\sigma _{1} \\ 
y_{1,1}\sigma _{2} & y_{1,2}\sigma _{2} & y_{1,3}\sigma _{2} \\ 
y_{3,1}\omega & y_{3,2}\omega & y_{3,3}\omega%
\end{array}%
\right)
\end{equation}

\subsubsection{Lepton sector}

First we consider the charged leptons; and then turn to neutrinos.

\  \  \  \  \ 

$\bullet $ \emph{Charged leptons}\newline
Charged leptons masses are determined by the same operators used in the case
of the down quark sector $10^{M}\otimes \overline{5}^{M}\otimes \overline{5}%
^{H_{d}}$; using spectrum eqs(\ref{3u},\ref{hu}), the appropriate operators
which provide mass to charged leptons are%
\begin{equation}
\begin{tabular}{lll}
& $(10_{+,-})_{0}\otimes (\overline{5}_{+,-}^{M})_{0}\otimes (\overline{5}%
_{+,+}^{H_{d}})_{t_{5}}\otimes \left( \vartheta _{+,+}\right) _{-t_{5}}$ & 
\\ 
& $(10_{0,0})_{0}\otimes (\overline{5}_{+,-}^{M})_{0}\otimes (\overline{5}%
_{+,+}^{H_{d}})_{t_{5}}\otimes \left( \vartheta _{0,0}\right) _{-t_{5}}$ & 
\end{tabular}%
\end{equation}%
giving the lepton mass term $m^{ij}e_{i}^{c}L_{j}$ with mass matrix%
\begin{equation}
m_{e,\mu ,\tau }=v_{d}\left( 
\begin{array}{ccc}
z_{1,1}\sigma _{1} & z_{1,2}\sigma _{1} & z_{1,3}\sigma _{1} \\ 
z_{1,1}\sigma _{2} & z_{1,2}\sigma _{2} & z_{1,3}\sigma _{2} \\ 
z_{3,1}\omega & z_{3,2}\omega & z_{3,3}\omega%
\end{array}%
\right)
\end{equation}

$\bullet $ \emph{Neutrinos}\newline
Right handed neutrinos are as in eq(\ref{nn}), they have negative R- parity.
Dirac neutrino term is embedded in the coupling $\left( \nu _{i}^{c}\otimes 
\overline{5}^{M}\right) \otimes 5^{H_{u}}$ where the right neutrino $\nu
_{i}^{c}$ is an $SU_{5}$ singlet; it allows a total neutrino mass matrix
using see-saw I mechanism \cite{B5}. The invariant operators that give the
Dirac neutrino in $SU_{5}\times \mathbb{D}_{4}\times U_{1}^{\perp }$ model
are%
\begin{equation}
\begin{tabular}{lll}
& $x_{1,i}\left( \vartheta _{+,-}\right) _{0}\otimes (\overline{5}%
_{+,-}^{M})_{0}\otimes (5_{-,+}^{H_{u}})_{0}\otimes \left( \vartheta
_{-,+}\right) _{0}$ &  \\ 
& $x_{2,i}\left( \vartheta _{0,0}\right) _{0}\otimes (\overline{5}%
_{+,-}^{M})_{0}\otimes (5_{-,+}^{H_{u}})_{0}\otimes \left( \vartheta
_{0,0}\right) _{0}^{\prime }$ & 
\end{tabular}%
\end{equation}%
Using the $\mathbb{D}_{4}$ algebra rules\ and flavon VEV's, these couplings
lead to%
\begin{equation}
\begin{tabular}{lll}
& $x_{1,i}v_{u}\rho _{0}L_{i}\nu _{3}^{c}$ &  \\ 
& $x_{2,i}v_{u}\rho _{2}L_{i}\nu _{1}^{c}-x_{2,i}v_{u}\rho _{1}L_{i}\nu
_{2}^{c}$ & 
\end{tabular}%
\end{equation}%
and then to a Dirac neutrino mass matrix as%
\begin{equation}
m_{D}=v_{u}\left( 
\begin{array}{ccc}
x_{2,1}\rho _{2} & -x_{2,1}\rho _{1} & x_{1,1}\rho _{0} \\ 
x_{2,2}\rho _{2} & -x_{2,2}\rho _{1} & x_{1,2}\rho _{0} \\ 
x_{2,3}\rho _{2} & -x_{2,3}\rho _{1} & x_{1,3}\rho _{0}%
\end{array}%
\right)
\end{equation}%
The Majorana neutrino term is given by $M\nu _{i}^{c}\otimes \nu _{j}^{c}$;
by using eqs(\ref{3u},\ref{hu}), the Majorana neutrino couplings in $%
SU_{5}\times \mathbb{D}_{4}\times U_{1}^{\perp }$ model are as follows%
\begin{eqnarray}
&&\left( \vartheta _{+,-}\right) _{0}\otimes \left( \vartheta _{+,-}\right)
_{0}  \notag \\
&&\left( \vartheta _{0,0}\right) _{0}\otimes \left( \vartheta _{0,0}\right)
_{0}  \label{Majo} \\
&&\left( \vartheta _{+,-}\right) _{0}\otimes \left( \vartheta _{0,0}\right)
_{0}\otimes \left( \vartheta _{0,0}\right) _{0}^{\prime }  \notag
\end{eqnarray}%
we can also add the singlet $\left( \vartheta _{-,+}\right) _{0}$ as a
correction of the last two operators. The operators in above (\ref{Majo})
lead to 
\begin{equation}
m\nu _{3}^{c}\nu _{3}^{c}\text{ \qquad ,\qquad }M\nu _{1}^{c}\nu _{2}^{c}%
\text{ \qquad ,\qquad }\lambda \text{\ }\nu _{3}^{c}(\nu _{1}^{c}\rho
_{1}+\nu _{2}^{c}\rho _{2})
\end{equation}%
and ends with a Majorana neutrino mass matrix like%
\begin{equation}
m_{_{M}}=\left( 
\begin{array}{ccc}
0 & M & \lambda \rho _{1} \\ 
M & 0 & \lambda \rho _{2} \\ 
\lambda \rho _{1} & \lambda \rho _{2} & m%
\end{array}%
\right)
\end{equation}%
The general neutrino mass matrix is calculated using see-saw I mechanism; it
reads as $\boldsymbol{M}_{\nu }=-m_{D}m_{M}^{-1}m_{D}^{\top }$; and leads to
the following effective neutrino mass matrix 
\begin{equation}
\boldsymbol{M}_{\nu }\simeq \xi _{0}\left( 
\begin{array}{ccc}
m_{1,1} & m_{1,2} & m_{1,3} \\ 
m_{1,2} & m_{2,2} & m_{2,3} \\ 
m_{1,3} & m_{2,3} & m_{3,3}%
\end{array}%
\right)  \label{neut}
\end{equation}%
with 
\begin{equation}
\begin{tabular}{lll}
$m_{1,1}$ & $=$ & $\lambda ^{2}x_{2,1}^{2}\rho _{2}^{4}-2x_{2,1}^{2}\rho
_{1}\rho _{2}mM+2\lambda x_{2,1}\rho _{2}^{2}(\lambda x_{2,1}\rho
_{1}^{2}-x_{1,1}M\rho _{0})$ \\ 
&  & $+(\lambda x_{2,1}\rho _{1}^{2}+x_{1,1}M\rho _{0})^{2}$ \\ 
$m_{2,2}$ & $=$ & $\lambda ^{2}x_{2,2}^{2}\rho _{2}^{4}-2x_{2,2}^{2}\rho
_{1}\rho _{2}mM+2\lambda x_{2,2}\rho _{2}^{2}(\lambda x_{2,2}\rho
_{1}^{2}-x_{1,2}M\rho _{0})$ \\ 
&  & $+(\lambda x_{2,2}\rho _{1}^{2}+x_{1,2}M\rho _{0})^{2}$ \\ 
$m_{3,3}$ & $=$ & $\lambda ^{2}x_{2,3}^{2}\rho _{2}^{4}-2x_{2,3}^{2}\rho
_{1}\rho _{2}mM+2\lambda x_{2,3}\rho _{2}^{2}(\lambda x_{2,3}\rho
_{1}^{2}-x_{1,3}M\rho _{0})$ \\ 
&  & $+(\lambda x_{2,3}\rho _{1}^{2}+x_{1,3}M\rho _{0})^{2}$%
\end{tabular}%
\end{equation}%
and%
\begin{equation}
\begin{tabular}{lll}
$m_{1,2}$ & $=$ & $\lambda ^{2}x_{2,1}x_{2,2}\rho
_{2}^{4}-2x_{2,1}x_{2,2}\rho _{1}\rho _{2}mM$ \\ 
&  & $+(\lambda x_{2,1}\rho _{1}^{2}+x_{1,1}\rho _{0}M)(\lambda x_{2,2}\rho
_{1}^{2}+x_{1,2}\rho _{0}M)$ \\ 
&  & $+\lambda \rho _{2}^{2}[2\lambda x_{2,1}x_{2,2}\rho _{1}^{2}-\rho
_{0}M(x_{1,1}x_{2,2}+x_{2,1}x_{1,2})]$ \\ 
$m_{1,3}$ & $=$ & $\lambda ^{2}x_{2,1}x_{2,3}\rho
_{2}^{4}-2x_{2,1}x_{2,3}\rho _{1}\rho _{2}mM$ \\ 
&  & $+(\lambda x_{2,1}\rho _{1}^{2}+x_{1,1}\rho _{0}M)(\lambda x_{2,3}\rho
_{1}^{2}+x_{1,3}\rho _{0}M)$ \\ 
&  & $+\lambda \rho _{2}^{2}[2\lambda x_{2,1}x_{2,3}\rho _{1}^{2}-\rho
_{0}M(x_{1,1}x_{2,3}+x_{2,1}x_{1,3})]$ \\ 
$m_{2,3}$ & $=$ & $\lambda ^{2}x_{2,2}x_{2,3}\rho
_{2}^{4}-2x_{2,2}x_{2,3}\rho _{1}\rho _{2}mM$ \\ 
&  & $+(\lambda x_{2,2}\rho _{1}^{2}+x_{1,2}\rho _{0}M)(\lambda x_{2,3}\rho
_{1}^{2}+x_{1,3}\rho _{0}M)$ \\ 
&  & $+\lambda \rho _{2}^{2}[2\lambda x_{2,2}x_{2,3}\rho _{1}^{2}-\rho
_{0}M(x_{1,3}x_{2,2}+x_{2,3}x_{1,2})]$%
\end{tabular}%
\end{equation}%
and where we have set 
\begin{equation}
\xi _{0}=\frac{v_{u}^{2}}{M(mM-2\lambda ^{2}\rho _{1}\rho _{2})}
\end{equation}%
\begin{equation*}
\end{equation*}%
To obtain neutrino mixing compatible with experiments we need a particular
parametrization and some approximations on\emph{\ }$\boldsymbol{M}_{\nu }$.
To that purpose, recall that there are three approaches to mixing using: $%
\left( i\right) $ the well know \textit{Tribimaximal} (TBM) mixing matrix, $%
\left( ii\right) $\  \textit{Bimaximal} (BM) and $\left( iii\right) $ \textit{%
Democratic} (DC); all of the TBM, BM and DC mixing matrices predict a zero
value for the angle $\theta _{13}$. However recent results reported by MINOS 
\cite{C4}, Double Chooz \cite{C5},T2K \cite{C6}, Daya Bay \cite{C7}, and
RENO \cite{C8} collaborations reaveled a non-zero $\theta _{13}$; such
non-zero $\theta _{13}$ has been recently subject of great interest; in
particular by perturbation of the TBM mixing matrix \textrm{\cite{C9}}.%
\newline
To estimate the proper masses of the $\boldsymbol{M}_{\nu }$ matrix; we
diagonalise it by using the unitary $U_{TBM}$ TBM mixing matrix; we use the $%
\mu $-$\tau $ symmetry requiring $m_{2,2}=m_{3,3},\ m_{1,2}=m_{1,3}$; as
well as the condition $m_{2,3}=m_{1,1}+m_{1,2}-m_{2,2}$. So we have $%
\boldsymbol{M}_{\nu }^{diag}=U_{TBM}^{\top }\boldsymbol{M}_{\nu
}U_{TBM}^{{}} $ with%
\begin{equation}
U_{TBM}=\left( 
\begin{array}{ccc}
\sqrt{\frac{2}{3}} & \frac{1}{\sqrt{3}} & 0 \\ 
-\frac{1}{\sqrt{6}} & \frac{1}{\sqrt{3}} & -\frac{1}{\sqrt{2}} \\ 
-\frac{1}{\sqrt{6}} & \frac{1}{\sqrt{3}} & \frac{1}{\sqrt{2}}%
\end{array}%
\right)  \label{M2}
\end{equation}%
and therefore%
\begin{equation}
\boldsymbol{M}_{\nu }^{diag}\simeq \xi _{0}\left( 
\begin{array}{ccc}
\lambda _{1} & 0 & 0 \\ 
0 & \lambda _{2} & 0 \\ 
0 & 0 & \lambda _{3}%
\end{array}%
\right)
\end{equation}%
with eigenvalues as%
\begin{equation}
\begin{tabular}{lll}
$\lambda _{1}$ & $=$ & $\xi _{0}(m_{1,1}-m_{1,2})$ \\ 
$\lambda _{2}$ & $=$ & $\xi _{0}\left( m_{1,1}+2m_{1,2}\right) $ \\ 
$\lambda _{3}$ & $=$ & $\xi _{0}\left( 2m_{3,2}-m_{1,1}-m_{1,2}\right) $%
\end{tabular}%
\end{equation}

\section{Conclusion and discussions}

In this paper, we have developed a method based on characters of discrete
group representations to study $SU_{5}\times \mathbb{D}_{4}\times
U_{1}^{\perp }$- GUT models with dihedral monodromy symmetry. After having
revisited the construction of $SU_{5}\times \mathbb{S}_{4}\times
U_{1}^{\perp }$ and $SU_{5}\times \mathbb{S}_{3}\times \left( U_{1}^{\perp
}\right) ^{2}$ models from the character representation view, we have
derived three $SU_{5}\times \mathbb{D}_{4}\times U_{1}^{\perp }$ models
(referred here to as I, II and III) with curves spectrum respectively given
by eqs(\ref{m1}-\ref{4t}), (\ref{m2}-\ref{5t}) and (\ref{m3}-\ref{6t}).
These models follow from the three different ways of decomposing the
irreducible $\mathbb{S}_{4}$- triplets in terms of irreducible
representations of $\mathbb{D}_{4}$; see eqs (\ref{w1},\ref{w2},\ref{w3});
such richness may be interpreted as due to the fact that $\mathbb{D}_{4}$
has four kinds of singlets with generator group characters given by the $%
\left( p,q\right) $ pairs with $p,q=\pm 1$. \newline
Then we have focussed on the curve spectrum (\ref{m1}-\ref{4t}) of the first 
$SU_{5}\times \mathbb{D}_{4}\times U_{1}^{\perp }$ model; and studied the
derivation of a MSSM- like spectrum by using particular multiplicity values
and turning on adequate fluxes. We have found that with the choice of: $%
\left( i\right) $ top-quark family $\mathbf{10}_{3}$ as $\left( \mathbf{10}%
_{+-}\right) _{0}$, transforming into a $\mathbb{D}_{4}$- singlet with $\chi
^{\left( a,b\right) }$ character equal to $\left( 1,-1\right) $; and $\left(
ii\right) $ a $5^{H_{u}}$ up-Higgs as $\left( 5_{{\small -,+}}\right) _{0}$,
transforming into a different $\mathbb{D}_{4}$- singlet with character equal
to $\left( -1,1\right) $; there is no tri-Yukawa couplings of the form 
\begin{equation*}
\left( 10_{+,-}\right) _{0}\otimes \left( 10_{+,-}\right) _{0}\otimes \left(
5^{H_{u}}\right) _{++}
\end{equation*}%
as far as $\mathbb{D}_{4}\times U_{1}^{\perp }$ invariance is required; this
makes $SU_{5}\times \mathbb{D}_{4}\times U_{1}^{\perp }$ model with two
quark generations accommodated into a $\mathbb{D}_{4}$- doublet non
interesting phenomenologically.\ Monodromy invariant couplings require
implementation of flavons $\vartheta _{p,q}$ by thinking of $5^{H_{u}}\sim
\left( 5_{{\small -,+}}\right) _{0}\otimes \left( \vartheta _{-,+}\right)
_{0}$ leading therefore to a superpotential of order 4. The same property
appears with the down-Higgs couplings where $\mathbb{D}_{4}\times
U_{1}^{\perp }$ invariance of $10_{3}\otimes \bar{5}_{3}^{M}\otimes \bar{5}%
^{H_{d}}$ requires: $\left( \alpha \right) $ a matter $\bar{5}_{3}^{M}\equiv
(\bar{5}_{-,{\small +}}^{M})_{0}$ in a $U_{1}^{\perp }$ chargeless $\mathbb{D%
}_{4}$- singlet with character $\left( -1,1\right) $; and $\left( \beta
\right) $ a curve $\bar{5}^{H_{d}}$ with a $\mathbb{D}_{4}$- character like $%
(\bar{5}_{-{\small ,-}})_{+t_{5}}$ composed with a charged flavon $%
(\vartheta _{++})_{-t_{5}}$; that is as 
\begin{equation*}
(\bar{5}_{-{\small ,-}})_{+t_{5}}\otimes (\vartheta _{++})_{-t_{5}}
\end{equation*}%
By analysing the conditions that a $\mathbb{D}_{4}\times U_{1}^{\perp }$%
-spectrum has to fulfill in order to have a tri- Yukawa coupling for
top-quark family $\mathbf{10}_{3}$, we end with the constraint that the
character of $5^{H_{u}}$ up-Higgs should be equal to $\left( 1,1\right) $ as
clearly seen on $10_{+,-}\otimes 10_{+,-}\otimes 5^{H_{u}}$. This constraint
is valid even if $10_{3}$ was chosen like $10_{+,+}$. By inspecting the
spectrum of the three studied $SU_{5}\times \mathbb{D}_{4}\times
U_{1}^{\perp }$ models; it results that the spectrum of the third model
given by eqs(\ref{m3}-\ref{6t}) which allow tri- Yukawa coupling; for
details on contents and couplings of models II and III; see\textrm{\
appendix B.}

\section{Appendix A: Characters in $\mathbb{S}_{4}$-models}

In this appendix, we give details on some useful properties of $\Gamma $%
-models studied in this paper; in particular on the representations of $%
\mathbb{S}_{4}$ and their characters.

\subsection{Irreducible representations of $\mathbb{S}_{4}$}

First, recall that $\mathbb{S}_{4}$ has five \emph{irreducible}
representations; as shown on the character formula $24=1^{2}+1^{\prime
2}+2^{2}+3^{2}+3^{\prime 2}$; these are the 1-$\dim $ representations
including the trivial $\mathbf{1}$ and the sign $\epsilon =\mathbf{1}%
^{\prime }$; a 2-$\dim $ representation $\mathbf{2}$; and the 3-$\dim $
representations $\mathbf{3}$ and $\mathbf{3}^{\prime }$, obeying some
"duality relation". This duality may be stated in different manners; but, in
simple words, it may be put in parallel with polar and axial vectors of
3-dim euclidian space. In the language of Young diagrams; these five
irreducible representations are given by%
\begin{equation}
\mathbf{1}:\text{ \  \ }%
\begin{tabular}{|l|l|l|l|}
\hline
&  &  &  \\ \hline
\end{tabular}%
\qquad ,\qquad \mathbf{2}:\text{ \  \ }%
\begin{tabular}{|l|l|}
\hline
&  \\ \hline
&  \\ \hline
\end{tabular}%
\qquad ,\qquad \mathbf{3}:\text{ \  \ }%
\begin{tabular}{|l|ll}
\hline
&  & \multicolumn{1}{|l|}{} \\ \hline
&  &  \\ \cline{1-1}
\end{tabular}%
\end{equation}%
and%
\begin{equation}
\mathbf{3}^{\prime }:\text{ \  \ }%
\begin{tabular}{|l|l}
\hline
& \multicolumn{1}{|l|}{} \\ \hline
&  \\ \cline{1-1}
&  \\ \cline{1-1}
\end{tabular}%
\qquad ,\qquad \mathbf{1}^{\prime }:\text{ \  \ }%
\begin{tabular}{|l|}
\hline
\\ \hline
\\ \hline
\\ \hline
\\ \hline
\end{tabular}%
\end{equation}%
This diagrammatic description is very helpful in dealing with $\mathbb{S}%
_{4} $ representation theory \textrm{\cite{E9,E10,E11}}; it teaches us a set
of useful information; in particular helpful data on the three following:

$i)$ \emph{Expressions of} (\ref{m})\newline
In the representation 3 of the permutation group $\mathbb{S}_{4}$, the three 
$x_{i}$- weights in (\ref{m}) read in terms of the t$_{i}$'s as 
\begin{equation}
\vec{x}=\frac{1}{2}\left( 
\begin{array}{c}
t_{1}-t_{2}-t_{3}+t_{4} \\ 
t_{1}+t_{2}-t_{3}-t_{4} \\ 
t_{1}-t_{2}+t_{3}-t_{4}%
\end{array}%
\right) =\left( 
\begin{array}{c}
x_{4}-t_{2}-t_{3} \\ 
x_{4}-t_{3}-t_{4} \\ 
x_{4}-t_{4}-t_{2}%
\end{array}%
\right)  \label{2}
\end{equation}%
where $x_{4}=\frac{1}{2}\left( t_{1}+t_{2}+t_{3}+t_{4}\right) $ is the
completely symmetric term. The normalisation coefficient $\frac{1}{2}$ is
fixed by requiring the transformation $x_{i}=U_{ij}t_{j}$ as follows 
\begin{equation}
U=\frac{1}{2}\left( 
\begin{array}{cccc}
1 & -1 & -1 & 1 \\ 
1 & 1 & -1 & -1 \\ 
1 & -1 & 1 & -1 \\ 
1 & 1 & 1 & 1%
\end{array}%
\right) \qquad ,\qquad \det U=1  \label{tr}
\end{equation}
For the the representation $3^{\prime }$, we have%
\begin{equation}
\vec{x}^{\prime }=\frac{1}{\sqrt{8}}\left( 
\begin{array}{c}
t_{1}-3t_{2}+t_{3}+t_{4} \\ 
t_{1}+t_{2}-3t_{3}+t_{4} \\ 
t_{1}+t_{2}+t_{3}-3t_{4}%
\end{array}%
\right) =\frac{1}{\sqrt{2}}\left( 
\begin{array}{c}
x_{4}-2t_{2} \\ 
x_{4}-2t_{3} \\ 
x_{4}-2t_{4}%
\end{array}%
\right)  \label{3}
\end{equation}%
The entries of these triplets are cyclically rotated by the $\left(
234\right) $ permutation.

\  \ 

$ii)$ $\mathbb{S}_{4}$- \emph{triplets as 3-cycle} $\left( 234\right) $%
\newline
The $\left \{ \left \vert t_{i}\right \rangle \right \} $ and $\left \{
\left \vert x_{i}\right \rangle \right \} $ weight bases are related by the
orthogonal 5$\times $5 matrix%
\begin{equation}
\left( 
\begin{array}{cc}
U & 0 \\ 
0 & 1%
\end{array}%
\right) \qquad ,\qquad \left \vert x_{i}\right \rangle =U_{ij}\left \vert
t_{j}\right \rangle
\end{equation}%
with $U$ as in (\ref{tr}); and then 
\begin{equation}
\begin{array}{ccc}
t_{1} & = & \frac{1}{2}\left( x_{4}+x_{1}+x_{2}+x_{3}\right) \\ 
t_{2} & = & \frac{1}{2}\left( x_{4}-x_{1}+x_{2}-x_{3}\right) \\ 
t_{3} & = & \frac{1}{2}\left( x_{4}-x_{1}-x_{2}+x_{3}\right) \\ 
t_{4} & = & \frac{1}{2}\left( x_{4}+x_{1}-x_{2}-x_{3}\right)%
\end{array}%
\  \qquad ,\qquad
\end{equation}%
\begin{equation*}
\end{equation*}%
From these transformations, we learn $t_{i}=U_{ki}x_{k}$; and then $t_{i}\pm
t_{j}=\left( U_{ki}\pm U_{kj}\right) x_{k}$ which can be also expressed $%
t_{i}\pm t_{j}=V_{ij}^{\pm kl}X_{kl}^{\pm }$. Similar relations can be
written down for $\left \{ \left \vert x_{i}^{\prime }\right \rangle
\right
\} $.

\subsection{Characters}

The discrete symmetry group $\mathbb{S}_{4}$ model has \emph{24} elements
arranged into five conjugacy classes $\mathfrak{C}_{1},...,\mathfrak{C}_{5}$
as on table (\ref{ca}); it has five irreducible representations $\boldsymbol{%
R}_{1},...,\boldsymbol{R}_{5}$ with dimensions given by the relation $%
24=1^{2}+1^{2\prime }+2^{2}+3^{2}+3^{2\prime }$; their character table $%
\mathrm{\chi }_{ij}=\mathrm{\chi }_{\boldsymbol{R}_{j}}\left( \mathfrak{C}%
_{i}\right) $ is as given below

\begin{equation}
\begin{tabular}{|l|l|l|l|l|l|l|}
\hline
$\mathfrak{C}_{i}$\TEXTsymbol{\backslash}irrep $\boldsymbol{R}_{j}$ & $\  \ 
\mathrm{\chi }_{_{\boldsymbol{I}}}$ & $\  \  \mathrm{\chi }_{_{\boldsymbol{3}%
^{\prime }}}$ & $\  \  \mathrm{\chi }_{_{\boldsymbol{2}}}$ & $\  \  \mathrm{\chi 
}_{_{\boldsymbol{3}}}$ & $\  \  \mathrm{\chi }_{_{\epsilon }}$ & {\small number%
} \\ \hline
$\mathfrak{C}_{1}\equiv \mathrm{\  \ e}$ & $\  \ 1\  \ $ & $\  \ 3\  \ $ & $\  \
2\  \ $ & $\  \ 3\  \ $ & $\  \ 1\  \ $ & $\  \ 1\  \ $ \\ \hline
$\mathfrak{C}_{2}\equiv \mathrm{(\alpha \beta )}$ & $\  \ 1$ & $-1$ & $\  \ 0$
& $\  \ 1$ & $-1$ & $\  \ 6$ \\ \hline
$\mathfrak{C}_{3}\equiv \mathrm{(\alpha \beta )(\gamma \delta )}$ & $\  \ 1$
& $-1$ & $\  \ 2$ & $-1$ & $\  \ 1$ & $\  \ 3$ \\ \hline
$\mathfrak{C}_{4}\equiv \mathrm{(\alpha \beta \gamma )}$ & $\  \ 1$ & $\  \ 0$
& $-1$ & $\  \ 0$ & $\  \ 1$ & $\  \ 8$ \\ \hline
$\mathfrak{C}_{5}\equiv \mathrm{(\alpha \beta \gamma \delta )}$ & $\  \ 1$ & $%
\  \ 1$ & $\  \ 0$ & $-1$ & $-1$ & $\  \ 6$ \\ \hline
\end{tabular}
\label{ca}
\end{equation}%
The $\mathbb{S}_{4}$ group has \emph{3} non commuting generators $\left(
a,b,c\right) $ which can be chosen as given by the 2- , 3- and 4- cycles
obeying amongst others the cyclic relations $a^{2}=b^{3}=c^{4}=I_{id}$. In
our approach the character of these generators have been used in the
engineering of GUT models with $\mathbb{S}_{4}$ monodromy; they are as
follows%
\begin{equation}
\begin{tabular}{|l|l|l|l|l|l|}
\hline
$\mathrm{\chi }_{ij}$ & $\mathrm{\chi }_{_{\boldsymbol{I}}}$ & $\mathrm{\chi 
}_{_{\boldsymbol{3}^{\prime }}}$ & $\mathrm{\chi }_{_{\boldsymbol{2}}}$ & $%
\mathrm{\chi }_{_{\boldsymbol{3}}}$ & $\mathrm{\chi }_{_{\epsilon }}$ \\ 
\hline
$a$ & $1$ & $-1$ & $\  \ 0$ & $\  \ 1$ & $-1$ \\ \hline
$b$ & $1$ & $\  \ 0$ & $-1$ & $\  \ 0$ & $\  \ 1$ \\ \hline
$c$ & $1$ & $\  \ 1$ & $\  \ 0$ & $-1$ & $-1$ \\ \hline
\end{tabular}%
\end{equation}%
In the SU$_{5}\times \mathbb{S}_{4}$ theory considered in paper, the various
curves of the spectrum of the GUT- model belong to $\mathbb{S}_{4}$-
multiplets which can be decomposed into irreducible representation of $%
\mathbb{S}_{4}$. In doing so, one ends with curves indexed by the characters
of the generators of $\mathbb{S}_{4}$ as follows 
\begin{equation}
\begin{tabular}{lll}
$\mathbf{4}$ & $\mathbf{=}$ & $\mathbf{1}_{\left( 1,1,1\right) }\oplus 
\mathbf{3}_{\left( 1,0,-1\right) }$ \\ 
$\mathbf{6}$ & $\mathbf{=}$ & $\mathbf{3}_{\left( 1,0,-1\right) }\oplus 
\mathbf{3}_{\left( -1,0,1\right) }^{\prime }$%
\end{tabular}%
\end{equation}

\section{Appendix B: Results on $SU_{5}\times \mathbb{D}_{4}$ models II \&
III}

In this appendix, we collect results regarding the $SU_{5}\times \mathbb{D}%
_{4}\times U_{1}^{\perp }$ models II and III of subsections \S \ 4.2.2 and 
\S \ 4.2.3. In addition to higher order terms, we also study when couplings
like 
\begin{equation}
\begin{tabular}{|l|l|l|l|l|}
\hline
\  \  \  \  \ Couplings & $SU_{5}$ & $\mathbb{D}_{4}$ & $U_{1}^{\bot }$ & Parity
\\ \hline
$\left. 
\begin{array}{c}
10_{i}\otimes 10_{j}\otimes 5_{H_{u}} \\ 
10_{i}\otimes \overline{5}_{j}\otimes \overline{5}_{H_{d}} \\ 
\nu _{i}^{c}\otimes \overline{5}_{M}\otimes 5_{H_{u}} \\ 
m_{{}}\nu _{i}^{c}\otimes \nu _{j}^{c}%
\end{array}%
\right. $ & $\left. 
\begin{array}{c}
1 \\ 
1 \\ 
1 \\ 
1%
\end{array}%
\right. $ & $\left. 
\begin{array}{c}
1_{+,+} \\ 
1_{+,+} \\ 
1_{+,+} \\ 
1_{+,+}%
\end{array}%
\right. $ & $\left. 
\begin{array}{c}
0 \\ 
0 \\ 
0 \\ 
0%
\end{array}%
\right. $ & $\left. 
\begin{array}{c}
+ \\ 
+ \\ 
+ \\ 
+%
\end{array}%
\right. $ \\ \hline
\end{tabular}%
\end{equation}%
can be generated.

\subsection{$SU_{5}\times \mathbb{D}_{4}$ model II}

The spectrum of the $SU_{5}\times \mathbb{D}_{4}\times U_{1}^{\bot }$ model
II under breaking $SU_{5}\times \mathbb{D}_{4}\times U_{1}^{\bot }$ to MSSM
is given by:

\begin{equation}
\begin{tabular}{|l|l|l|}
\hline
Curve in $D_{4}$ model II & U$_{1}^{\perp }$ & Spectrum in MSSM \\ \hline
$10_{1}=10_{+,-}$ & $0$ & ${\small M}_{1}{\small Q}_{L}{\small +u}_{L}^{c}%
{\small (M}_{1}{\small -N-P)+e}_{L}^{c}{\small (M}_{1}{\small +N+P)}$ \\ 
\hline
$10_{2}=10_{-,+}$ & $0$ & ${\small M}_{2}{\small Q}_{L}{\small +u}_{L}^{c}%
{\small M}_{2}{\small +e}_{L}^{c}{\small M}_{2}$ \\ \hline
$10_{3}=10_{+,-}$ & $0$ & ${\small M}_{3}{\small Q}_{L}{\small +u}_{L}^{c}%
{\small M}_{3}{\small +e}_{L}^{c}{\small M}_{3}$ \\ \hline
$10_{4}=10_{+,+}$ & $0$ & ${\small M}_{4}{\small Q}_{L}{\small +u}_{L}^{c}%
{\small (M}_{4}{\small +P)+e}_{L}^{c}{\small (M}_{4}{\small -P)}$ \\ \hline
$10_{5}=10_{+,+}$ & $1$ & ${\small M}_{5}{\small Q}_{L}{\small +u}_{L}^{c}%
{\small (M}_{5}{\small +N)+e}_{L}^{c}{\small (M}_{5}{\small -N)}$ \\ \hline
$5_{1}=5_{+,-}$ & $0$ & ${\small M}_{1}^{\prime }\overline{d}_{L}^{c}{\small %
+(M}_{1}^{\prime }{\small +N+P)}\overline{L}$ \\ \hline
$5_{2}=5_{-,+}$ & $0$ & ${\small M}_{2}^{\prime }\overline{d}_{L}^{c}{\small %
+M}_{2}^{\prime }\overline{L}$ \\ \hline
$5_{3}=5_{-,+}$ & $0$ & ${\small M}_{3}^{\prime }\overline{D}_{d}{\small +(M}%
_{3}^{\prime }{\small -N)}\overline{H}_{d}$ \\ \hline
$5_{4}=5_{+,-}$ & $0$ & ${\small M}_{4}^{\prime }{\small D}_{u}{\small +(M}%
_{4}^{\prime }{\small -P)H}_{u}$ \\ \hline
$5_{5}=5_{-,+}$ & $0$ & ${\small M}_{5}^{\prime }\overline{d}_{L}^{c}{\small %
+M}_{5}^{\prime }\overline{L}$ \\ \hline
$5_{6}=5_{+,-}$ & $0$ & ${\small M}_{6}^{\prime }\overline{d}_{L}^{c}{\small %
+M}_{6}^{\prime }\overline{L}$ \\ \hline
$5_{7}=5_{+,-}^{t_{5}}$ & $-1$ & ${\small M}_{7}^{\prime }\overline{d_{L}^{c}%
}{\small +(M}_{7}^{\prime }{\small -P)}\overline{L}$ \\ \hline
$5_{8}=5_{-,+}^{t_{5}}$ & $-1$ & ${\small M}_{8}^{\prime }\overline{d_{L}^{c}%
}{\small +M}_{8}^{\prime }\overline{L}$ \\ \hline
$5_{9}=5_{+,-}^{t_{5}}$ & $-1$ & $M_{9}^{\prime }\overline{d}%
_{L}^{c}+(M_{9}^{\prime }-N)\overline{L}$ \\ \hline
$5_{10}=5_{+,+}^{t_{5}}$ & $-1$ & $M_{10}^{\prime }\overline{d}%
_{L}^{c}+(M_{10}^{\prime }+N+P)\overline{L}$ \\ \hline
\end{tabular}%
\end{equation}%
\begin{equation*}
\end{equation*}%
To get 3 generations of matter curves and 2 Higgs doublets of MSSM, taking
into account the constraints in subsection (5.1), we make the following
choice of the flux parameters; $P=-N=1,$ and%
\begin{equation}
\begin{tabular}{lll}
$M_{1}=M_{2}=M_{3}=M_{4}=-M_{5}$ & $=$ & $1$ \\ 
$M_{1}^{\prime }=M_{3}^{\prime }=M_{4}^{\prime }=M_{8}^{\prime
}=M_{10}^{\prime }$ & $=$ & $0$ \\ 
$M_{2}^{\prime }=M_{5}^{\prime }=M_{6}^{\prime }=M_{9}^{\prime
}=-M_{7}^{\prime }$ & $=$ & $-1$%
\end{tabular}%
\end{equation}%
Using the property $\sum_{i}M_{5}^{i}=-\sum_{i}M_{10}^{i}=-3$, the
localization of Higgs curves are as $5^{H_{u}}=5_{-,+}$, $\overline{5}%
^{H_{d}}=5_{+,-}$; and the third generation like $10_{1}=10^{M_{3}}$; and $%
5_{2}=5^{M_{3}}$. The distribution of the matter curves is collected in the
following table: 
\begin{equation}
\begin{tabular}{|l|l|l|l|}
\hline
{\small Curve in }${\small D}_{4}${\small \ model II} & {\small U}$%
_{1}^{\perp }$ & {\small Spectrum in MSSM} & $Z_{2}${\small \ parity} \\ 
\hline
${\small 10}_{1}{\small =10}^{M_{3}}{\small =(10}_{+,-}{\small )}_{0}$ & $%
{\small 0}$ & ${\small Q}_{L}{\small +u}_{L}^{c}{\small +e}_{L}^{c}$ & $%
{\small -}$ \\ \hline
${\small 10}_{2}{\small =(10}_{-,+}{\small )}_{0}$ & ${\small 0}$ & ${\small %
Q}_{L}{\small +u}_{L}^{c}{\small +e}_{L}^{c}$ & ${\small -}$ \\ \hline
${\small 10}_{3}{\small =(10}_{+,-}{\small )}_{0}$ & ${\small 0}$ & ${\small %
Q}_{L}{\small +u}_{L}^{c}{\small +e}_{L}^{c}$ & ${\small -}$ \\ \hline
${\small 10}_{4}{\small =(10}_{+,+}{\small )}_{0}$ & ${\small 0}$ & ${\small %
Q}_{L}{\small +2u}_{L}^{c}$ & ${\small +}$ \\ \hline
${\small 10}_{5}{\small =(10}_{+,+}{\small )}_{t_{5}}$ & ${\small 1}$ & $%
{\small -Q}_{L}{\small -2u}_{L}^{c}$ & ${\small -}$ \\ \hline
${\small 5}_{1}{\small =(5}_{+,-}{\small )}_{0}$ & ${\small 0}$ & ${\small -}
$ & ${\small +}$ \\ \hline
${\small 5}_{2}{\small =5}^{M_{3}}{\small =(5}_{-,+}{\small )}_{0}$ & $%
{\small 0}$ & ${\small -}\overline{d}_{L}^{c}{\small -}\overline{L}$ & $%
{\small -}$ \\ \hline
${\small 5}_{3}{\small =(5}_{-,+}^{H_{u}}{\small )}_{0}$ & ${\small 0}$ & $%
{\small H}_{u}$ & ${\small +}$ \\ \hline
${\small 5}_{4}{\small =(5}_{+,-}^{H_{d}}{\small )}_{0}$ & ${\small 0}$ & $%
{\small -}\overline{H}_{d}$ & ${\small +}$ \\ \hline
${\small 5}_{5}{\small =5}^{M_{1}}{\small =(5}_{-,+}{\small )}_{0}$ & $%
{\small 0}$ & ${\small -}\overline{d}_{L}^{c}{\small -}\overline{L}$ & $%
{\small -}$ \\ \hline
${\small 5}_{6}{\small =5}^{M_{2}}{\small =(5}_{+,-}{\small )}_{0}$ & $%
{\small 0}$ & ${\small -}\overline{d}_{L}^{c}{\small -}\overline{L}$ & $%
{\small -}$ \\ \hline
${\small 5}_{7}{\small =(5}_{+,-}{\small )}_{-t_{5}}$ & ${\small -1}$ & $%
\overline{d}_{L}^{c}$ & ${\small +}$ \\ \hline
${\small 5}_{8}{\small =(5}_{-,+}{\small )}_{-t_{5}}$ & ${\small -1}$ & $%
{\small -}$ & ${\small +}$ \\ \hline
${\small 5}_{9}{\small =(5}_{+,-}{\small )}_{-t_{5}}$ & ${\small -1}$ & $%
{\small -}\overline{d}_{L}^{c}$ & ${\small +}$ \\ \hline
${\small 5}_{10}{\small =(5}_{+,+}{\small )}_{-t_{5}}$ & ${\small -1}$ & $%
{\small -}$ & ${\small +}$ \\ \hline
\end{tabular}
\label{xx}
\end{equation}%
\begin{equation*}
\end{equation*}%
From this spectrum, we learn that we have three families of fermions, an
extra vector like pairs, $d_{L}^{c}+\overline{d}_{L}^{c}$, $Q_{L}+\overline{Q%
}_{L}$; and two $2(u_{L}^{c}+\overline{u}_{L}^{c})$ which are expected to
get a large mass if some of the singlet states acquire large VEV's. In this $%
\mathbb{D}_{4}$ model; there are only singlet flavons transforming in the
representations $1_{+,+},$ $1_{+,-},$ $1_{-,+};$ with and without $t_{5\text{
}}$charges, they are classified as $\ (\vartheta _{p,q})_{0,\pm t_{5}}$\
with $p,q=\pm 1;$ they lead to the following order 4-couplings

$\bullet $\textbf{\ }\emph{Up-type quark Yukawa couplings}\newline
The allowed Yukawa couplings that are invariant under $\mathbb{D}_{4}\times
U_{1}^{\perp }$\ are:%
\begin{eqnarray}
&&(10_{+,-})_{0}\otimes (10_{+,-})_{0}\otimes (5_{-,+}^{H_{u}})_{0}\otimes
(\vartheta _{-,+})_{0}  \notag \\
&&(10_{-,+})_{0}\otimes (10_{-,+})_{0}\otimes (5_{-,+}^{H_{u}})_{0}\otimes
(\vartheta _{-,+})_{0}  \notag \\
&&(10_{+,-})_{0}\otimes (10_{+,-})_{0}\otimes (5_{-,+}^{H_{u}})_{0}\otimes
(\vartheta _{-,+})_{0} \\
&&(10_{+,-})_{0}\otimes (10_{-,+})_{0}\otimes (5_{-,+}^{H_{u}})_{0}\otimes
(\vartheta _{+,-})_{0}  \notag \\
&&(10_{-,+})_{0}\otimes (10_{+,-})_{0}\otimes (5_{-,+}^{H_{u}})_{0}\otimes
(\vartheta _{+,-})_{0}  \notag
\end{eqnarray}

$\bullet $\textbf{\ }\emph{Down-type quark Yukawa couplings}\newline
The Yukawa couplings down-type are:%
\begin{eqnarray}
&&(10_{+,-})_{0}\otimes (\overline{5}_{-,+})_{0}\otimes
(5_{+,-}^{H_{d}})_{0}\otimes (\vartheta _{-,+})_{0}  \notag \\
&&(10_{+,-})_{0}\otimes (\overline{5}_{-,+})_{0}\otimes
(5_{+,-}^{H_{d}})_{0}\otimes (\vartheta _{-,+})_{0}  \notag \\
&&(10_{+,-})_{0}\otimes (\overline{5}_{+,-})_{0}\otimes
(5_{+,-}^{H_{d}})_{0}\otimes (\vartheta _{+,-})_{0} \\
&&(10_{-,+})_{0}\otimes (\overline{5}_{-,+})_{0}\otimes
(5_{+,-}^{H_{d}})_{0}\otimes (\vartheta _{+,-})_{0}  \notag \\
&&(10_{-,+})_{0}\otimes (\overline{5}_{-,+})_{0}\otimes
(5_{+,-}^{H_{d}})_{0}\otimes (\vartheta _{+,-})_{0}  \notag \\
&&(10_{-,+})_{0}\otimes (\overline{5}_{+,-})_{0}\otimes
(5_{+,-}^{H_{d}})_{0}\otimes (\vartheta _{-,+})_{0}  \notag \\
&&(10_{+,-})_{0}\otimes (\overline{5}_{-,+})_{0}\otimes
(5_{+,-}^{H_{d}})_{0}\otimes (\vartheta _{-,+})_{0}  \notag \\
&&(10_{+,-})_{0}\otimes (\overline{5}_{-,+})_{0}\otimes
(5_{+,-}^{H_{d}})_{0}\otimes (\vartheta _{-,+})_{0}  \notag \\
&&(10_{+,-})_{0}\otimes (\overline{5}_{+,-})_{0}\otimes
(5_{+,-}^{H_{d}})_{0}\otimes (\vartheta _{+,-})_{0}  \notag
\end{eqnarray}

\subsection{$SU_{5}\times \mathbb{D}_{4}$ model III}

The spectrum of the model $SU_{5}\times \mathbb{D}_{4}\times U_{1}^{\bot }$
Model III is as follows%
\begin{equation}
\begin{tabular}{|l|l|l|}
\hline
{\small Curves in }${\small D}_{4}${\small \ model III} & {\small U}$%
_{1}^{\perp }$ & {\small Spectrum in MSSM} \\ \hline
${\small 10}_{1}{\small =10}_{+,+}$ & ${\small 0}$ & ${\small M}_{1}{\small Q%
}_{L}{\small +u}_{L}^{c}{\small (M}_{1}{\small -N-P)+e}_{L}^{c}{\small (M}%
_{1}{\small +N+P)}$ \\ \hline
${\small 10}_{2}{\small =10}_{-,-}$ & ${\small 0}$ & ${\small M}_{2}{\small Q%
}_{L}{\small +u}_{L}^{c}{\small M}_{2}{\small +e}_{L}^{c}{\small M}_{2}$ \\ 
\hline
${\small 10}_{3}{\small =10}_{+,-}$ & ${\small 0}$ & ${\small M}_{3}{\small Q%
}_{L}{\small +u}_{L}^{c}{\small M}_{3}{\small +e}_{L}^{c}{\small M}_{3}$ \\ 
\hline
${\small 10}_{4}{\small =10}_{+,+}$ & ${\small 0}$ & ${\small M}_{4}{\small Q%
}_{L}{\small +u}_{L}^{c}{\small (M}_{4}{\small +P)+e}_{L}^{c}{\small (M}_{4}%
{\small -P)}$ \\ \hline
${\small 10}_{5}{\small =10}_{+,+}$ & ${\small 1}$ & ${\small M}_{5}{\small Q%
}_{L}{\small +u}_{L}^{c}{\small (M}_{5}{\small +N)+e}_{L}^{c}{\small (M}_{5}%
{\small -N)}$ \\ \hline
${\small 5}_{1}{\small =5}_{+,+}$ & ${\small 0}$ & ${\small M}_{1}^{\prime }%
\overline{d}_{L}^{c}{\small +(M}_{1}^{\prime }{\small +N+P)}\overline{L}$ \\ 
\hline
${\small 5}_{2}{\small =5}_{-,-}$ & ${\small 0}$ & ${\small M}_{2}^{\prime }%
\overline{d}_{L}^{c}{\small +(M}_{2}^{\prime }{\small -\kappa }_{1}{\small P)%
}\overline{L}$ \\ \hline
${\small 5}_{3}{\small =5}_{-,+}$ & ${\small 0}$ & ${\small M}_{3}^{\prime }%
{\small D}_{u}{\small +(M}_{3}^{\prime }{\small -N)}\overline{H}_{d}$ \\ 
\hline
${\small 5}_{4}{\small =5}_{+,+}$ & ${\small 0}$ & ${\small M}_{4}^{\prime }%
\overline{D}_{d}{\small +(M}_{4}^{\prime }{\small -\kappa }_{2}{\small P)H}%
_{u}$ \\ \hline
${\small 5}_{5}{\small =5}_{-,-}$ & ${\small 0}$ & ${\small M}_{5}^{\prime }%
\overline{d}_{L}^{c}{\small +M}_{5}^{\prime }\overline{L}$ \\ \hline
${\small 5}_{6}{\small =5}_{+,-}$ & ${\small 0}$ & ${\small M}_{6}^{\prime }%
\overline{d}_{L}^{c}{\small +M}_{6}^{\prime }\overline{L}$ \\ \hline
${\small 5}_{7}{\small =5}_{+,+}^{t_{5}}$ & ${\small -1}$ & ${\small M}%
_{7}^{\prime }\overline{d}_{L}^{c}{\small +(M}_{7}^{\prime }{\small -\kappa }%
_{1}{\small P)}\overline{L}$ \\ \hline
${\small 5}_{8}{\small =5}_{-,-}^{t_{5}}$ & ${\small -1}$ & ${\small M}%
_{8}^{\prime }\overline{d}_{L}^{c}{\small +}^{(}{\small M}_{8}^{\prime }%
{\small -\kappa }_{2}{\small P)}\overline{L}$ \\ \hline
${\small 5}_{9}{\small =5}_{+,-}^{t_{5}}$ & ${\small -1}$ & ${\small M}%
_{9}^{\prime }\overline{d}_{L}^{c}{\small +(M}_{9}^{\prime }{\small -N)}%
\overline{L}$ \\ \hline
${\small 5}_{10}{\small =5}_{+,+}^{t_{5}}$ & ${\small -1}$ & ${\small M}%
_{10}^{\prime }\overline{d}_{L}^{c}{\small +(M}_{10}^{\prime }{\small +N+P)}%
\overline{L}$ \\ \hline
\end{tabular}%
\end{equation}%
\begin{equation*}
\end{equation*}%
The 3 generations of fermions and the 2 Higgs $H_{u},$ $H_{d}$ are obtained
by taking the fluxes like $N=-P=-1$ with $\kappa _{1}=0$, $\kappa _{2}=1$;
and 
\begin{equation}
\begin{tabular}{lll}
$M_{1}=M_{2}=M_{3}=M_{4}=-M_{5}$ & $=$ & $1$ \\ 
$M_{1}^{\prime }=M_{3}^{\prime }=M_{4}^{\prime }=M_{7}^{\prime
}=M_{10}^{\prime }$ & $=$ & $0$ \\ 
$M_{2}^{\prime }=M_{5}^{\prime }=M_{6}^{\prime }=M_{8}^{\prime
}=-M_{9}^{\prime }$ & $=$ & $-1$%
\end{tabular}%
\end{equation}%
We choose the Higgs curves as $5^{H_{u}}=(5_{+,+}^{H_{u}})_{0}$, $%
5^{H_{d}}=(5_{-,+}^{H_{d}})_{0}$\ and the third $10^{M_{3}}$, $\overline{5}%
^{M_{3}}$\ generation as follow

\begin{equation}
\begin{tabular}{|l|l|l|l|}
\hline
{\small Curves in }${\small D}_{4}${\small \ model III} & {\small U}$%
_{1}^{\perp }$ & {\small Spectrum in MSSM} & $Z_{2}${\small \ parity} \\ 
\hline
${\small 10}_{1}{\small =10}^{M_{3}}{\small =(10}_{+,+}{\small )}_{0}$ & $%
{\small 0}$ & ${\small Q}_{L}{\small +u}_{L}^{c}{\small +e}_{L}^{c}$ & $%
{\small -}$ \\ \hline
${\small 10}_{2}{\small =(10}_{-,-}{\small )}_{0}$ & ${\small 0}$ & ${\small %
Q}_{L}{\small +u}_{L}^{c}{\small +e}_{L}^{c}$ & ${\small -}$ \\ \hline
${\small 10}_{3}{\small =(10}_{+,-}{\small )}_{0}$ & ${\small 0}$ & ${\small %
Q}_{L}{\small +u}_{L}^{c}{\small +e}_{L}^{c}$ & ${\small -}$ \\ \hline
${\small 10}_{4}{\small =(10}_{+,+}{\small )}_{0}$ & ${\small 0}$ & ${\small %
Q}_{L}{\small +2e}_{L}^{c}$ & ${\small +}$ \\ \hline
${\small 10}_{5}{\small =(10}_{+,+}{\small )}_{t_{5}}$ & ${\small 1}$ & $%
{\small -Q}_{L}{\small -2e}_{L}^{c}$ & ${\small -}$ \\ \hline
${\small 5}_{1}{\small =(5}_{+,+}{\small )}_{0}$ & ${\small 0}$ & ${\small -}
$ & ${\small +}$ \\ \hline
${\small 5}_{2}{\small =5}^{M_{3}}{\small =(5}_{-,-}{\small )}_{0}$ & $%
{\small 0}$ & ${\small -}\overline{d}_{L}^{c}{\small -}\overline{L}$ & $%
{\small -}$ \\ \hline
${\small 5}_{3}{\small =(5}_{-,+}^{H_{d}}{\small )}_{0}$ & ${\small 0}$ & $%
{\small -}\overline{H}_{d}$ & ${\small +}$ \\ \hline
${\small 5}_{4}{\small =(5}_{+,+}^{H_{u}}{\small )}_{0}$ & ${\small 0}$ & $%
{\small H}_{u}$ & ${\small +}$ \\ \hline
${\small 5}_{5}{\small =5}^{M_{1}}{\small =(5}_{-,-}{\small )}_{0}$ & $%
{\small 0}$ & ${\small -}\overline{d}_{L}^{c}{\small -}\overline{L}$ & $%
{\small -}$ \\ \hline
${\small 5}_{6}{\small =5}^{M_{2}}{\small =(5}_{+,-}{\small )}_{0}$ & $%
{\small 0}$ & ${\small -}\overline{d}_{L}^{c}{\small -}\overline{L}$ & $%
{\small -}$ \\ \hline
${\small 5}_{7}{\small =(5}_{+,+}{\small )}_{-t_{5}}$ & ${\small -1}$ & $%
{\small -}$ & ${\small +}$ \\ \hline
${\small 5}_{8}{\small =(5}_{-,-}{\small )}_{-t_{5}}$ & ${\small -1}$ & $%
{\small -}\overline{d}_{L}^{c}$ & ${\small +}$ \\ \hline
${\small 5}_{9}{\small =(5}_{+,-}{\small )}_{-t_{5}}$ & ${\small -1}$ & $%
\overline{d}_{L}^{c}$ & ${\small +}$ \\ \hline
${\small 5}_{10}{\small =(5}_{+,+}{\small )}_{-t_{5}}$ & ${\small -1}$ & $%
{\small -}$ & ${\small +}$ \\ \hline
\end{tabular}
\label{xy}
\end{equation}%
\begin{equation*}
\end{equation*}

$\bullet $\textbf{\ }\emph{Up-type quark Yukawa couplings}\newline
The allowed Yukawa couplings that are invariant under $\mathbb{D}_{4}\times
U_{1}^{\perp }$\ and preserving parity symmetry are:%
\begin{equation}
\begin{tabular}{l}
$(10_{+,-})_{0}\otimes (10_{+,-})_{0}\otimes (5_{+,+}^{H_{u}})_{0}$%
\end{tabular}
\label{33}
\end{equation}%
for third generation; and%
\begin{equation}
\begin{tabular}{l}
$(10_{-,-})_{0}\otimes (10_{-,-})_{0}\otimes (5_{+,+}^{H_{u}})_{0}$ \\ 
$(10_{+,+})_{0}\otimes (10_{+,+})_{0}\otimes (5_{+,+}^{H_{u}})_{0}$ \\ 
$(10_{+,+})_{0}\otimes (10_{-,-})_{0}\otimes (5_{+,+}^{H_{u}})_{0}\otimes
(\vartheta _{-,-})_{0}$ \\ 
$(10_{+,+})_{0}\otimes (10_{+,-})_{0}\otimes (5_{+,+}^{H_{u}})_{0}\otimes
(\vartheta _{+,-})_{0}$ \\ 
$(10_{-,-})_{0}\otimes (10_{+,-})_{0}\otimes (5_{+,+}^{H_{u}})_{0}\otimes
(\vartheta _{-,+})_{0}$%
\end{tabular}%
\end{equation}

$\bullet $\textbf{\ }\emph{Down-type quark Yukawa couplings}\newline
The Yukawa coupling down-type are:%
\begin{equation}
\begin{tabular}{ll}
$(10_{+,+})_{0}\otimes (\overline{5}_{-,-})_{0}\otimes
(5_{-,+}^{H_{d}})_{0}\otimes (\vartheta _{+,-})_{0}$ & 
\end{tabular}%
\end{equation}%
for third generation; and%
\begin{equation}
\begin{tabular}{ll}
$(10_{+,+})_{0}\otimes (\overline{5}_{-,-})_{0}\otimes
(5_{-,+}^{H_{d}})_{0}\otimes (\vartheta _{+,-})_{0}$ &  \\ 
$(10_{+,+})_{0}\otimes (\overline{5}_{+,-})_{0}\otimes
(5_{-,+}^{H_{d}})_{0}\otimes (\vartheta _{-,-})_{0}$ &  \\ 
$(10_{-,-})_{0}\otimes (\overline{5}_{-,-})_{0}\otimes
(5_{-,+}^{H_{d}})_{0}\otimes (\vartheta _{-,+})_{0}$ &  \\ 
$(10_{-,-})_{0}\otimes (\overline{5}_{-,-})_{0}\otimes
(5_{-,+}^{H_{d}})_{0}\otimes (\vartheta _{-,+})_{0}$ &  \\ 
$(10_{-,-})_{0}\otimes (\overline{5}_{+,-})_{0}\otimes (5_{-,+}^{H_{d}})_{0}$
&  \\ 
$(10_{+,-})_{0}\otimes (\overline{5}_{-,-})_{0}\otimes (5_{-,+}^{H_{d}})_{0}$
&  \\ 
$(10_{+,-})_{0}\otimes (\overline{5}_{-,-})_{0}\otimes (5_{-,+}^{H_{d}})_{0}$
&  \\ 
$(10_{+,-})_{0}\otimes (\overline{5}_{+,-})_{0}\otimes
(5_{-,+}^{H_{d}})_{0}\otimes (\vartheta _{-,+})_{0}$ & 
\end{tabular}%
\end{equation}%
\begin{equation*}
\end{equation*}%
For the neutrino sectors in both models II and III, the couplings are
embedded in the Dirac and Majorana operators as for model I; their mass
matrix depend on the choice of the localization of right neutrino in the
singlet curves $\vartheta _{\pm ,\pm }.$

\section{Appendix C: Monodromy and flavor symmetry}

We begin by recalling that in F-theory GUTs, quantum numbers of particle
fields and their gauge invariant interactions descend from an affine $E_{8}$
singularity in the internal Calabi-Yau Geometry: $CY4\sim \mathcal{E}%
\rightarrow \mathcal{B}_{3}$. The observed gauge bosons, the 4D matter
generations and the Yukawa couplings of standard model arise from symmetry
breaking of the underlying $E_{8}$ gauge symmetry of compactification of F-
theory to 4D space time. \newline
In this appendix, we use known results on F-theory GUTs to exhibit the link
between non abelian monodromy and flavor symmetry which relates the three
flavor generations of SM. First, we briefly describe how abelian monodromy
like $\mathbb{Z}_{p}$ appear in F-GUT models; then we study the extension to
non abelian discrete symmetries such the dihedral $\mathbb{D}_{4}$ we have
considered in present study.\ 

\subsection{ Abelian monodromy}

One of the interesting field realisations of the F-theory approach to GUT is
given by the remarkable $SU_{5}\times SU_{5}^{\perp }$ model with basic
features encoded in the internal geometry; in particular the two following
useful ones: $\left( i\right) $ the $SU_{5}\times SU_{5}^{\perp }$
invariance follows from a particular breaking way of $E_{8}$; and $\left(
ii\right) $ the full spectrum of the field representations of the model is
as in eq(\ref{dec}). From the internal CY4 geometry view, $SU_{5}$ and $%
SU_{5}^{\perp }$ have interpretation in terms of singularities; the $SU_{5}$
lives on the so called GUT surface $\mathcal{S}_{GUT}$; it appears in terms
of the singular locus of the following Tate form of the elliptic fibration $%
y^{2}=x^{3}+b_{5}xy+b_{4}x^{2}z+b_{3}yz^{2}+b_{2}xz^{3}+b_{0}z^{5}$; it is
the gauge symmetry visible in 4D space time of the GUT model. Quite
similarly, the $SU_{5}^{\perp }$ may be also imagined to have an analogous
geometric representation in the internal geometry; but with different
physical interpretation; it lives as well on a complex surface $\mathcal{S}%
^{\prime }$; another divisor of the base $\mathcal{B}_{3}$ of the complex
four dimensional elliptic CY4 fibration. Obviously these two divisors are
different, but intersect. Here, we want to focus on aspects of the
representations of $SU_{5}^{\perp }$ appearing in eq(\ref{dec}) and too
particulary on the associated matter curves $\Sigma _{t_{i}}$, $\Sigma
_{t_{i}+t_{j}}$, $\Sigma _{t_{i}-t_{j}}$; which are nicely described in the
spectral cover method using an extra spectral parameter $s$. If thinking of
the hidden $SU_{5}^{\perp }$ in terms of a broken symmetry by an abelian
flux or Higgsing down to its Cartan subgroup, the resulting symmetry of the
GUT model becomes $U\left( 1\right) ^{4}\times SU_{5}$ with\textrm{\footnote{%
\ Recall the three useful relations: $\left( a\right) $ Let $\vec{H}=\left(
H_{1},...,H_{4}\right) $ the generators of the U$\left( 1\right) _{i}$
charge factors and $E_{\pm \alpha _{i}}$ the step operators associated with
the simple roots $\vec{\alpha}_{i}$, then we have $\left[ E_{+\alpha
_{i}},E_{-\alpha _{i}}\right] =\vec{\alpha}_{i}.\vec{H}.$ $\left( b\right) $%
\ If denoting by $\left \vert \vec{\mu}\right \rangle $ a weight vector of
the fundamental representation of SU$_{5}^{\perp }$, then we have $\vec{%
\alpha}_{i}.\vec{H}\left \vert \vec{\mu}\right \rangle =\lambda
_{i}\left
\vert \vec{\mu}\right \rangle $ with $\lambda _{i}=\vec{\alpha}%
_{i}.\vec{\mu} $. $\left( c\right) $ using the 4 usual fundamental weight
vectors $\vec{\omega}_{i}$ dual to the 4 simple roots, the 5 weight vectors $%
\left \{ \vec{\mu}_{k}\right \} $ of the representation are: $\vec{\mu}_{1}=%
\vec{\omega}_{1},$ $\vec{\mu}_{2}=\vec{\omega}_{2}-\vec{\omega}_{1},$ $\vec{%
\mu}_{3}=\vec{\omega}_{3}-\vec{\omega}_{2},$ $\vec{\mu}_{4}=\vec{\omega}_{4}-%
\vec{\omega}_{3},$ $\vec{\mu}_{5}=-\vec{\omega}_{4}.$}} 
\begin{equation}
\begin{tabular}{lll}
$U\left( 1\right) ^{4}$ & $=$ & $U\left( 1\right) _{1}\times U\left(
1\right) _{2}\times U\left( 1\right) _{3}\times U\left( 1\right) _{4}$ \\ 
& $\equiv $ & $\dprod \nolimits_{i=1}^{4}U\left( 1\right) _{i}$%
\end{tabular}
\label{91}
\end{equation}%
The extra $U\left( 1\right) $'s in the breaking $U\left( 1\right) ^{4}\times
SU_{5}$ put constraints on the superpotential couplings of the effective low
energy model; the simultaneous existence of $U\left( 1\right) ^{4}$ is
phenomenologically undesirable since it does not allow a tree-level Yukawa
coupling for the top quark. This ambiguity is overcome by imposing abelian
monodromies among the $U\left( 1\right) $'s allowing the emergence of a rank
one fermion mass matrix structure; see eqs(\ref{c1}-\ref{c2}) given below. 
\newline
Following the presentation of section 2 of this paper, the spectral covers
describing the above invariance are given by polynomials with an\textrm{\ }%
affine variable $s$ as in eq(\ref{22}); see also (\ref{C5},\ref{c10},\ref%
{c11}). To fix the ideas, we consider monodromy properties of 10-plets $%
\Sigma _{t_{i}}$ encoded in the spectral cover equation%
\begin{equation}
\mathcal{C}_{5}:b_{5}+b_{3}s^{2}+b_{2}s^{3}+b_{4}s^{4}+b_{0}s^{5}=0
\end{equation}%
The location of the seven branes on GUT surface associated to this $SU_{5}$
representation is given by $b_{5}=0$. Using the method of \textrm{\cite%
{B5,D2,E1,E2},} the possible abelian monodromies are $\mathbb{Z}_{2}$, $%
\mathbb{Z}_{3}$, $\mathbb{Z}_{4},$ $\mathbb{Z}_{2}\times \mathbb{Z}_{3}$ and 
$\mathbb{Z}_{2}\times \mathbb{Z}_{2}$; they lead to factorizations of the $%
\mathcal{C}_{5}$ spectral cover as%
\begin{equation}
\mathcal{C}_{2}\times \left( \mathcal{C}_{1}\right) ^{3}\text{ \ },\text{ \ }%
\mathcal{C}_{3}\times \left( \mathcal{C}_{1}\right) ^{2}\text{ \ },\text{ \ }%
\mathcal{C}_{4}\times \mathcal{C}_{1}\text{ \ },\text{ \ }\mathcal{C}%
_{3}\times \mathcal{C}_{2}\text{ \ },\text{ \ }\left( \mathcal{C}_{2}\right)
^{2}\times \mathcal{C}_{1}
\end{equation}%
and to the respective identification of the weights $\left \{
t_{1},t_{2}\right \} $, $\left \{ t_{1},t_{2},t_{3}\right \} $, $\left \{
t_{1},t_{2},t_{3},t_{4}\right \} $, $\left \{ t_{1},t_{2}\right \} \cup
\left \{ t_{3},t_{4},t_{5}\right \} $ and $\left \{ t_{1},t_{2}\right \}
\cup \left \{ t_{3},t_{4}\right \} $.\newline
The algebraic equations for the matter curves $\Sigma _{t_{i}}$, $\Sigma
_{t_{i}+t_{j}}$, $\Sigma _{t_{i}-t_{j}}$ in terms of the $t_{i}$ weights
associated with the $SU_{5}^{\perp }$ fundamental representation are
respectively given by $t_{i}=0$; $\left( t_{i}+t_{j}\right) _{i<j}=0$ and $%
\pm \left( t_{i}-t_{j}\right) _{i<j}=0$; they are denoted like $10_{t_{i}},$ 
$\bar{5}_{t_{i}+t_{j}}$ and $1_{\pm \left( t_{i}-t_{j}\right) }$; see eq(\ref%
{matter}). \newline
As a first step to approach non abelian monodromies we are interested in
here, it is helpful to notice the two useful following things: $\left(
a\right) $ the homology 2-cycles in the CY4 underlying $SU_{5}\times U\left(
1\right) ^{4}$ invariance has monodromies captured by a finite discrete
group that can be used as a constraint in the modeling. $\left( b\right) $
from the view of phenomenology, these monodromies must be at least $\mathbb{Z%
}_{2}$ in order to have top- quark Yukawa coupling at tree level as noticed
before. Notice moreover that under this $\mathbb{Z}_{2}$, matter multiplets
of the $SU_{5}$ model\ split into two $\mathbb{Z}_{2}$ sectors\textrm{%
\footnote{%
\ In general we have two $\mathbb{Z}_{2}$ eigenstates: $t_{\pm }=\frac{1}{2}%
\left( t_{1}\pm t_{2}\right) $ with eigenvalues $\pm 1.$ While any function
of $t_{+}$ is $\mathbb{Z}_{2}$ invariant, only those functions depending on $%
\left( t_{-}\right) ^{2}$ which are symmetric with respect to $\mathbb{Z}%
_{2} $.}: }even and odd; for example the two tenplets $\left \{
10_{t_{1}},10_{t_{2}}\right \} $ are interchanged under $t_{1}%
\leftrightarrow t_{2}$; the corresponding eigenstates are given by $%
10_{t_{\pm }}$ with eigenvalues $\pm 1$. By requiring the identification $%
t_{1}\leftrightarrow t_{2},$ naively realised by setting $t_{1}=t_{2}=t$,
matter couplings in the model get restricted; therefore the off diagonal
tree level Yukawa coupling%
\begin{equation}
10_{t_{1}}.10_{t_{2}}.5_{-t_{1}-t_{2}}  \label{c1}
\end{equation}%
which is invariant under $SU_{5}\times U\left( 1\right) ^{4}$, becomes after 
$t_{1}\leftrightarrow t_{2}$ identification a diagonal top-quark interaction
invariant under $\mathbb{Z}_{2}$ monodromy. The resulting Yukawa coupling
reads as follows\textrm{\  \cite{D2,E1,E2}} 
\begin{equation}
10_{t}.10_{t}.5_{-2t}  \label{c2}
\end{equation}%
the other diagonal coupling $10_{0}.10_{0}.5_{-2t}$ is forbidden by the U$%
\left( 1\right) $ symmetry; \emph{see footnote 5}. Notice that for bottom-
quark the typical Yukawa coupling $10_{t}.\bar{5}_{t_{i}+t_{j}}.\bar{5}%
_{t_{k}+t_{l}}$ is allowed by $\mathbb{Z}_{2}$ while $10_{0}.\bar{5}%
_{t_{i}+t_{j}}.\bar{5}_{t_{k}+t_{l}}$ is forbidden.\newline
In this monodromy invariant theory, the symmetry of the model is given by $%
SU_{5}\times U\left( 1\right) ^{3}\times \mathbb{Z}_{2}$; it may be
interpreted as the invariance that remains after taking the coset with
respect to $\mathbb{Z}_{2}$; that is by a factorisation of type $G=H\times 
\mathbb{Z}_{2}$ with $H=G/\mathbb{Z}_{2}$. Indeed, starting from $%
SU_{5}\times U\left( 1\right) ^{4}$ and performing the two following
operations: $\left( i\right) $ use the traceless property of the \emph{%
fundamental} representation of $SU_{5}^{\bot }$ to think of (\ref{91}) like 
\begin{equation}
U\left( 1\right) ^{4}=\left( \dprod \nolimits_{i=1}^{5}U\left( 1\right)
_{t_{i}}\right) /\mathcal{J}
\end{equation}%
with $\mathcal{J}=\left \{ t_{i}\text{ \ }|\text{ \ }%
t_{1}+t_{2}+t_{3}+t_{4}+t_{5}=0\right \} \simeq U\left( 1\right) _{diag}$;
this property is a rephrasing of the usual $U\left( 5\right) $
factorisation; i.e $SU\left( 5\right) =\frac{U\left( 5\right) }{U\left(
1\right) }$. $\left( ii\right) $ substitute the product $U\left( 1\right)
_{t_{1}}\times U\left( 1\right) _{t_{2}}$ by the reduced abelian group $%
U\left( 1\right) _{t}\times \mathbb{Z}_{2}$ where monodromy group has been
explicitly exhibited. In this way of doing, one disposes of a discrete group
that may be promoted to a symmetry of the fields spectrum. To that purpose,
we need two more steps: first explore all allowed discrete monodromy groups;
and second study how to link these groups to flavor symmetry. For the
extension of above $\mathbb{Z}_{2}$, a similar method can be used to build
other prototypes; in particular models with abelian discrete symmetries like 
$SU_{5}\times U\left( 1\right) ^{5-k}\times \mathbb{Z}_{k}$ with $k=3,4,5;$\
or more generally as 
\begin{equation}
SU_{5}\times U\left( 1\right) ^{5-p-q}\times \mathbb{Z}_{p}\times \mathbb{Z}%
_{q}  \label{v}
\end{equation}%
where $1<p+q\leq 5$ and $\mathbb{Z}_{1}\equiv I_{id}$, $\mathbb{Z}_{0}\equiv
I_{id}$. Notice that the discrete groups in eq(\ref{v}) are natural
extensions of those of the theories with $SU_{5}\times U\left( 1\right)
^{5-k}\times \mathbb{Z}_{k}$ symmetry; and that the condition $p+q\leq 5$\
on allowed abelian monodromies is intimately related with the Weyl symmetry $%
\mathcal{W}_{SU_{5}^{\perp }}$ of $SU_{5}^{\perp }$. Therefore, we end with
the conclusion that the $\mathbb{Z}_{p}\times \mathbb{Z}_{q}$ abelian
discrete groups in above relation are in fact particular subgroups of the
non abelian symmetric group $\mathcal{W}_{SU_{5}^{\perp }}\simeq \mathbb{S}%
_{5}$.\  \ 

\subsection{Non abelian monodromy and flavor symmetry}

To begin notice that the appearance of abelian discrete symmetry in the $%
SU_{5}$ based GUT models with invariance (\ref{v}) is remarkable and
suggestive. It is remarkable because these finite discrete symmetries have a
geometric interpretation in the internal CY4; and constitutes then a
prediction of F- theory GUT. It is suggestive since such kind of discrete
groups, especially their non abelian generalisation, are highly desirable in
phenomenology; particularly in playing the role of a flavor symmetry. In
this regards, it is interesting to recall that it is quite well established
that neutrino flavors are mixed; and this property requires non abelian
discrete group symmetries like the alternating $\mathbb{A}_{4}$ group which
has been subject to intensive research during last decade \textrm{\cite%
{U1,U2,U3,U4,U5}}. \newline
Following\textrm{\ }the conjecture of\textrm{\  \cite{B3,B4}, }non-abelian
discrete symmetries may be reached in F- theory GUT by assuming the
existence of a non abelian flux breaking the $SU_{5}^{\perp }$\ down to a
non abelian group\textrm{\ }$\Gamma \subset \mathcal{W}_{SU_{5}^{\perp }}$.
In this view, one may roughly think about the $\mathbb{Z}_{p}\times \mathbb{Z%
}_{q}$ group of (\ref{v}) as special symmetries of a family of $SU_{5}$
based GUT models with invariance given by 
\begin{equation}
SU_{5}\times U\left( 1\right) ^{5-k}\times \Gamma _{k}  \label{gk}
\end{equation}%
where now $\Gamma _{k}$ is a subgroup of $\mathbb{S}_{5}$ that can be a non
abelian discrete group. In this way of doing, one then distinguishes several 
$SU_{5}$ GUT models with non abelian discrete symmetries classified by the
number of surviving $U\left( 1\right) $'s. In presence of no $U\left(
1\right) $ symmetry, we have prototypes like $SU_{5}\times \mathbb{S}_{5}$
and $SU_{5}\times \mathbb{A}_{5}$; while for a theory with one $U\left(
1\right) $, we have symmetries as follows%
\begin{eqnarray}
&&SU_{5}\times U\left( 1\right) \times \mathbb{S}_{4}  \notag \\
&&SU_{5}\times U\left( 1\right) \times \mathbb{A}_{4}  \label{mo} \\
&&SU_{5}\times U\left( 1\right) \times \mathbb{D}_{4}  \notag
\end{eqnarray}%
where the alternating $\mathbb{A}_{4}$ and dihedral $\mathbb{D}_{4}$ are the
usual subgroups of $\mathbb{S}_{4}$ itself contained in $\mathbb{S}_{5}$. In
the case with two $U\left( 1\right) $'s, monodromy gets reduced like $%
SU_{5}\times U\left( 1\right) ^{2}\times \mathbb{S}_{3}$. \newline
Moreover, by using non abelian discrete monodromy groups $\Gamma _{k}$, one
ends with an important feature; these discrete groups have, in addition to
trivial representations, higher dimensional representations that are
candidates to host more than one matter generation. Under transformations of 
$\Gamma _{k}$; the generations get in general mixed. Therefore the non
abelian $\Gamma _{k}$'s in particular those having 3- and/or 2-dimensional
irreducible representations may be naturally interpreted in terms of flavor
symmetry.\newline
In the end of this section, we would like to add a comment on the splitting
spectral cover construction regarding non abelian discrete monodromy groups
like $\mathbb{A}_{4}$ and $\mathbb{D}_{4}$. In the models\ (\ref{mo}), the
spectral cover for the fundamental $\mathcal{C}_{5}$ is factorised like $%
\mathcal{C}_{5}=\mathcal{C}_{4}\times \mathcal{C}_{1}$ and similarly for $%
\mathcal{C}_{10}$ and $\mathcal{C}_{20}$ respectively associated with the
antisymmetric and the adjoint of $SU_{5}^{\perp }$. In the $\mathcal{C}%
_{4}\times \mathcal{C}_{1}$ splitting, we have%
\begin{eqnarray}
\mathcal{C}_{4} &=&a_{5}s^{4}+a_{4}s^{3}+a_{3}s^{2}+a_{2}s+a_{1}  \notag \\
\mathcal{C}_{1} &=&a_{7}s+a_{6}
\end{eqnarray}%
where the $a_{i}$'s are complex holomorphic sections. For the generic case
where the coefficients $a_{i}$ are free, the splitted spectral cover $%
\mathcal{C}_{4}\times \mathcal{C}_{1}$ has an $\mathbb{S}_{4}$ monodromy. To
have splitted spectral covers with monodromies given by the subgroups $%
\mathbb{A}_{4}$ and $\mathbb{D}_{4}$, one needs to put constraints on the $%
a_{i}$'s; these conditions have been studied in \textrm{\cite{B2,B4}}; they
are non linear relations given by Galois theory. Indeed, starting from $%
SU_{5}\times SU_{5}^{\perp }$ model and borrowing tools from \textrm{\cite%
{B4}, }the breaking of\textrm{\ }$SU_{5}\times SU_{5}^{\perp }$\textrm{\ }%
down to $SU_{5}\times \mathbb{D}_{4}\times U\left( 1\right) $ model
considered in this paper may be imagined in steps as follows: first breaking%
\textrm{\ }$SU_{5}^{\perp }$\textrm{\ }to\textrm{\ }subgroup\textrm{\ }$%
SU_{4}^{\perp }\times U\left( 1\right) $\textrm{\ }by an abelian flux; then
breaking the $SU_{4}^{\perp }$\ part to the discrete group $\mathbb{S}_{4}$%
\textrm{\ }by a non-abelian flux as conjectured in\textrm{\  \cite{B3,B4};}
deformations of this flux lead to subgroups of\textrm{\ }$\mathbb{S}_{4}$%
\textrm{.\ }To obtain the constraints describing the $\mathbb{D}_{4}$
splitted spectral cover\textrm{\ }descending from\textrm{\ }$\mathcal{C}%
_{4}\times \mathcal{C}_{1}$\textrm{, }we use Galois theory; they are given
by a set of two constraints on the holomorphic sections of $\mathcal{C}%
_{4}\times \mathcal{C}_{1}$; and are obtained as follows:\newline
$\left( i\right) $ the first constraint comes from the discriminant $\Delta
_{\mathcal{C}_{4}}$ of the spectral cover $\mathcal{C}_{4}$ which should not
be a perfect square; that is $\Delta _{\mathcal{C}_{4}}\neq \delta ^{2}$.
The explicit expression of the discriminant of $\mathcal{C}_{4}$ has been
computed in literature; so we have%
\begin{equation}
108a_{0}(\lambda a_{6}^{2}+4a_{1}a_{7})(\kappa ^{2}a_{7}^{2}+a_{0}(\lambda
a_{6}^{2}+4a_{1}a_{7}))^{2}\neq \delta ^{2}
\end{equation}%
where dependence into $a_{6}$ and $a_{7}$ is due to solving the traceless
condition $b_{1}=0$ in $\mathcal{C}_{5}=\mathcal{C}_{4}\times \mathcal{C}%
_{1} $. $\left( ii\right) $ the second constraint is given by a condition on
the cubic resolvent which should be like $\left. R_{\mathcal{C}_{4}}\left(
s\right) \right \vert _{s=0}=0$. The expression of $R_{\mathcal{C}%
_{4}}\left( s\right) $ is known; it leads to 
\begin{equation}
a_{2}^{2}a_{7}=a_{1}\left( a_{0}a_{6}^{2}+4a_{3}a_{7}\right)
\end{equation}%
where $a_{0}$\ is a parameter introduced by the solving the traceless
condition $b_{1}=0$; for explicit details see \textrm{\cite{B4}.}

\begin{acknowledgement}
\ Saidi thanks ICTP Trieste- Italy for kind hospitality where part of this
work has been done.
\end{acknowledgement}


\begin{thebibliography}{99}
\bibitem{A1} C. Vafa, \emph{Evidence for F theory}, Nucl. Phys. B 469 (1996)
403 arXiv:hep-th/9602022.

\bibitem{A2} R.~Donagi and M.~Wijnholt, \emph{Model Building with F-Theory}%
,\ Adv.\ Theor.\ Math.\ Phys.\  \textbf{15} (2011) 1237 [arXiv:0802.2969
[hep-th]].

\bibitem{A3} C.~Beasley, J.~J.~Heckman and C.~Vafa, \emph{GUTs and
Exceptional Branes in F-theory - I},\ JHEP \textbf{0901} (2009) 058
[arXiv:0802.3391 [hep-th]].

\bibitem{A4} C.~Beasley, J.~J.~Heckman and C.~Vafa, \emph{GUTs and
Exceptional Branes in F-theory - II: Experimental Predictions},\ JHEP 
\textbf{0901} (2009) 059 [arXiv:0806.0102 [hep-th]].

\bibitem{A5} T.~Weigand, \emph{Lectures on F-theory compactifications and
model building},\ Class.\ Quant.\ Grav.\  \textbf{27} (2010) 214004
[arXiv:1009.3497 [hep-th]].

\bibitem{A6} R.~Donagi and M.~Wijnholt, \emph{Higgs Bundles and UV
Completion in F-Theory},\ Commun.\ Math.\ Phys.\  \textbf{326} (2014) 287
[arXiv:0904.1218 [hep-th]].

\bibitem{A7} R.~Donagi and M.~Wijnholt, \emph{Breaking GUT Groups in F-Theory%
},\ Adv.\ Theor.\ Math.\ Phys.\  \textbf{15} (2011) 1523 [arXiv:0808.2223
[hep-th]].

\bibitem{A8} R.~Blumenhagen, T.~W.~Grimm, B.~Jurke and T.~Weigand, \emph{%
Global F-theory GUTs},\ Nucl.\ Phys.\ B \textbf{829} (2010) 325
[arXiv:0908.1784].

\bibitem{A9} J. Tate, \textquotedblleft Algorithm for Determining the Type
of a Singular Fiber in an Elliptic Pencil,\textquotedblright \ in Modular
Functions of One Variable IV, Lecture Notes in Math. vol. 476,
Springer-Verlag, Berlin (1975).

\bibitem{A91} Sven Krippendorf, Sakura Schafer-Nameki, Jin-Mann Wong,
arXiv:1507.05961,

\bibitem{A10} F.~Denef, \emph{Les Houches Lectures on Constructing String
Vacua},\ arXiv:0803.1194.

\bibitem{B1} J.~Marsano, N.~Saulina and S.~Schafer-Nameki, \emph{%
Monodromies, Fluxes, and Compact Three-Generation F-theory GUTs},\ JHEP 
\textbf{0908} (2009) 046 [arXiv:0906.4672 [hep-th]].

\bibitem{B11} Florent Baume, Eran Palti, Sebastian Schwieger, \emph{On E}$%
_{8}$\emph{\ and F-Theory GUTs}, arXiv:1502.03878,

\bibitem{B2} Athanasios Karozasy, Stephen F. King, George K. Leontaris and
Andrew Meadowcroft, \emph{Discrete Family Symmetry from F-Theory GUTs},\
[arXiv:1406.6290v4 [hep-ph]],

\bibitem{B3} I.~Antoniadis and G.~K.~Leontaris, \emph{Neutrino mass textures
from F-theory},\ Eur.\ Phys.\ J.\ C \textbf{73} (2013) 2670 [arXiv:1308.1581
[hep-th]].

\bibitem{B4} Athanasios Karozasy, Stephen F. King, George K. Leontaris,
Andrew K.Meadowcroft, \emph{Phenomenological implications of a minimal
F-theory GUT with discrete symmetry}, [arXiv1505.009337v3 [hep-th]].

\bibitem{B12} Anshuman Maharana, Eran Palti, \emph{Models of Particle
Physics from Type IIB String Theory and F-theory}: A Review, arXiv:1212.0555,

\bibitem{B5} J.~J.~Heckman, A.~Tavanfar and C.~Vafa, \emph{The Point of E(8)
in F-theory GUTs},\ JHEP \textbf{1008} (2010) 040 [arXiv:0906.0581 [hep-th]].

\bibitem{B51} Hirotaka Hayashi, Teruhiko Kawano, Yoichi Tsuchiya, Taizan
Watari, HEP 1008:036, 2010, [arXiv:0910.2762 [hep-th]].

\bibitem{C0} A.~Font, L.~E.~Ibanez, F.~Marchesano and D.~Regalado, \emph{%
Non-perturbative effects and Yukawa hierarchies in F-theory SU(5) Unification%
},\ JHEP \textbf{1303} (2013) 140 [Erratum-ibid.\  \textbf{1307} (2013) 036]
[arXiv:1211.6529 [hep-th]].

\bibitem{C1} S. Cecotti, M. C. N. Cheng, J. J. Heckman and C.Vafa,
arXiv:0910.0477[hep-th].

\bibitem{C2} L. Aparicio, A. Font, L. E. Ibanez and F. Marchesano, JHEP 1108
(2011)152 [arXiv:1104.2609[hep-th]].

\bibitem{C3} Asan Damanik, \emph{Non zero }$\theta _{13}$\emph{\ and
Neutrino Masses from Modified TBM},\ EJTP 11, No. 31 (2014)125--130P.

\bibitem{C4} Adamson et. al. (MINOS Collab.), Phys. Rev. Lett. 107, 181802
(2011), arXiv:1108.0015[hep-ex]. doi: 10.1103/Phys Rev Lett.107.181802

\bibitem{C5} Y. Abe et al. [DOUBLE-CHOOZ Collaboration], Phys. Rev. Lett.
108, 131801 (2012) [arxiv: 1112.6353 [hep-ex]].

\bibitem{D1} H.~Hayashi, T.~Kawano, Y.~Tsuchiya and T.~Watari, \emph{Flavor
Structure in F-theory Compactifications},\ JHEP \textbf{1008} (2010) 036
[arXiv:0910.2762 [hep-th]]

\bibitem{D2} I.~Antoniadis and G.~K.~Leontaris, \emph{Building SO(10) models
from F-theory},\ JHEP \textbf{1208} (2012) 001 [arXiv:1205.6930 [hep-th]].

\bibitem{D3} Patrick Morandi, \emph{Field and Galois Theory}, Springer,1996.

\bibitem{D4} Michael Artin, \emph{Algebra}, Prentice-Hall Inc.1991.

\bibitem{E1} J. C. Callaghan, S. F. King, G. K. Leontaris and G. G. Ross, 
\emph{Towards a Realistic F-theory GUT},\ [arXiv:1109.1297[hep-th]].

\bibitem{E2} Emilian Dudas and Eran Palti, \emph{On hypercharge flux and
exotics in F-theory GUTs},\ [arXiv:1007.1297[hep-th]].

\bibitem{U1} S.~F.~King and C.~Luhn, \emph{Neutrino Mass and Mixing with
Discrete Symmetry},\ Rept.\ Prog.\ Phys.\  \textbf{76} (2013) 056201
[arXiv:1301.1340 [hep-ph]].

\bibitem{U2} Guido Altarelli, Ferruccio Feruglio, \emph{Discrete Flavor
Symmetries and Models of Neutrino Mixing}, [arXiv:1000.0211[hep-th]].

\bibitem{U3} S.~F.~King, A.~Merle, S.~Morisi, Y.~Shimizu and M.~Tanimoto, 
\emph{Neutrino Mass and Mixing: from Theory to Experiment},\ New J.\ Phys.\ 
\textbf{16} (2014) 045018 [arXiv:1402.4271 [hep-ph]].

\bibitem{E4} H. Hayashi, T. Kawano, R. Tatar and T. Watari, \emph{%
Codimension-3 Singularities and Yukawa Couplings in F-theory},\
[arXiv:0901.4941 [hep-th]].

\bibitem{E5} S.~F.~King, G.~K.~Leontaris and G.~G.~Ross, \emph{Family
symmetries in F-theory GUTs},\ Nucl.\ Phys.\ B \textbf{838} (2010) 119
[arXiv:1005.1025 [hep-ph]].

\bibitem{E6} G.~K.~Leontaris, \emph{Aspects of F-Theory GUTs},\ PoS CORFU 
\textbf{2011} (2011) 095 [arXiv:1203.6277 [hep-th]].

\bibitem{E7} J.~Marsano, \emph{Hypercharge Flux, Exotics, and Anomaly
Cancellation in F-theory GUTs},\ Phys.\ Rev.\ Lett.\  \textbf{106} (2011)
081601 [arXiv:1011.2212 [hep-th]].

\bibitem{E8} Wiliam Fulton, Joe Harris, \emph{Young Tabeaux with
Applications to Representation Theory and Geometry}, Springer-Verlag (1991).

\bibitem{E9} H. Ishimori, T. Kobayashi, H. Ohki, H. Okada, Y. Shimizu and M.
Tanimoto, \emph{Non-Abelian Discrete Symmetries in Particle Physics},\
[arXiv:1003.3552 [hep-th]],

\bibitem{E10} R. Ahl Laamara, M. Miskaoui, E.H Saidi, \emph{Building SO}$%
_{10}$ \emph{models with} $\mathbb{D}_{4}$\emph{\ symmetry},
[arXiv:1511.03166[hep-th]].

\bibitem{E11} El Hassan Saidi, \emph{On Building superpotentials in F- GUTs}%
, Prog. Theor. Exp. Phys. (2016) 013 B07, arXiv:1512.02530.

\bibitem{EE0} G.~K.~Leontaris and G.~G.~Ross, \emph{Yukawa couplings and
fermion mass structure F-theory GUTs}, JHEP 02 (2011) 108 [arXiv:1009.6000
[hep-ph]].

\bibitem{EE1} C. Ludeling, H. P. Nilles and C. C. Stephan, \emph{The
Potential Fate of Local Model Building}, [arXiv:1101.3346 [hep-ph]].

\bibitem{X0} J.~J.~Heckman, \emph{Particle Physics Implications of F-theory}%
,\ Ann.\ Rev.\ Nucl.\ Part.\ Sci.\  \textbf{60} (2010) 237 [arXiv:1001.0577
[hep-th]].

\bibitem{X1} T.~W.~Grimm, \emph{The N=1 effective action of F-theory
compactifications},\ Nucl.\ Phys.\ B \textbf{845} (2011) 48 [arXiv:1008.4133
[hep-th]].

\bibitem{X2} C. M. Chen, J. Knapp, M. Kreuzer and C. Mayrhofer, \emph{Global
SO(10) F-theory GUTs}, JHEP 1010 (2010)057 [arXiv:1005.5735[hep-th]].

\bibitem{X3} J.~C.~Callaghan and S.~F.~King, \emph{E6 Models from F-theory,}%
\ JHEP \textbf{1304} (2013) 034 [arXiv:1210.6913 [hep-ph]].

\bibitem{X4} R.~Tatar and W.~Walters, \emph{GUT theories from Calabi-Yau
4-folds with SO(10) Singularities},\ JHEP \textbf{1212} (2012) 092
[arXiv:1206.5090 [hep-th]].

\bibitem{X5} J.~C.~Callaghan, S.~F.~King and G.~K.~Leontaris, \emph{Gauge
coupling unification in E6 F-theory GUTs with matter and bulk exotics from
flux breaking},\ JHEP \textbf{1312} (2013) 037 [arXiv:1307.4593 [hep-ph]].

\bibitem{S1} El Hassan Saidi, \emph{Breaking discrete symmetries in F-GUT},
LPHE-MS -1511,

\bibitem{U4} Fredrik Bj\"{o}rkeroth, Francisco J. de Anda, Ivo de Medeiros
Varzielas, Stephen F. King, Journal-ref: JHEP 06 (2015) 141,
arXiv:1503.03306,

\bibitem{U5} S.~F.~King, \emph{Neutrino mass models},\ Rept.\ Prog.\ Phys.\ 
\textbf{67} (2004) 107 [hep-ph/0310204].

\bibitem{C6} K. Abe et al. (T2K Collab.), Phys. Rev. Lett. 107, 041801
(2011), arXiv:1106.2822 [hep-ph]. doi: 10.1103/PhysRevLett.107.041801

\bibitem{C7} F. P. An et al, Phys. Rev. Lett. 108, 171803 (2012),
arXiv:1203.1669v2 [hep-ex]. doi:10.1103/PhysRevLett.108.171803

\bibitem{C8} J. K. Ahn et al. (RENO Collab.), Phys. Rev. Lett. 108, 191802
(2012), arXiv: 1204.0626v2 [hep-ex]. doi: 10.1103/PhysRevLett.108.191802

\bibitem{C9} S. Boudjemaa and S.F.King, \emph{Deviations from Tri-bimaximal
Mixing: Charged Lepton Corrections and Renormalization Group Running},
[arXiv:0808.2782v3 [hep-th]],

\bibitem{F1} J. Marsano, N. Saulina and Sakura Schafer Nameki,
\textquotedblleft \emph{Compact F-theory GUTs with U(1)}$_{PQ}$%
,\textquotedblright \ [arXiv:0912.0272v2 [hep-th]].

\bibitem{F2} J. Marsano, N. Saulina and Sakura Schafer Nameki,
\textquotedblleft \emph{F-theory Compactifications for Supersymmetric GUTs}%
,\textquotedblright \ [arXiv:0904.3932v3 [hep-th]],

\bibitem{F3} S. Krippendorf, D. K. M. Penaa, P. K. Oehlmanna and F. Ruehle,
\textquotedblleft \emph{Rational F-theory GUTs without exotics}%
,\textquotedblright \ [arXiv:1401.5084v1 [hep-th]],

\bibitem{F4} T. W. Grimm and H. Hayashi, \textquotedblleft \emph{F-theory
fluxes, Chirality and Chern-Simons theories},\textquotedblright \
[arXiv:1111.1232v2 [hep-th]],

\bibitem{F5} E. Dudas and E. Palti, \textquotedblleft \emph{Froggatt-Nielsen
models from E(8) in F-theory GUTs},\textquotedblright \ [arXiv:0912.0853
[hep-th]],

\bibitem{F6} M. Cvetic, T. W. Grimm and D. Klevers, \textquotedblleft \emph{%
Anomaly Cancellation And Abelian Gauge Symmetries In F-theory}%
,\textquotedblright \ [arXiv:1210.6034v2 [hep-th]].

\bibitem{Z0} R. N. Mohapatra, \emph{Neutrino mass and Grand Unification of
flavor}, [arXiv:1007.1633 [hep-ph]].

\bibitem{Z1} K. S. Babu and S. M. Barr, \emph{Natural Gauge Hierarchy in
SO(10)}, [arXiv:9402291v1 [hep-ph]].

\bibitem{Z2} Z. Berezhiani, M. Chianese, G. Mielec and S. Morisi, \emph{%
Chances for SUSY-GUT in the LHC Epoch}, [arXiv:1505.04950 [hep-ph]].
\end{thebibliography}
\end{document}